\def\ra{\rangle}
\def\la{\langle}
\def\be{\begin{equation}}
\def\ee{\end{equation}}
\def\ba{\begin{array}}
\def\ea{\end{array}}
\def\Cb{{\Bbb C}}

\documentclass[12pt]{article}
\usepackage{amssymb,epsfig,graphicx}
\begin{document}
\newtheorem{theorem}{\textbf Theorem}[section]
\def\thetheorem{\thesection.\arabic{theorem}}
\newtheorem{definition}{\textbf Definition}[section]
\def\thedefinition{\thesection.\arabic{definition}}
\newtheorem{example}{\textbf Example}[section]
\def\theexample{\thesection.\arabic{example}}
\newtheorem{lemma}{\textbf Lemma}[section]
\def\thelemma{\thesection.\arabic{lemma}}
\def\theequation{\thesection.\arabic{equation}}
\newtheorem{corollary}{\textbf Corollary}[section]
\def\thecorollary{\thesection.\arabic{theorem}}

\baselineskip=18pt

\begin{center}
{\LARGE \textbf{Quantum Entanglement: Separability, Measure, Fidelity of Teleportation and Distillation}}
 \vspace{4ex}
 \end{center}
\begin{center}
Ming Li$^{1}$, Shao-Ming Fei$^{2,3}$ and Xianqing Li-Jost$^{3}$\vspace{2ex}

\begin{minipage}{5.5in}

\small $~^{1}$ {\small College of Mathematics and Computational
Science, China University of Petroleum, 257061 Dongying, China}

\small $~^{2}$ {\small Department of Mathematics, Capital Normal
University, 100037 Beijing, China}

\small $~^{3}$ {\small Max-Planck-Institute for Mathematics in the
Sciences, 04103 Leipzig, Germnay}

\end{minipage}
\end{center}

\vspace{0.3cm} Quantum entanglement plays crucial roles in quantum
information processing. Quantum entangled states have become the key
ingredient in the rapidly expanding field of quantum information
science. Although the nonclassical nature of entanglement has been
recognized for many years, considerable efforts have been taken to
understand and characterize its properties recently. In this review,
we introduce some recent results in the theory of quantum entanglement.
In particular separability criteria based on the Bloch representation,
covariance matrix, normal form and entanglement witness;
lower bounds, subadditivity property of concurrence and tangle;
fully entangled fraction related to the
optimal fidelity of quantum teleportation and entanglement
distillation will be discussed in detail.

\tableofcontents

\section {\bf Introduction}

Entanglement is the characteristic trait of quantum mechanics, and
it reflects the property that a quantum system can simultaneously
appear in two or more different states $\cite{einstein}$. This
feature implies the existence of global states of composite system
which cannot be written as a product of the states of individual
subsystems. This phenomenon \cite{peres1}, now known as ``quantum
entanglement", plays crucial roles in quantum information processing
\cite{nielsen}. Quantum entangled states have become the key
ingredient in the rapidly expanding field of quantum information
science, with remarkable prospective applications such as quantum
computation \cite{nielsen, di}, quantum teleportation
\cite{teleportation,fefandtel}, dense coding \cite{dense}, quantum
cryptographic schemes \cite{schemes}, entanglement swapping
\cite{swapping} and remote states preparation (RSP)
\cite{RSP1,RSP2,RSP3,RSP4}. All such effects are based on
entanglement and have been demonstrated in pioneering
experiments.

It has become clear that entanglement is not only the subject of
philosophical debates, but also a new quantum resource for tasks
which can not be performed by means of classical resources.
Although considerable efforts have been taken to
understand and characterize the properties of quantum entanglement recently,
the physical character and mathematical structure of entangled
states have not been satisfactorily understood yet \cite{rmp,phyreport}.
In this review we mainly introduce some recent results related to our researches
on several basic questions in this subject:

\begin{enumerate}
{\sf \item[{(1)}] Separability of quantum states}
\end{enumerate}

We first discuss the separability of a quantum
states, namely, for a given quantum state, how can we know whether
or not it is entangled.

For pure quantum states, there are many ways to verify the separability.
For instance for a bipartite pure quantum state the separability is easily
determined in terms of its Schmidt numbers. For multipartite pure states, the
generalized concurrence given in \cite{anote} can be used to
judge if the state is separable or not. In addition
separable states must satisfy all possible Bell inequalities \cite{wer}.

For mixed states we still have no general criterion. The well-known
PPT (partial positive transposition) criterion was proposed by Peres in 1996
\cite{ppt}. It says that for any bipartite separable quantum state
the density matrix must be positive under partial transposition.
By using the method of positive maps Horodeckis
\cite{ho96} showed that the Peres' criterion is also sufficient for
$2\times 2$ and $2\times3$ bipartite systems. And for higher dimensional
states, the PPT criterion is only necessary. Horodecki \cite{ho232}
has constructed some classes entangled states with
positive partial transposes for $3\times 3$ and $2\times 4$ systems.
States of this kind are said to be bound entangled (BE). Another
powerful operational criterion is the realignment
criterion \cite{ru02,ChenQIC03}. It demonstrates a remarkable
ability to detect many bound entangled states and even genuinely
tripartite entanglement \cite{3h0206008}. Considerable
efforts have been made in finding stronger variants and
multipartite generalizations for this criterion
\cite{generalizerealignment,chenkai}. It was shown that PPT
criterion and realignment criterion are equivalent to the
permutations of the density matrix's indices \cite{3h0206008}.
Another important criterion for separability is the reduction
criterion \cite{2hPRA99,cag99}. This criterion is equivalent to the PPT criterion for
$2\times N$ composite systems. Although it is generally weaker than
the PPT, the reduction criteria has tight relation to the distillation
of quantum states.

There are also some other necessary criteria for separability.
Nielsen et al. \cite{nielson01} presented a necessary criterion
called majorization: the decreasing ordered vector of
the eigenvalues for $\rho$ is majorized by that of $\rho^{A_1}$ or
$\rho^{A_2}$ alone for a separable state. i.e. if a state $\rho$ is
separable, then
$\lambda^\downarrow_{\rho}\prec\lambda^\downarrow_{\rho^{A_1}}$,
$\lambda^\downarrow_{\rho}\prec\lambda^\downarrow_{\rho^{A_2}}$.
Here $\lambda^\downarrow_{\rho}$ denotes the decreasing ordered
vector of the eigenvalues of $\rho$. A $d$-dimensional vector
$x^\downarrow$ is majorized by $y^\downarrow$, $x^\downarrow\prec
y^\downarrow$, if $\sum_{j=1}^k x^\downarrow_j\leq \sum_{j=1}^k
y^\downarrow_j$ for $k=1,\ldots,d-1$ and the equality holds for
$k=d$. Zeros are appended to the vectors
$\lambda^\downarrow_{\rho^{A_1,A_2}}$ such that their dimensions
equal to the one of $\lambda^\downarrow_{\rho}$.

In Ref. \cite{ho232}, another necessary criterion called range criterion was
given. If a bipartite state $\rho$ acting on the space ${\cal H}_{A}\otimes{\cal H}_{B}$ is
separable, then there exists a family of product vectors
$\psi_i\otimes\phi_i$ such that: (i) they span the range of $\rho$;
(ii) the vector $\{\psi_i\otimes \phi^*_i\}^k_{i=1}$ span the range of
 $\rho^{T_{B}}$, where $*$ denotes complex conjugation in the basis
 in which partial transposition was performed, $\rho^{T_{B}}$ is the partially transposed
 matrix of $\rho$ with respect to the subspace $B$.
In particular, any of the vectors $\psi_i\otimes \phi^*_i$
belongs to the range of $\rho$.

Recently, some elegant results for the separability problem have
been derived. In \cite{hofmann,guhne1,117903}, a
separability criteria based on the local uncertainty relations (LUR)
was obtained. The authors show that for any separable state
$\rho\in{\cal{H}}_A\otimes{\cal{H}}_B$, $$1-\sum\limits_{k}\la
G_{k}^{A}\otimes G_{k}^{B}\ra-\frac{1}{2}\la G_{k}^{A}\otimes I -
I\otimes G_{k}^{B}\ra^{2}\geq 0,$$ where $G_{k}^{A}$ or $G_{k}^{B}$
are arbitary local orthogonal and normalized operators (LOOs) in
${\cal{H}}_A\otimes{\cal{H}}_B$. This
criterion is strictly stronger than the realignment criterion. Thus
more bound entangled quantum states can be recognized by the LUR
criterion. The criterion is optimized in \cite{optloo} by choosing
the optimal LOOs. In \cite{vicente1} a
criterion based on the correlation matrix of a state has been presented. The
correlation matrix criterion is shown to be independent of PPT
and realignment criterion \cite{vicente2}, i.e. there exist
quantum states that can be recognized by correlation criterion while
the PPT and realignment criterion fail. The covariance matrix of a
quantum state is also used to study separability in \cite{guhne}. It
has been shown that the LUR criterion,
including the optimized one, can be derived from the covariance
matrix criterion \cite{guhnecov2}.

\begin{enumerate}
{\sf \item[{(2)}] Measure of quantum entanglement}
\end{enumerate}

One of the most difficult and fundamental problems in entanglement
theory is to quantify entanglement.  The initial idea to quantify
entanglement was connected with its usefulness in terms of
communication \cite{fm}. A good entanglement measure has to fulfill
some conditions \cite{db01}. For bipartite quantum systems, we have
several good entanglement measures such as Entanglement of
Formation(EOF), Concurrence, Tangle ctc. For two-quibt systems it
has been proved that EOF is a monotonically increasing function of
the concurrence and an elegant formula for the concurrence was
derived analytically by Wootters \cite{wotters}. However with the
increasing dimensions of the subsystems the computation of EOF and
concurrence become formidably difficult. A few explicit analytic
formulae for EOF and concurrence have been found only for some
special symmetric states
\cite{Terhal-Voll2000,fjlw,fl,fwz,Rungta03}.

The first analytic lower bound of concurrence for arbitrary
dimensional bipartite quantum states was derived by Mintert et al.
in \cite{167902}. By using the positive partial
transposition (PPT) and realignment separability criterion analytic
lower bounds on EOF and concurrence for any dimensional mixed
bipartite quantum states have been derived in
\cite{Chen-Albeverio-Fei1,chen}. These bounds are exact for some
special classes of states and can be used to detect many bound
entangled states. In \cite{breuer} another lower bound on EOF for
bipartite states has been presented from a new separability
criterion \cite{breuerprl}. A lower bound of concurrence based on
local uncertainty relations (LURs) criterion is derived in
\cite{vicente}. This bound is further optimized in \cite{optloo}. The
lower bound of concurrence for tripartite systems has been studied
in \cite{gao}. In \cite{edward,ou} the authors presented lower
bounds of concurrence for bipartite systems by considering the
``two-qubit" entanglement of bipartite quantum states with arbitrary
dimensions. It has been shown that this lower bound has a tight
relationship with the distillability of bipartite quantum states.
Tangle is also a good entanglement measure that has a close relation
with concurrence, as it is defined by the square of the concurrence
for a pure state. It is also meaningful to derive tight lower and
upper bounds for tangle \cite{vicentejpa}.

In \cite{260502} Mintert {\sl et al.} proposed an experimental method to
measure the concurrence directly by using joint measurements on two copies of a pure
state. Then S. P. Walborn {\sl et al.}
presented an experimental determination of concurrence
for two-qubit states \cite{nature}, where only one-setting measurement
is needed, but two copies of the state have to be prepared in every measurement.
In \cite{zmj-c} another way of experimental
determination of concurrence for two-qubit and multi-qubit states has been presented, in which
only one-copy of the state is needed in every measurement. To determine the concurrence
of the two-qubit state used in \cite{nature},
also one-setting measurement is needed, which avoids the preparation of the twin states
or the imperfect copy of the unknown state, and the experimental difficulty is dramatically reduced.

\begin{enumerate}
{\sf \item[{(3)}] Fidelity of quantum teleportation and distillation}
\end{enumerate}

Quantum teleportation, or entanglement-assisted teleportation, is a
technique used to transfer information on a quantum level, usually
from one particle (or series of particles) to another particle (or
series of particles) in another location via quantum entanglement.
It does not transport energy or matter, nor does it allow
communication of information at super luminal (faster than light)
speed.

In \cite{teleportation}, Bennett et. al. first
presented a protocol to teleport an unknown qubit state by using
a pair of maximally entangled pure qubit state.
The protocol is generalized to transmit high dimensional quantum states
\cite{fefandtel}. The optimal fidelity of teleportation
is shown to be determined by the fully entangled fraction
of the entangled resource which is generally a mixed state.
Nevertheless similar to the estimation of concurrence,
the computation of the fully entangled fraction for a given mixed state
is also very difficult.

The distillation protocol has been presented to get maximally
entangled pure states from many entangled mixed states.
by means of local quantum operations and classical communication (LQCC) between the
parties sharing the pairs of particles in this mixed state
\cite{distill1,distill2,distill3,bennett}. Bennett et. al.
first derived a protocol to distill one maximally entangled pure
bell state from many copies of not maximally entangled quantum mixed
states in \cite{distill1} in 1996. The protocol is then generalized
to distill any bipartite quantum state with higher dimension by
Horodeckis in 1999 \cite{gdp}. It is proven that a quantum state
can be always distilled if it violates the
reduced matrix separability criterion \cite{gdp}.

This review mainly contains three parts. In section $\ref{sepsec}$
we investigate the separability of quantum states. We first introduce
several important separability criteria. Then we
discuss the criterions by using the Bloch representation of the
density matrix of a quantum state. We also study the covariance
matrix of a quantum density matrix and derive separability criterion
for multipartite systems. We investigate the normal forms for multipartite
quantum states at the end of this section and show that the normal
form can be used to improve the power of these criteria. In
section $\ref{consec}$ we mainly consider the entanglement measure
concurrence. We investigate
the lower and upper bounds of concurrence for
both bipartite and multipartite systems. We also show that the
concurrence and tangle of two entangled quantum states will be
always larger than that of one, even both the two states are bound
entangled (not distillable). In section $\ref{fefsec}$ we study the
fully entangled fraction of an arbitrary bipartite quantum state.
We derive precise formula of fully entangled fraction for two qubits
system. For bipartite system with higher dimension we obtain tight
upper bounds which can not only be used to estimate the optimal
teleportation fidelity but also helps to improve the distillation
protocol. We further investigate the evolution of the fully
entangled fraction when
one of the bipartite system undergoes a noisy channel. We give
a summary and conclusion in the last section.

\section{Separability criteria and normal form}\label{sepsec}

A multipartite pure quantum state $\rho_{12\ldots
N}\in{\mathcal{H}}_{1}\otimes{\mathcal
{H}}_{2}\otimes\cdots\otimes{\mathcal {H}}_{N}$ is said to be fully
separable if it can be written as
\be\label{mps}
\rho_{12\ldots N}=
\rho_{1}\otimes\rho_{2} \otimes\cdots\otimes \rho_{N},
\ee
where
${\rho}_{1}$ and ${\rho}_{2}$, $\cdots$, ${\rho}_{N}$ are reduced
density matrices defined as
$\rho_{1}={\rm Tr}_{23\ldots N}[\rho_{12\ldots N}]$,
$\rho_{2}={\rm Tr}_{13\ldots N}[\rho_{12\ldots N}]$, ...,
$\rho_{N}={\rm Tr}_{12\ldots N-1}[\rho_{12\ldots N}]$.
This is equivalent to the condition
$$\rho_{12\ldots N}=|\psi_{1}\ra\la\psi_{1}|\otimes
|\phi_{2}\ra\la\phi_{2}|\otimes\cdots\otimes|\mu_{N}\ra\la\mu_{N}|,$$
where $|\psi_{1}\ra\in {\cal H}_{1},\ \  |\phi_{2}\ra\in {\cal
H}_{2},\ \  \ldots,\ \  |\mu_{N}\ra\in {\cal H}_{N}$.

A multipartite quantum mixed state $\rho_{12\ldots N}\in{\cal
{H}}_{1}\otimes{\cal {H}}_{2}\otimes\cdots\otimes{\cal {H}}_{N}$ is said to be
fully separable if it can be written as
\be\label{sec}
\rho_{12\ldots N}=\sum_i q_i {\rho_i}^{1}\otimes{\rho_i}^{2}
\otimes\cdots\otimes {\rho_i}^{N},
\ee
where ${\rho_i}^{1},{\rho_i}^{2} \ldots, {\rho_i}^{N}$ are the reduced
density matrices with respect to the systems
$1, 2, \ldots, N$ respectively, $q_i>0$ and $\sum_i q_i=1$. This is
equivalent to the condition
$$\rho_{12\ldots N}=\sum_i p_i |\psi_i^{1}\ra\la\psi_i^{1}|\otimes
|\phi_i^{2}\ra\la\phi_i^{2}|\otimes\cdots\otimes|\mu_i^{N}\ra\la\mu_i^{N}|,$$
where $|\psi_i^{1}\ra, |\phi_i^{2}\ra, \ldots, |\mu_i^{N}\ra$ are
normalized pure states of systems $1, 2, \ldots, N$ respectively,
$p_i>0$ and $\sum_i p_i=1$.

For pure states, the definition (\ref{mps}) itself is an
operational separability criterion. In particular, for bipartite case,
there are Schmidt decompositions:

\begin{theorem}(Schmidt decomposition)\label{SD}:
Suppose $|\psi\ra\in{\mathcal{H}}_{A}\otimes{\mathcal {H}}_{B}$
is a pure state of a composite system, $AB$, then there exist
orthonormal states $|i_A\ra$ for system $A$, and  orthonormal states
$|i_B\ra$ for system $B$ such that $$|\psi\ra=\sum_{i}
\lambda_{i}|i_{A}\ra |i_{B}\ra,$$ where $\lambda_{i}$ are
non-negative real numbers satisfying $\sum_i {\lambda_i}^2=1$, known
as Schmidt coefficients.
\end{theorem}

$|i_{A}\ra$ and $|i_B\ra$ are called Schmidt bases with respect to
${\cal H}_A$ and ${\cal H}_B$. The number of non-zero values
$\lambda_i$ is called Schmidt number, also known as Schmidt rank, which is
invariant under unitary
transformations on system $A$ or system $B$.
For a bipartite pure state $|\psi\ra$, $|\psi\ra$ is separable if and only
if the Schmidt number of $|\psi\ra$ is one.

For multipartite pure states, one has no such Schmidt  decomposition.
In \cite{acin} it has been verified that any pure three-qubit
state $|\Psi\rangle$ can be uniquely written as
\begin{eqnarray} |\Psi\rangle &=& \lambda_0
|000\rangle+\lambda_1 e^{i\psi} |100\rangle+\lambda_2 |101\rangle
+\lambda_3 |110\rangle+\lambda_4 |111\rangle \label{gschmidt}
\end{eqnarray}
with normalization condition $\lambda_i \geq 0,\ \ 0 \leq \psi \leq
\pi$, where $\sum_i \mu_i=1$, $\mu_i \equiv\lambda_i^2$.
Eq.~(\ref{gschmidt}) is called generalized Schmidt decomposition.

For mixed states it is generally very hard to verify
if a decomposition like $(\ref{sec})$ exists. For a
given generic separable density matrix, it is also not easy to
find the decomposition $(\ref{sec})$ in detail.

\subsection{Separability criteria for mixed states}\label{sec3}

In this section we introduce several separability criteria and the
relations among themselves. These criteria have also tight relations with
lower bounds of entanglement measures and distillation that will be discussed in the next section.

\subsubsection{Partial positive transpose criterion}

The positive partial transpose (PPT) criterion provided by Peres \cite{ppt} says
that if a bipartite state $\rho_{AB}\in{\mathcal{H}}_{A}\otimes{\mathcal {H}}_{B}$ is separable,
then the new matrix $\rho_{AB}^{T_B}$ with matrix elements defined in some fixed product
basis as:
$$\la m|\la
\mu|\rho_{AB}^{T_B}|n\ra|\nu\ra\equiv \la m|\la
\nu|\rho_{AB}|n\ra|\mu\ra$$ is also a density matrix (i.e. has
nonnegative spectrum). The operation $T_B$,
called partial transpose, just corresponds to the transposition of
the indices with respect to the second subsystem $B$. It has
an interpretation as a partial time reversal \cite{san98}.

Afterwards the Horodeckis showed that the
Peres' criterion is also sufficient for $2\times 2$ and $2\times3$
bipartite systems \cite{ho96}. This
criterion is now called PPT or Peres-Horodecki (P-H) criterion. For
high-dimensional states, the P-H criterion is only necessary.
Horodecki has constructed some classes of families of
entangled states with positive partial transposes for $3\times 3$
and $2\times 4$ systems \cite{ho232}. States of this kind are said to be bound
entangled (BE).

\subsubsection{Reduced density matrix criterion}

Cerf et al. \cite{cag} and Horodecki \cite{ho54} independently,
introduced a map $\Gamma: \rho\rightarrow {\rm
Tr}_{B}[\rho_{AB}]\otimes I-\rho_{AB}$ ($I\otimes {\rm
Tr}_{A}[\rho_{AB}]-\rho_{AB}$),  which gives rise to a simple
necessary condition for separability in arbitrary dimensions, called
the reduction criterion: If $\rho_{AB}$ is separable, then
$$\rho_{A}\otimes I-\rho_{AB}\geq 0,\ \ I\otimes\rho_{B}-\rho_{AB}\geq
0,$$ where $\rho_{A}={\rm Tr}_{B}[\rho_{AB}]$, $\rho_{B}={\rm
Tr}_{A}[\rho_{AB}]$. This criterion is simply equivalent to the P-H
criterion for $2\times n$ composite systems. It is also sufficient
for $2\times 2$ and $2\times 3$ systems. In higher dimensions the
reduction criterion is weaker than the P-H criterion.

\subsubsection{Realignment criterion}

There is yet another class of criteria based on linear
contractions on product states. They stem from the new criterion
discovered in \cite{rudolph, ChenQIC03} called computable cross norm
(CCN) criterion or matrix realignment criterion which is operational
and independent on PPT test \cite{ppt}. If a state $\rho_{AB}$ is separable
then the realigned matrix ${\cal R}(\rho)$ with elements
${\cal{R}}(\rho)_{ij,kl}=\rho_{ik,jl}$  has trace norm not greater than
one,
\be||{\cal{R}}(\rho)||_{KF}\leq 1.
\ee
Quite remarkably, the realignment criterion can detect some
PPT entangled (bound entangled) states \cite{rudolph, ChenQIC03} and can be used for
construction of some nondecomposable maps. It also provides nice
lower bound for concurrence \cite{chen}.

\subsubsection{Criteria based on Bloch representations}

Any Hermitian operator on an $N$-dimensional Hilbert space ${\cal
H}$ can be expressed according to the generators of the special
unitary group $SU(N)$ {\cite{Hioe}}. The generators of $SU(N)$ can
be introduced according to the transition-projection operators
$P_{jk}=|j\rangle\langle k|$,
where $|i\rangle$, $i=1,...,N$, are the orthonormal eigenstates of a
linear Hermitian operator on ${\cal H}$. Set
\begin{eqnarray*}
&&\omega_{l}=-\sqrt{\frac{2}{l(l+1)}}({P}_{11}+{P}_{22}
+\cdots+{P}_{ll}-l {P}_{l+1,l+1}),\\[3mm]
&&{u}_{jk}={P}_{jk}+{P}_{kj}, ~~~{v}_{jk}=i({P}_{jk}-{P}_{kj}),
\end{eqnarray*}
where  $1\leq l \leq N-1$ and $1\leq j < k \leq N$. We get a set of
$N^2-1$ operators
$$\Gamma\equiv\{{\omega}_{l},{\omega}_{2},\cdots, {\omega}_{N-1}, {u}_{12},{u}_{13},\cdots, {v}_{12},{v}_{13},
\cdots \},$$ which satisfy the relations
$${\rm Tr}[\lambda_{i}]=0,\quad\quad
{\rm
Tr}[\lambda_{i}\lambda_{j}]=2{\delta}_{ij},~~~\forall~\lambda_i\in\Gamma$$
and thus generate the $SU(N)$ ${\cite{duichenxing}}$.

Any Hermitian operator $\rho$ in ${\cal {H}}$ can be represented in
terms of these generators of $SU(N)$,
\begin{eqnarray}\label{1bloch}
\rho=\frac{1}{N}I_{N}+\frac{1}{2}\sum^{N^{2}-1}_{j=1}r_{j}\lambda_{j},
\end{eqnarray}
where $I_{N}$ is a unit matrix and ${\bf{r}}=(r_{1}, r_{2},\cdots
r_{N^{2}-1})\in {\Bbb{R}}^{N^{2}-1}$. ${\bf{r}}$ is called Bloch
vector. The set of all the Bloch vectors that constitute a density
operator is known as the Bloch vector space
$B({\Bbb{R}}^{N^{2}-1})$.

A matrix of the form $(\ref{1bloch})$ is of unit trace and
Hermitian, but it might not be positive. To guarantee the positivity
restrictions must be imposed on the Bloch vector. It is shown that
$B({\Bbb{R}}^{N^{2}-1})$ is a subset of the ball
$D_{R}({\Bbb{R}}^{N^{2}-1})$ of radius $R=\sqrt{2(1-\frac{1}{N})}$,
which is the minimum ball containing it, and that the ball
$D_{r}({\Bbb{R}}^{N^{2}-1})$ of radius $r=\sqrt{\frac{2}{N(N-1)}}$
is included in $B({\Bbb{R}}^{N^{2}-1})$ {\cite{Harriman}}, that is,
$$D_{r}({\Bbb{R}}^{N^{2}-1})\subseteq B({\Bbb{R}}^{N^{2}-1})\subseteq D_{R}({\Bbb{R}}^{N^{2}-1}).$$

Let the dimensions of systems A, B and C be $d_A=N_{1}, d_B=N_{2}$
and
 $d_C=N_{3}$ respectively. Any tripartite quantum states $\rho_{ABC} \in {\cal {H}}_{A}
  \otimes {\cal {H}}_{B} \otimes {\cal {H}}_{C}$ can
  be written as:
 \begin{eqnarray}\label{322}
  \rho_{ABC}&=&I_{N_{1}}\otimes I_{N_{2}}\otimes M_{0}+\sum_{i=1}^{N_{1}^{2}-1}\lambda_{i}(1)\otimes I_{N_{2}}\otimes
  M_{i}+\sum_{j=1}^{N_{2}^{2}-1}I_{N_{1}}\otimes \lambda_{j}(2)\otimes \widetilde{M}_{j}\nonumber \\[3mm]
 &&+\sum_{i=1}^{N_{1}^{2}-1}\sum_{j=1}^{N_{2}^{2}-1}\lambda_{i}(1)\otimes \lambda_{j}(2)\otimes
  M_{ij},  \end{eqnarray}
where $\lambda_{i}(1)$, $\lambda_{j}(2)$ are  the generators of
$SU(N_{1})$ and $SU(N_{2})$; $M_{i}, \widetilde{M}_{j}$ and $M_{ij}$
are operators of ${\cal{H}}_{C}$.

\begin{theorem} \ Let ${\bf{r}} \in {\Bbb{R}}^{N_{1}^{2}-1}$,
${\bf{s}} \in {\Bbb{R}}^{N_{2}^{2}-1}$ and $|{\bf{r}}| \leq
\sqrt{\frac{2}{N_{1}(N_{1}-1)}}$, $|{\bf{s}}| \leq
\sqrt{\frac{2}{N_{2}(N_{2}-1)}}$. For a tripartite quantum state
$\rho \in {\cal {H}}_{A}
  \otimes {\cal {H}}_{B} \otimes {\cal {H}}_{C}$
with representation (\ref{322}), we have \be\label{t1112}
M_{0}-\sum_{i=1}^{N_{1}^{2}-1}r_{i}M_{i}-\sum_{j=1}^{N_{2}^{2}-1}s_{j}\widetilde{M}_{j}
+\sum_{i=1}^{N_{1}^{2}-1}\sum_{j=1}^{N_{2}^{2}-1}r_{i}s_{j}M_{ij}\geq
0. \ee \end{theorem}

{\sf[Proof]}\ Since ${\bf{r}} \in {\Bbb{R}}^{N_{1}^{2}-1}$,
${\bf{s}} \in {\Bbb{R}}^{N_{2}^{2}-1}$ and $|{\bf{r}}| \leq
\sqrt{\frac{2}{N_{1}(N_{1}-1)}}$, $|{\bf{s}}| \leq
\sqrt{\frac{2}{N_{2}(N_{2}-1)}}$, we have that $A_{1}\equiv
\frac{1}{2}(\frac{2}{N_{1}}I-\sum\limits_{i=1}^{N_{1}^{2}-1}r_{i}\lambda_{i}(1))$
and $A_{2}\equiv
\frac{1}{2}(\frac{2}{N_{2}}I-\sum\limits_{j=1}^{N_{2}^{2}-1}s_{j}\lambda_{j}(2))$
are positive Hermitian operators. Let $A=\sqrt{A_{1}}\otimes
\sqrt{A_{2}} \otimes I_{N_{3}}$. Then $A\rho A \geq 0$ and $(A\rho
A)^{\dag}=A\rho A$. The partial trace of $A\rho A$ over ${\cal
{H}}_{A}$ (and ${\cal {H}}_{B}$) should be also positive. Hence
\begin{eqnarray*}
0 &\leq& {\rm Tr}_{AB}[A\rho A] \nonumber \\[3mm]
  &=& {\rm Tr}_{AB}[A_{1}
\otimes A_{2}\otimes M_{0}+\sum\limits_{i} \sqrt{A_{1}}
\lambda_{i}(1) \sqrt{A_{1}}\otimes A_{2} \otimes
M_{i}\nonumber \\[3mm]
&+&\sum\limits_{j} A_{1} \otimes
\sqrt{A_{2}}\lambda_{j}(2)\sqrt{A_{2}} \otimes
\widetilde{M}_{j}+\sum\limits_{ij}
\sqrt{A_{1}}\lambda_{i}(1)\sqrt{A_{1}} \otimes
\sqrt{A_{2}}\lambda_{j}(2)\sqrt{A_{2}} \otimes M_{ij}] \nonumber \\[3mm]
  &=&M_{0}-\sum_{i=1}^{N_{1}^{2}-1}r_{i}M_{i}-\sum_{j=1}^{N_{2}^{2}-1}s_{j}\widetilde{M}_{j}
+\sum_{i=1}^{N_{1}^{2}-1}\sum_{j=1}^{N_{2}^{2}-1}r_{i}s_{j}M_{ij}.\nonumber
\end{eqnarray*}
$\hfill\Box$

Formula (\ref{t1112}) is valid for any tripartite states. By setting
${\bf s}=0$ in (\ref{t1112}), one can get a result for bipartite
systems:

\begin{corollary}
Let $\rho_{AB}\in {\cal {H}}_{A}\otimes{\cal {H}}_{B}$, which can be
generally written as $\rho_{AB}=I_{N_{1}}\otimes
M_{0}+\sum\limits_{j=1}^{N_{1}^{2}-1}\lambda_{j}\otimes M_{j}$, then
for any ${\bf{r}} \in {\Bbb{R}}^{N_{1}^{2}-1}$ with $|{\bf{r}}| \leq
\sqrt{\frac{2}{N_{1}(N_{1}-1)}}$,
$M_{0}-\sum\limits_{j=1}^{N_{1}^{2}-1}r_{j}M_{j}\geq 0$.
\end{corollary}

A separable tripartite state $\rho_{ABC}$ can be written as
$$\rho_{ABC}=\sum\limits_{i} p_{i}| \psi_{i}^{A}\rangle\langle
\psi_{i}^{A}|\otimes| \phi_{i}^{B}\rangle\langle
\phi_{i}^{B}|\otimes| \omega_{i}^{C}\rangle\langle
\omega_{i}^{C}|.$$
From (\ref{1bloch}) it can also be represented as:
\begin{eqnarray}\label{324}
\rho_{ABC}&=&\sum\limits_{i} p_{i}
\frac{1}{2}(\frac{2}{N_{1}}I_{N_{1}}+\sum\limits_{k=1}^{N_{1}^{2}-1}a_{i}^{(k)}\lambda_{k}(1))
\otimes\frac{1}{2}(\frac{2}{N_{2}}I_{N_{2}}+\sum\limits_{l=1}^{N_{2}^{2}-1}b_{i}^{(l)}\lambda_{l}(2))
\otimes| \omega_{i}^{C}\rangle\langle \omega_{i}^{C}| \nonumber \\
&=&I_{N_{1}}\otimes
I_{N_{2}}\otimes\frac{1}{N_{1}N_{2}}\sum\limits_{i} p_{i}|
\omega_{i}^{C}\rangle\langle \omega_{i}^{C}|\nonumber\\[3mm]
&&+\sum\limits_{k=1}^{N_{1}^{2}-1}\lambda_{k}(1)\otimes
I_{N_{2}}\otimes\frac{1}{2N_{2}}\sum\limits_{i} a_{i}^{(k)}p_{i}|
\omega_{i}^{C}\rangle\langle \omega_{i}^{C}| \nonumber \\[3mm]
&&+\sum\limits_{l=1}^{N_{2}^{2}-1}I_{N_{1}}\otimes
\lambda_{l}(2)\otimes\frac{1}{2N_{1}}\sum\limits_{i}
b_{i}^{(l)}p_{i}|
\omega_{i}^{C}\rangle\langle \omega_{i}^{C}| \nonumber \\[3mm]
&&+\sum\limits_{k}^{N_{1}^{2}-1}\sum\limits_{l}^{N_{2}^{2}-1}\lambda_{k}(1)\otimes
\lambda_{l}(2)\otimes\frac{1}{4}\sum\limits_{i}
a_{i}^{(k)}b_{i}^{(l)}p_{i}| \omega_{i}^{C}\rangle\langle
\omega_{i}^{C}|,
\end{eqnarray}
where $(a_{i}^{(1)}, a_{i}^{(2)} \cdots, a_{i}^{(N_{1}^{2}-1)})$ and
$(b_{i}^{(1)}, b_{i}^{(2)} \cdots, b_{i}^{(N_{2}^{2}-1)})$ are real
vectors on the Bloch sphere satisfying
$|\overrightarrow{a_{i}}|^{2}=\sum\limits_{j=1}^{N_{1}^{2}-1}(a_{i}^{(j)})^{2}=2(1-\frac{1}{N_{1}})$
and
$|\overrightarrow{b_{i}}|^{2}=\sum\limits_{j=1}^{N_{2}^{2}-1}(b_{i}^{(j)})^{2}=2(1-\frac{1}{N_{2}})$.

Comparing $(\ref{322})$ with $(\ref{324})$, we have
\begin{eqnarray}\label{325}
&M_{0}=\frac{1}{N_{1}N_{2}}\sum\limits_{i} p_{i}|
\omega_{i}^{C}\rangle\langle \omega_{i}^{C}|, \quad\quad
M_{k}=\frac{1}{2N_{2}}\sum\limits_{i} a_{i}^{(k)} p_{i}|
\omega_{i}^{C}\rangle\langle \omega_{i}^{C}|, \nonumber \\[3mm]
&\widetilde{M}_{l}=\frac{1}{2N_{1}}\sum\limits_{i} b_{i}^{(l)}
p_{i}| \omega_{i}^{C}\rangle\langle \omega_{i}^{C}|, \quad
M_{kl}=\frac{1}{4}\sum\limits_{i} a_{i}^{(k)}b_{i}^{(l)} p_{i}|
\omega_{i}^{C}\rangle\langle \omega_{i}^{C}|.
\end{eqnarray}

For any $(N_{1}^{2}-1) \times (N_{1}^{2}-1)$ real matrix $R(1)$ and
$(N_{2}^{2}-1)\times (N_{2}^{2}-1)$ real matrix $R(2)$ satisfying
$\frac{1}{(N_{1}-1)^{2}}I-R(1)^{T}R(1)\geq 0$ and
$\frac{1}{(N_{2}-1)^{2}}I-R(2)^{T}R(2)\geq 0$, we define a new
matrix
\begin{eqnarray}\label{5}
\cal {R}=\left(%
    \begin{array}{ccc}
      R(1) & 0 & 0 \\
      0 & R(2) & 0 \\
      0 & 0 & T \\
    \end{array}%
    \right),
\end{eqnarray}
where $T$ is a transformation acting on an $(N_{1}^{2}-1) \times
(N_{2}^{2}-1)$ matrix $M$ by $$T(M)=R(1) M R^{T}(2).$$ Using $\cal
{R}$ we define a new operator $\gamma_{\cal {R}}$,
\begin{eqnarray}\label{327}
  \gamma_{\cal {R}}(\rho_{ABC})&=&I_{N_{1}}\otimes I_{N_{2}}\otimes M_{0}^{'}+\sum_{i=1}^{N_{1}^{2}-1}\lambda_{i}(1)
  \otimes I_{N_{2}}\otimes   M_{i}^{'}
  +\sum_{j=1}^{N_{2}^{2}-1}I_{N_{1}}\otimes \lambda_{j}(2)\otimes \widetilde{M}_{j}^{'}\nonumber \\[3mm]
  &&+\sum_{i=1}^{N_{1}^{2}-1}\sum_{j=1}^{N_{2}^{2}-1}\lambda_{i}(1)\otimes \lambda_{j}(2)\otimes
  M_{ij}^{'},
  \end{eqnarray}
where $M_{0}^{'}=M_{0}, \quad
M_{k}^{'}=\sum\limits_{m=1}^{N_{1}^{2}-1}R_{km}(1)M_{m},\quad
\widetilde{M}_{l}^{'}=\sum\limits_{n=1}^{N_{2}^{2}-1}R_{ln}(2)\widetilde{M}_{n}$
and
$M_{ij}^{'}=(T(M))_{ij}=(R(1)MR^{T}(2))_{ij}$.\\

\begin{theorem}\label{dingli} If $\rho_{ABC}$ is separable, then
$\gamma_{\cal {R}}(\rho_{ABC})\geq 0$.\\
\end{theorem}

{\sf[Proof]}\ From $(\ref{325})$ and $(\ref{327})$ we get
\begin{eqnarray*}
M_{0}^{'}&=&M_{0}=\frac{1}{N_{1}N_{2}}\sum\limits_{i} p_{i}|
\omega_{i}^{C}\rangle\langle \omega_{i}^{C}|,~
M_{k}^{'}=\frac{1}{2N_{2}}\sum\limits_{mi} R_{km}(1)a_{i}^{(m)}
p_{i}|\omega_{i}^{C}\rangle\langle \omega_{i}^{C}|,  \\[3mm]
\widetilde{M}_{l}^{'}&=&\frac{1}{2N_{1}}\sum\limits_{ni}
R_{ln}(2)b_{i}^{(n)} p_{i}| \omega_{i}^{C}\rangle\langle
\omega_{i}^{C}|,~ M_{kl}^{'}=\frac{1}{4}\sum\limits_{mni}
R_{km}(1)a_{i}^{(m)}R_{ln}(2)b_{i}^{(n)} p_{i}|
\omega_{i}^{C}\rangle\langle \omega_{i}^{C}|.
\end{eqnarray*}
A straightforward calculation gives rise to
\begin{eqnarray*}\label{3}
  \gamma_{\cal {R}}(\rho_{ABC})
  &=&\sum\limits_{i} p_{i}\frac{1}{2}\left(\frac{2}{N_{1}}I_{N_{1}}
             +\sum\limits_{k=1}^{N_{1}^{2}-1}\sum\limits_{m=1}^{N_{1}^{2}-1}R_{km}(1)a_{i}^{(m)}\lambda_{k}(1)\right)\\[3mm]
&&\quad\quad \otimes\frac{1}{2}\left(\frac{2}{N_{2}}I_{N_{2}}
          +\sum\limits_{l=1}^{N_{2}^{2}-1}\sum\limits_{n=1}^{N_{2}^{2}-1}R_{ln}(2)b_{i}^{(n)}\lambda_{l}(2)\right)
\otimes| \omega_{i}^{C}\rangle\langle \omega_{i}^{C}|.
  \end{eqnarray*}
As $\frac{1}{(N_{1}-1)^{2}}I-R(1)^{T}R(1)\geq 0$ and
$\frac{1}{(N_{2}-1)^{2}}I-R(2)^{T}R(2)\geq 0$, we get
$$|\overrightarrow{a_{i}^{'}}|^{2}=|R(1)\overrightarrow{a_{i}}|^{2}\leq\frac{1}{(N_{1}-1)^{2}}|\overrightarrow{a_{i}}|^{2}=
\frac{2}{N_{1}(N_{1}-1)},$$
$$|\overrightarrow{b_{i}^{'}}|^{2}=|R(2)\overrightarrow{b_{i}}|^{2}\leq\frac{1}{(N_{2}-1)^{2}}|\overrightarrow{b_{i}}|^{2}=
\frac{2}{N_{2}(N_{2}-1)}.$$ Therefore $\gamma_{\cal
{R}}(\rho_{ABC})$ is still a density operator, i.e. $\gamma_{\cal
{R}}(\rho_{ABC})\geq 0$.\hfill$\Box$

Theorem $\ref{dingli}$ gives a necessary separability criterion for
general tripartite systems. The result can be also applied to
bipartite systems. Let $\rho_{AB}\in{\cal {H}}_{A}\otimes {\cal
{H}}_{B}$, $\rho_{AB}=I_{N_{1}}\otimes
M_{0}+\sum\limits_{j=1}^{N_{1}^{2}-1}\lambda_{j}\otimes M_{j}$. For
any real $(N_{1}^{2}-1)\times (N_{1}^{2}-1)$ matrix $\cal {R}$
satisfying $\frac{1}{(N_{1}-1)^{2}}I-{\cal {R}}^{T}{\cal {R}}\geq 0$
and any state $\rho_{AB}$, we define $$\gamma_{\cal
{R}}(\rho_{AB})=I_{N_{1}}\otimes
M_{0}+\sum\limits_{j=1}^{N_{1}^{2}-1}\lambda_{j}\otimes M_{j}^{'},$$
where $M_{j}^{'}=\sum\limits_{k}{\cal {R}}_{jk}M_{k}$.

\begin{corollary} For $\rho_{AB}\in{\cal {H}}_{A}\otimes {\cal {H}}_{B}$,
if there exists an ${\cal {R}}$ with $\frac{1}{(N_{1}-1)^{2}}I-{\cal
{R}}^{T}{\cal {R}}\geq 0$ such that $\gamma_{\cal {R}}(\rho_{AB})<
0$, then $\rho_{AB}$ must be entangled.
\end{corollary}

For $2\times N$ systems, the above corollary is reduced to the results in
\cite{S.J.Wu}. As
an example we consider the $3\times 3$ istropic states,
$$
\rho_{I}=\frac{1-p}{9}I_{3}\otimes I_{3}+ \frac{p}{3}\sum\limits_{i,j=1}^{3}|ii\rangle\langle
jj|=I_{3}\otimes (\frac{1}{9}I_{3})+\sum\limits_{i=1}^{5}\lambda_{i}\otimes(\frac{p}{6}\lambda_{i})
    -\sum\limits_{i=6}^{8}\lambda_{i}\otimes(\frac{p}{6}\lambda_{i}).
$$
If we choose ${\cal {R}}$ to be $\rm{Diag}\{\frac{1}{2},
\frac{1}{2}, \frac{1}{2}, \frac{1}{2}, \frac{1}{2}, -\frac{1}{2},
-\frac{1}{2}, -\frac{1}{2}\}$, we get that
$\rho_{I}$ is entangled for $0.5< p \leq 1$.

For tripartite case, we take the following $3\times3\times3$ mixed state as an example:
\begin{eqnarray*}
\rho=\frac{1-p}{27}I_{27}+p|\psi\ra\la\psi|,
\end{eqnarray*}
where
$|\psi\ra=\frac{1}{\sqrt{3}}(|000\ra+|111\ra+|222\ra)(\la000|+\la111|+\la222|)$.
Taking $R(1)=R(2)=\rm{Diag}\{\frac{1}{2}, \frac{1}{2}, \frac{1}{2},
\frac{1}{2}, \frac{1}{2}, -\frac{1}{2}, -\frac{1}{2},
-\frac{1}{2}\}$, we have that $\rho$ is entangled for
$0.6248<p\leq1$.

In fact the criterion for $2\times N$ systems \cite{S.J.Wu} is
equivalent to the PPT criterion \cite{Rudolph}. Similarly theorem
$\ref{dingli}$ is also equivalent to the PPT criterion for $2\times
2\times N$ systems.

\subsubsection{Covariance matrix criterion}

In this subsection we study the separability problem by using the
covariance matrix approach. We first give a brief
review of covariance matrix criterion proposed in \cite{guhne}.
Let ${\cal {H}}^{A}_{d}$ and ${\cal {H}}^{B}_{d}$ be $d$-dimensional
complex vector spaces, and $\rho_{AB}$ a bipartite quantum state in
${\cal {H}}^{A}_{d}\otimes{\cal {H}}^{B}_{d}$. Let $A_{k}$ (resp.
$B_{k}$) be $d^{2}$ observables on ${\cal{H}}^{A}_{d}$ (resp.
${\cal{H}}^{B}_{d}$) such that they form an orthonormal normalized
basis of the observable space, satisfying ${\rm
Tr}[A_{k}A_{l}]=\delta_{k,l}$ (resp. ${\rm
Tr}[B_{k}B_{l}]=\delta_{k,l}$). Consider the total set
$\{M_{k}\}=\{A_{k}\otimes I, I\otimes B_{k}\}$. It can be proven
that {\cite{guhne1}},
\begin{eqnarray}\label{lemma1}
\sum\limits_{k=1}^{N^{2}}(M_{k})^{2}=dI,\quad\quad\quad
\sum\limits_{k=1}^{N^{2}}\langle M_{k} \rangle^{2}={\rm
Tr}[\rho^{2}_{AB}].
\end{eqnarray}

The covariance matrix $\gamma$ is defined with entries
\begin{eqnarray}\label{gammae}
\label{c} \gamma_{ij}(\rho_{AB}, \{M_{k}\})=\frac{\langle
M_{i}M_{j}\rangle+\langle M_{j}M_{i}\rangle}{2}-\langle
M_{i}\rangle\langle M_{j}\rangle,
\end{eqnarray}
which has a block structure \cite{guhne}:
\begin{eqnarray}
\label{defd}
\gamma=\left(%
    \begin{array}{cc}
      A & C \\
      C^{T} & B \\
    \end{array}%
    \right),
\end{eqnarray}
where $A=\gamma(\rho_{A},\{A_{k}\}), B=\gamma(\rho_{B},\{B_{k}\}),
C_{ij}=\langle A_{i}\otimes B_{j} \rangle_{\rho_{AB}}-\langle A_{i}
\rangle_{\rho_{A}}\langle B_{j} \rangle_{\rho_{B}}$, $\rho_{A}={\rm
Tr}_B[\rho_{AB}]$, $\rho_{B}={\rm Tr}_A[\rho_{AB}]$. Such covariance
matrix has a concavity property: for a mixed density matrix
$\rho=\sum\limits_{k}p_{k}\rho_{k}$ with $p_{k}\geq 0$ and
$\sum\limits_{k}p_{k}=1$, one has $\gamma(\rho)\geq
\sum\limits_{k}p_{k}\gamma(\rho_{k})$.

For a bipartite product state $\rho_{AB}=\rho_{A}\otimes\rho_{B}$,
$C$ in $(\ref{defd})$ is zero. Generally if $\rho_{AB}$ is
separable, then there exist states $|a_{k}\rangle\langle a_{k}|$ on
${\cal {H}}^{A}_{d}$, $|b_{k}\rangle\langle b_{k}|$ on ${\cal
{H}}^{B}_{d}$ and $p_{k}$ such that \be\label{p1} \gamma(\rho)\geq
\kappa_{A}\oplus \kappa_{B}, \ee where $\kappa_{A}=\sum
p_{k}\gamma(|a_{k}\rangle\langle a_{k}|,\{A_{k}\})$,
$\kappa_{B}=\sum p_{k}\gamma(|b_{k}\rangle\langle
b_{k}|,\{B_{k}\})$.

For a separable bipartite state, it has been shown that \cite{guhne}
\begin{eqnarray}\label{p2}
\sum\limits_{i=1}^{d^{2}}|C_{ii}|\leq\frac{ (1-{\rm
Tr}[\rho_{A}^{2}])+(1-{\rm Tr}[\rho_{B}^{2}])}{2}.
\end{eqnarray}

Criterion (\ref{p2}) depends on the choice of the orthonormal
normalized basis of the observables. In fact the term
$\sum\limits_{i=1}^{d^{2}}|C_{ii}|$ has an upper bound $||C||_{KF}$
which is invariant under unitary transformation and can be attained
by choosing proper local orthonormal observable basis, where
$||C||_{KF}$ stands for the Ky Fan norm of $C$, $||C||_{KF}={\rm
Tr}[\sqrt{CC^{\dag}}]$, with $\dag$ denoting the transpose and
conjugation. It has been shown in \cite{optloo} that if $\rho_{AB}$
is separable, then
\begin{eqnarray}
\label{th2} ||C||_{KF}\leq \frac{(1-{\rm Tr}[\rho_{A}^{2}])+(1-{\rm
Tr}[\rho_{B}^{2}])}{2}.
\end{eqnarray}

From the covariance matrix approach, we can also get an alternative
criterion. From (\ref{defd}) and (\ref{p1}) we have that if
$\rho_{AB}$ is separable, then
\begin{eqnarray}
X\equiv\left(
    \begin{array}{cc}
      A-\kappa_{A} & C \\
      C^{T} & B-\kappa_{B} \\
          \end{array}
    \right)\geq 0.
\end{eqnarray}
Hence all the $2\times 2$ minor submatrices of X must be positive.
Namely one has
$$ \left| \begin{array}{cc}
(A-\kappa_{A})_{ii} & C_{ij}\\
C_{ji} & (B-\kappa_{B})_{jj}\\
\end{array}
\right|\geq 0,
$$
i.e. $(A-\kappa_{A})_{ii}(B-\kappa_{B})_{jj}\geq C_{ij}^{2}$.
Summing over all i, j and using (\ref{lemma1}), we get
\begin{eqnarray*}
\sum\limits_{i,j=1}^{d^{2}}C_{i,j}^{2}&\leq&({\rm Tr} [A]-{\rm Tr}[
\kappa_{A}])({\rm Tr} [B]-{\rm Tr} [\kappa_{B}])\\
&=&(d-{\rm Tr}[\rho_{A}^{2}]-d+1)(d-{\rm Tr}[\rho_{B}^{2}]-d+1)
=(1-{\rm Tr}[\rho_{A}^{2}])(1-{\rm Tr}[\rho_{B}^{2}]).
\end{eqnarray*}
That is \be\label{th1} ||C||_{HS}^{2}\leq (1-{\rm
Tr}[\rho_{A}^{2}])(1-{\rm Tr}[\rho_{B}^{2}]), \ee where $||C||_{HS}$
stands for the Euclid norm of $C$, i.e. $||C||_{HS}=\sqrt{{\rm
Tr}[CC^{\dag}]}$.

Formulae (\ref{th2}) and (\ref{th1}) are independent and could be
complement. When
$$
\sqrt{(1-{\rm Tr}[\rho_{A}^{2}])(1-{\rm Tr}[\rho_{B}^{2}])}<
||C||_{HS} \leq ||C||_{KF} \leq \frac{(1-{\rm
Tr}[\rho_{A}^{2}])+(1-{\rm Tr}[\rho_{B}^{2}])}{2},
$$
(\ref{th1}) can recognize the entanglement but (\ref{th2}) can not.
When
$$
||C||_{HS} \leq \sqrt{(1-{\rm Tr}[\rho_{A}^{2}])(1-{\rm
Tr}[\rho_{B}^{2}])} \leq \frac{(1-{\rm Tr}[\rho_{A}^{2}])+(1-{\rm
Tr}[\rho_{B}^{2}])}{2} < ||C||_{KF},
$$
(\ref{th2}) can recognize the entanglement while (\ref{th1}) not.

The separability criteria based on covariance matrix approach can be
generalized to multipartite systems. We first consider the
tripartite case, $\rho_{ABC}\in {\cal {H}}^{A}_{d}\otimes{\cal
{H}}^{B}_{d}\otimes{\cal {H}}^{C}_{d}$. Take $d^{2}$ observables
$A_{k}$ on ${\cal {H}}_{A}$ resp. $B_{k}$ on ${\cal {H}}_{B}$ resp.
$C_{k}$ on ${\cal {H}}_{C}$. Set $\{M_{k}\}=\{A_{k}\otimes I\otimes
I, I\otimes B_{k}\otimes I, I\otimes I \otimes C_{k}\}$. The
covariance matrix defined by $(\ref{gammae})$ has then the following
block structure:
\begin{eqnarray}\label{d3}
\gamma=\left(%
    \begin{array}{ccc}
      A & D & E \\
      D^{T} & B & F \\
      E^{T} & F^{T} & C \\
    \end{array}
    \right),
\end{eqnarray}
where $A=\gamma(\rho_{A},\{A_{k}\})$,
$B=\gamma(\rho_{B},\{B_{k}\})$, $C=\gamma(\rho_{C},\{C_{k}\})$,
$D_{ij}=\langle A_{i}\otimes B_{j} \rangle_{\rho_{AB}}-\langle A_{i}
\rangle_{\rho_{A}}\langle B_{j} \rangle_{\rho_{B}}$, $E_{ij}=\langle
A_{i}\otimes C_{j} \rangle_{\rho_{AC}}-\langle A_{i}
\rangle_{\rho_{A}}\langle C_{j} \rangle_{\rho_{C}}$, $F_{ij}=\langle
B_{i}\otimes C_{j} \rangle_{\rho_{BC}}-\langle B_{i}
\rangle_{\rho_{B}}\langle C_{j} \rangle_{\rho_{C}}$.

\begin{theorem}\label{tbloch} If $\rho_{ABC}$ is fully separable, then
\begin{eqnarray}
\label{t112} ||D||_{HS}^{2}&\leq&
(1-{\rm Tr}[\rho_{A}^{2}])(1-{\rm Tr}[\rho_{B}^{2}]),\\
\label{t113} ||E||_{HS}^{2}&\leq&
(1-{\rm Tr}[\rho_{A}^{2}])(1-{\rm Tr}[\rho_{C}^{2}]),\\
\label{t123} ||F||_{HS}^{2}&\leq& (1-{\rm Tr}[\rho_{B}^{2}])(1-{\rm
Tr}[\rho_{C}^{2}]),
\end{eqnarray}
and
\begin{eqnarray}
\label{t212}2||D||_{KF}\leq (1-{\rm Tr}[\rho_{A}^{2}])+(1-{\rm Tr}[\rho_{B}^{2}]),\\
\label{t213}2||E||_{KF}\leq (1-{\rm Tr}[\rho_{A}^{2}])+(1-{\rm Tr}[\rho_{C}^{2}]),\\
\label{t223}2||F||_{KF}\leq (1-{\rm Tr}[\rho_{B}^{2}])+(1-{\rm
Tr}[\rho_{C}^{2}]).
\end{eqnarray}
\end{theorem}

{\sf[Proof]}\ For a tripartite product state
$\rho_{ABC}=\rho_{A}\otimes\rho_{B}\otimes\rho_{C}$, $D$, $E$ and
$F$ in $(\ref{d3})$ are zero. If $\rho_{ABC}$ is fully separable,
then there exist states $|a_{k}\rangle\langle a_{k}|$ in ${\cal
{H}}^{A}_{d}$, $|b_{k}\rangle\langle b_{k}|$ in ${\cal {H}}^{B}_{d}$
and $|c_{k}\rangle\langle c_{k}|$ in ${\cal {H}}^{C}_{d}$, and
$p_{k}$ such that $\gamma(\rho)\geq \kappa_{A}\oplus \kappa_{B}
\oplus \kappa_{C}$, where $\kappa_{A}=\sum
p_{k}\gamma(|a_{k}\rangle\langle a_{k}|,\{A_{k}\})$,
$\kappa_{B}=\sum p_{k}\gamma(|b_{k}\rangle\langle b_{k}|,\{B_{k}\})$
and $\kappa_{C}=\sum p_{k}\gamma(|c_{k}\rangle\langle
c_{k}|,\{C_{k}\})$,
 i.e.
\begin{eqnarray}\label{th3}
Y\equiv\left(%
    \begin{array}{ccc}
      A-\kappa_{A} & D & E \\
      D^{T} & B-\kappa_{B} & F \\
      E^{T} & F^{T} & C-\kappa_{C} \\
    \end{array}%
    \right)\geq 0.
\end{eqnarray}
Thus all the $2 \times 2$ minor submatrices of Y must be positive.
Selecting one with two rows and columns from the first two block
rows and columns of Y, we have
\be\label{lllf} \left|
\begin{array}{cc}
(A-\kappa_{A})_{ii} & D_{ij}\\
D_{ji} & (B-\kappa_{B})_{jj}\\
\end{array}
\right|\geq 0, \ee i.e. $(A-\kappa_{A})_{ii}(B-\kappa_{B})_{jj}\geq
|D_{ij}|^{2}$. Summing over all i, j and using ($\ref{lemma1}$), we
get
\begin{eqnarray*}
||D||_{HS}^{2}&=&\sum\limits_{i,j=1}^{d^{2}}D_{i,j}^{2}\leq({\rm Tr}
[A]-{\rm Tr}
[\kappa_{A}])({\rm Tr} [B]-{\rm Tr} [\kappa_{B}]\\
&=&(d-{\rm Tr}[\rho_{A}^{2}]-d+1)(d-{\rm Tr}[\rho_{B}^{2}]-d+1)
=(1-{\rm Tr}[\rho_{A}^{2}])(1-{\rm Tr}[\rho_{B}^{2}]),
\end{eqnarray*}
which proves (\ref{t112}). (\ref{t113}) and (\ref{t123}) can be
similarly proved.

From (\ref{lllf}) we also have
$(A-\kappa_{A})_{ii}+(B-\kappa_{B})_{ii}\geq 2|D_{ii}|$. Therefore
\begin{eqnarray}\label{th3p}
\sum\limits_{i}|D_{ii}|&\leq&\frac{({\rm Tr} [A]-{\rm Tr}
[\kappa_{A}])+({\rm Tr} [B]-{\rm Tr}
[\kappa_{B}]}{2} \nonumber\\
&=&\frac{(d-{\rm Tr}[\rho_{A}^{2}]-d+1)+(d-{\rm Tr}[\rho_{B}^{2}]-d+1)}{2}\nonumber \\
 &=&\frac{(1-{\rm Tr}[\rho_{A}^{2}])+(1-{\rm Tr}[\rho_{B}^{2}])}{2}.
\end{eqnarray}
Note that
$\sum\limits_{i=1}^{d^{2}}D_{ii}\leq\sum\limits_{i=1}^{d^{2}}|D_{ii}|$.
By using that ${\rm Tr}[MU] \leq||M||_{KF}={\rm Tr}
[\sqrt{MM^{\dag}}]$ for any matrix $M$ and any unitary $U$
\cite{matrix}, we have $\sum\limits_{i=1}^{d^{2}}D_{ii}\leq
||D||_{KF}$.

Let $D=U^{\dag}\Lambda V$ be the singular value decomposition of
$D$. Make a transformation of the orthonormal normalized basis of
the local orthonormal observable space: ${\widetilde{A}}_{i}=
\sum\limits_{l}U_{il}A_{l}$ and ${\widetilde{B}}_{j}=
\sum\limits_{m}V_{jm}^{*}B_{m}$. In the new basis we have
\be\label{ppp8}
{\widetilde{D}}_{ij}=\sum\limits_{lm}U_{il}D_{lm}V_{jm}=(UDV^{\dag})_{ij}=\Lambda_{ij}.
\ee Then (\ref{th3p}) becomes
\begin{eqnarray*}
\sum\limits_{i=1}^{d^{2}}\widetilde{D}_{ii}=||D||_{KF}\leq
\frac{(1-{\rm Tr}[\rho_{A}^{2}])+(1-{\rm Tr}[\rho_{B}^{2}])}{2}
\end{eqnarray*}
which proves (\ref{t212}). (\ref{t213}) and (\ref{t223}) can
similarly treated. $\hfill\Box$

We consider now the case that $\rho_{ABC}$ is bi-partite separable.

\begin{theorem}\label{tbloch1} If $\rho_{ABC}$ is a bi-partite separable state
with respect to the bipartite partition of the sub-systems $A$ and
$BC$ (resp. $AB$ and $C$; resp. $AC$ and $B$), then $(\ref{t112})$,
$(\ref{t113})$ and $(\ref{t212})$, $(\ref{t213})$ (resp.
$(\ref{t113})$, $(\ref{t123})$ and $(\ref{t213})$, $(\ref{t223})$;
resp. $(\ref{t112})$, $(\ref{t123})$ and $(\ref{t212})$,
$(\ref{t223})$) must hold. \end{theorem}

{\sf[Proof]}\ We prove the case that $\rho_{ABC}$ is bi-partite
separable with respect to the $A$ system and $BC$ systems partition.
The other cases can be similarly treated. In this case the matrices
$D$ and $E$ in the covariance matrix $(\ref{d3})$ are zero.
$\rho_{ABC}$ takes the form
$\rho_{ABC}=\sum\limits_{m}p_{m}\rho_{A}^{m}\otimes\rho_{BC}^{m}$.
Define $\kappa_{A}=\sum p_{m}\gamma(\rho_{A}^{m},\{A_{k}\})$,
$\kappa_{BC}=\sum p_{m}\gamma(\rho_{BC}^{m},\{B_{k}\otimes I,
I\otimes C_{k}\})$. $\kappa_{BC}$ has a form
\begin{eqnarray*}
\kappa_{BC}=\left(%
    \begin{array}{cc}
      \kappa_{B} & F^{'} \\
      (F^{'})^{T} & \kappa_{C} \\
    \end{array}%
    \right),
\end{eqnarray*}
where $\kappa_{B}=\sum p_{k}\gamma(|b_{k}\rangle\langle
b_{k}|,\{B_{k}\})$ and $\kappa_{C}=\sum
p_{k}\gamma(|c_{k}\rangle\langle c_{k}|,\{C_{k}\})$,
$(F^{'})_{ij}=\sum\limits_{m}p_{m}(\langle B_{i}\otimes
C_{j}\rangle_{\rho_{BC}^{m}}-\langle
B_{i}\rangle_{\rho_{B}^{m}}\langle C_{j}\rangle_{\rho_{C}^{m}})$. By
using the concavity of covariance matrix we have
\begin{eqnarray*}
\gamma(\rho_{ABC})\geq\sum\limits_{m}p_{m}\gamma(\rho_{A}^{m}\otimes\rho_{BC}^{m})
=\left(%
    \begin{array}{ccc}
      \kappa_{A} & 0 & 0 \\
      0 & \kappa_{B} & F^{'} \\
      0 & (F^{'})^{T} & \kappa_{C} \\
    \end{array}
    \right).
\end{eqnarray*}
Accounting to the method used in proving Theorem 2, we get
$(\ref{t112})$, $(\ref{t113})$ and $(\ref{t212})$, $(\ref{t213})$.
$\hfill\Box$

From Theorem $\ref{tbloch}$ and $\ref{tbloch1}$ we have the
following corollary.

\begin{corollary}
If two of the inequalities $(\ref{t112})$, $(\ref{t113})$ and
$(\ref{t123})$ (or $(\ref{t212})$, $(\ref{t213})$ and
$(\ref{t223})$) are violated, the state must be fully entangled.
\end{corollary}

The result of Theorem $\ref{tbloch}$ can be generalized to general
multipartite case $\rho\in {\cal {H}}_{d}^{(1)}\otimes{\cal
{H}}_{d}^{(2)}\otimes \cdots \otimes{\cal {H}}_{d}^{(N)}$. Define
$\hat{A}^{i}_{\alpha}=I\otimes I\otimes \cdots
\lambda_{\alpha}\otimes I\otimes\cdots\otimes I$, where
$\lambda_{0}=I/d$, $\lambda_{\alpha}$ ($\alpha=1, 2, \cdots
d^{2}-1$) are the normalized generators of $SU(d)$ satisfying ${\rm
Tr}[\lambda_{\alpha}\lambda_{\beta}]=\delta_{\alpha\beta}$ and
acting on the $i^{th}$ system ${\cal {H}}_{d}^{(i)}$, $i=1, 2,
\cdots, N$. Denote $\{M_{k}\}$ the set of all $\hat{A}^i_{\alpha}$.
Then the covariance matrix of $\rho$ can be written as
\begin{eqnarray}
\label{dn}
\gamma(\rho)=\left(%
    \begin{array}{ccc}
      {\cal {A}}_{11} & {\cal {A}}_{12} \cdots & {\cal {A}}_{1N} \\
      {\cal {A}}_{12}^{T} & {\cal {A}}_{22} \cdots & {\cal {A}}_{2N} \\
      \vdots      & \vdots             & \vdots      \\
      {\cal {A}}_{1N}^{T} & {\cal {A}}_{2N}^{T} \cdots & {\cal {A}}_{NN} \\
    \end{array}%
    \right),
\end{eqnarray}
where ${\cal {A}}_{ii}=\gamma(\rho, \{\hat{A}^{i}_{k}\})$ and
$({\cal
{A}}_{ij})_{mn}=\langle\hat{A}^{i}_{m}\otimes\hat{A}^{j}_{n}\rangle-\langle\hat{A}^{i}_{m}
\rangle\langle\hat{A}^{j}_{n}\rangle$ for $i\neq j$.

For a product state $\rho_{12\cdots N}$, ${\cal {A}}_{ij}$, $i\neq
j$, in $(\ref{dn})$ are zero matrices.
Define
\begin{eqnarray}
\kappa_{{\cal
{A}}_{ii}}=\sum\limits_{k}p_{k}\gamma(|\psi^{i}_{k}\rangle\langle\psi^{i}_{k}|,
\{\hat{A}^{i}_{l}\}).
\end{eqnarray}
Then for a fully separable multipartite state $\rho=\sum\limits_{k}p_{k}|\psi^{1}_{k}\rangle\langle\psi^{1}_{k}|
\otimes |\psi^{2}_{k}\rangle\langle\psi^{2}_{k}|\otimes \cdots
\otimes |\psi^{N}_{k}\rangle\langle\psi^{N}_{k}|$ one has
\begin{eqnarray}
Z=\left(%
    \begin{array}{ccc}
      {\cal {A}}_{11}-\kappa_{{\cal {A}}_{11}} & {\cal {A}}_{12} \cdots & {\cal {A}}_{1N} \\
      {\cal {A}}_{12}^{T} & {\cal {A}}_{22}-\kappa_{{\cal {A}}_{22}} \cdots & {\cal {A}}_{2N} \\
      \vdots      & \vdots             & \vdots      \\
      {\cal {A}}_{1N}^{T} & {\cal {A}}_{2N}^{T} \cdots & {\cal {A}}_{NN}-\kappa_{{\cal {A}}_{NN}} \\
    \end{array}%
    \right)\geq 0.
\end{eqnarray}
From which we have the following separability criterion for
multipartite systems:

\begin{theorem} If a state $\rho\in {\cal
{H}}_{d}^{(1)}\otimes{\cal {H}}_{d}^{(2)}\otimes \cdots \otimes{\cal
{H}}_{d}^{(N)}$ is fully separable, the following inequalities
\begin{eqnarray}
||{\cal {A}}_{ij}||_{HS}^{2}&&\leq (1-{\rm Tr}[\rho_{i}^{2}])(1-{\rm Tr}[\rho_{j}^{2}]),\\[3mm]
||{\cal {A}}_{ij}||_{KF}&&\leq \frac{(1-{\rm
Tr}[\rho_{i}^{2}])+(1-{\rm Tr}[\rho_{j}^{2}])}{2}
\end{eqnarray}
must be fulfilled for any $i\neq j$. \end{theorem}

\subsection{Normal form of quantum states}

In this subsection we show that the correlation matrix (CM) criterion can be
improved from the normal form obtained under filtering
transformations. Based on CM criterion
entanglement witness in terms of local orthogonal observables (LOOs)
\cite{sixia} for both bipartite and multipartite systems can be also constructed.

For bipartite case, $\rho\in{\cal {H}}={\cal {H}}_{A}\otimes {\cal
{H}}_{B}$ with $dim\,{\cal {H}}_{A}=M$, $dim\,{\cal {H}}_{B}=N$,
$M\leq N$, is mapped to the following form under local filtering
transformations \cite{verstraete}:
\begin{eqnarray}
\label{FT}\rho\rightarrow \widetilde{\rho}=\frac{(F_{A}\otimes
F_{B})\rho(F_{A}\otimes F_{B})^{\dag}}{{\rm Tr}[(F_{A}\otimes
F_{B})\rho(F_{A}\otimes F_{B})^{\dag}]},
\end{eqnarray}
where $F_{A/B}\in GL(M/N, \Cb)$ are arbitrary invertible matrices.
This transformation is also known as stochastic local operations
assisted by classical communication (SLOCC). By the definition it is
obvious that filtering transformation will preserve the separability
of a quantum state.

It has been shown that under local filtering operations one can
transform a strictly positive $\rho$ into a normal form
\cite{leinaas},
\begin{eqnarray}
\label{NF2} \widetilde{\rho}=\frac{(F_{A}\otimes
F_{B})\rho(F_{A}\otimes F_{B})^{\dag}}{{\rm Tr}[(F_{A}\otimes
F_{B})\rho(F_{A}\otimes
F_{B})^{\dag}]}=\frac{1}{MN}(I+\sum\limits_{i=1}^{M^{2}-1}{\xi}_{i}
G_{i}^{A}\otimes G_{i}^{B}),
\end{eqnarray}
where ${\xi}_{i}\geq 0$, $G_{i}^{A}$ and $G_{i}^{B}$ are some
traceless orthogonal observables. The matrices $F_{A}$ and $F_{B}$
can be obtained by minimizing the function
\begin{eqnarray}
f(A,B)=\frac{{\rm Tr} [\rho(A\otimes B)]}{(\det A)^{1/M}(\det
B)^{1/N}},
\end{eqnarray}
where $A=F_{A}^{\dag}F_{A}$ and $B=F_{B}^{\dag}F_{B}$. In fact, one
can choose $F_{A}^{0}\equiv |\det
(\rho_{A})|^{1/2M}(\sqrt{\rho_{A}})^{-1}$, and $F_{B}^{0}\equiv
|\det (\rho_{B}^{'})|^{1/2N}(\sqrt{\rho_{B}^{'}})^{-1}$, where
$\rho_{B}^{'}={\rm Tr} _{A}[I\otimes (\sqrt{\rho_{A}})^{-1}\rho
I\otimes (\sqrt{\rho_{A}})^{-1}]$. Then by iterations one can get
the optimal A and B. In particular, there is a matlab code available
in \cite{verstraete2}.

For bipartite separable states $\rho$, the CM separability criterion
\cite{julio} says that
\begin{eqnarray}
\label{CM2}||T||_{KF}\leq \sqrt{MN(M-1)(N-1)},
\end{eqnarray}
where $T$ is an $(M^{2}-1)\times (N^{2}-1)$ matrix with
$T_{ij}=MN\cdot {\rm Tr} [\rho \lambda_{i}^{A}\otimes
\lambda_{j}^{B}]$, $||T||_{KF}$ stands for the trace norm of $T$,
$\lambda_{k}^{A/B}$s are the generators of $SU(M/N)$ and have been
chosen to be normalized, ${\rm Tr}
[\lambda_{k}^{(A/B)}\lambda_{l}^{(A/B)}]=\delta_{kl}$.

As the filtering transformation does not change the separability of
a state, one can study the separability of $\tilde{\rho}$ instead of
$\rho$. Under the normal form (\ref{NF2}) the criterion (\ref{CM2})
becomes
\begin{eqnarray}\label{g1}
\sum\limits_{i}\xi_{i}\leq\sqrt{MN(M-1)(N-1)}.
\end{eqnarray}

In \cite{guhne1} a separability criterion based on local uncertainty
relation (LUR) has been obtained. It says that for any separable
state $\rho$, \be\label{hh} 1-\sum\limits_{k}\la G_{k}^{A}\otimes
G_{k}^{B}\ra-\frac{1}{2}\la G_{k}^{A}\otimes I - I\otimes
G_{k}^{B}\ra^{2}\geq 0, \ee where $G_{k}^{A/B}$s are LOOs such as
the normalized generators of $SU(M/N)$ and $G_{k}^{A}=0$ for
$k=M^{2}+1, \cdots, N^{2}$. The criterion is shown to be strictly
stronger than the realignment criterion \cite{chen}. Under the
normal form ($\ref{NF2}$) criterion (\ref{hh}) becomes
\begin{eqnarray*}
1&-&\sum\limits_{k}\la G_{k}^{A}\otimes G_{k}^{B}\ra-\frac{1}{2}\la
G_{k}^{A}\otimes
I - I\otimes G_{k}^{B}\ra^{2} \\
&=&1-\frac{1}{MN}\sum\limits_{k}\xi_{k}-\frac{1}{2}(\frac{1}{M}+\frac{1}{N})\geq
0,
\end{eqnarray*}
i.e.
\begin{eqnarray}\label{g2}
\sum\limits_{k}\xi_{k}\leq MN - \frac{M+N}{2}.
\end{eqnarray}
As $\sqrt{MN(M-1)(N-1)}\leq MN - \frac{M+N}{2}$ holds for any $M$
and $N$, from (\ref{g1}) and (\ref{g2}) it is obvious that the CM
criterion recognizes entanglement better when the normal form is
taken into account.

We now consider multipartite systems. Let $\rho$ be a strictly
positive density matrix in ${\cal {H}}={\cal {H}}_{1} \otimes {\cal
{H}}_{2} \otimes \cdots \otimes {\cal {H}}_{N}$, $dim\,{\cal
{H}}_{i}=d_i$. $\rho$ can be generally expressed in terms of the
$SU(n)$ generators $\lambda_{\alpha_{k}}$ \cite{hassan},
\be\label{OS} \ba{rcl}
\rho&=&\displaystyle\frac{1}{\Pi_{i}^{N}d_{i}}\left(\otimes_{j}^{N}I_{d_{j}}
+\sum\limits_{\{\mu_{1}\}}\sum\limits_{\alpha_{1}}
{\cal{T}}_{\alpha_{1}}^{\{\mu_{1}\}}\lambda_{\alpha_{1}}^{\{\mu_{1}\}}
+\sum\limits_{\{\mu_{1}\mu_{2}\}}\sum\limits_{\alpha_{1}\alpha_{2}}
{\cal{T}}_{\alpha_{1}\alpha_{2}}^{\{\mu_{1}\mu_{2}\}}\lambda_{\alpha_{1}}
^{\{\mu_{1}\}}\lambda_{\alpha_{2}}^{\{\mu_{2}\}}\right.\\[6mm]
&&+\sum\limits_{\{\mu_{1}\mu_{2}\mu_{3}\}}\sum\limits_{\alpha_{1}\alpha_{2}\alpha_{3}}
{\cal{T}}_{\alpha_{1}\alpha_{2}\alpha_{3}}^{\{\mu_{1}\mu_{2}\mu_{3}\}}\lambda_{\alpha_{1}}
^{\{\mu_{1}\}}\lambda_{\alpha_{2}}^{\{\mu_{2}\}}\lambda_{\alpha_{3}}^{\{\mu_{3}\}}\\[4mm]
&&+\cdots
+\sum\limits_{\{\mu_{1}\mu_{2}\cdots\mu_{M}\}}\sum\limits_{\alpha_{1}\alpha_{2}\cdots\alpha_{M}}
{\cal{T}}_{\alpha_{1}\alpha_{2}\cdots\alpha_{M}}^{\{\mu_{1}\mu_{2}\cdots\mu_{M}\}}\lambda_{\alpha_{1}}
^{\{\mu_{1}\}}\lambda_{\alpha_{2}}^{\{\mu_{2}\}}\cdots\lambda_{\alpha_{M}}^{\{\mu_{M}\}}\\[3mm]
&&\left.+\cdots +\sum\limits_{\alpha_{1}\alpha_{2}\cdots\alpha_{N}}
{\cal{T}}_{\alpha_{1}\alpha_{2}\cdots\alpha_{M}}^{\{1,2,\cdots,N\}}\lambda_{\alpha_{1}}
^{\{1\}}\lambda_{\alpha_{2}}^{\{2\}}\cdots\lambda_{\alpha_{N}}^{\{N\}}\right),
\ea \ee where $\lambda_{\alpha_{k}}^{\{\mu_{k}\}}=I_{d_{1}}\otimes
I_{d_{2}}\otimes\cdots\otimes \lambda_{\alpha_{k}}\otimes
I_{d_{\mu_{k}+1}}\otimes\cdots\otimes I_{d_{N}}$ with
$\lambda_{\alpha_{k}}$ appears at the $\mu_k$th position and
\begin{eqnarray*}
{\cal{T}}_{\alpha_{1}\alpha_{2}\cdots\alpha_{M}}
^{\{\mu_{1}\mu_{2}\cdots\mu_{M}\}}=\frac{\prod_{i=1}^{M}
d_{\mu_{i}}}{2^{M}}{\rm Tr}[\rho\lambda_{\alpha_{1}}
^{\{\mu_{1}\}}\lambda_{\alpha_{2}}^{\{\mu_{2}\}}\cdots\lambda_{\alpha_{M}}^{\{\mu_{M}\}}].
\end{eqnarray*}

The generalized CM criterion says that: if $\rho$ in (\ref{OS}) is
fully separable, then
\begin{eqnarray}\label{hhhh}
||{\cal{T}}^{\{\mu_{1},\mu_{2}, \cdots, \mu_{M}\}}||_{KF}\leq
\sqrt{\frac{1}{2^{M}}\prod_{k=1}^{M}d_{\mu_{k}}(d_{\mu_{k}}-1)},
\end{eqnarray}
for $2\leq M \leq N, \{\mu_{1},\mu_{2}, \cdots, \mu_{M}\} \subset
\{1, 2, \cdots, N\}$. The KF norm is defined by
\begin{eqnarray*}
||{\cal{T}}^{\{\mu_{1},\mu_{2}, \cdots, \mu_{M}\}}||_{KF}=max_{m=1,
2, \cdots, M}||{\cal{T}}_{(m)}||_{KF},
\end{eqnarray*}
where ${\cal{T}}_{(m)}$ is a kind of matrix unfolding of
${\cal{T}}^{\{\mu_{1},\mu_{2}, \cdots, \mu_{M}\}}$.

The criterion (\ref{hhhh}) can be improved by investigating the
normal form of (\ref{OS}).

\begin{theorem} By filtering transformations of the form
\begin{eqnarray}
\label{LFM} \widetilde{\rho}=F_{1}\otimes F_{2}\otimes\cdots\otimes
F_{N}\rho F_{1}^{\dag}\otimes F_{2}^{\dag}\otimes F_{N}^{\dag},
\end{eqnarray}
where $F_{i}\in GL(d_{i},\Bbb{C}), i=1, 2, \cdots N$, followed by
normalization, any strictly positive state $\rho$ can be transformed
into a normal form \be\label{NF} \ba{rcl}
\rho&=&\displaystyle\frac{1}{\Pi_{i}^{N}d_{i}}\left(\otimes_{j}^{N}I_{d_{j}}
+\sum\limits_{\{\mu_{1}\mu_{2}\}}\sum\limits_{\alpha_{1}\alpha_{2}}
{\cal{T}}_{\alpha_{1}\alpha_{2}}^{\{\mu_{1}\mu_{2}\}}\lambda_{\alpha_{1}}
^{\{\mu_{1}\}}\lambda_{\alpha_{2}}^{\{\mu_{2}\}}\right.\\[6mm]
&&\left.+\sum\limits_{\{\mu_{1}\mu_{2}\mu_{3}\}}\sum\limits_{\alpha_{1}\alpha_{2}\alpha_{3}}
{\cal{T}}_{\alpha_{1}\alpha_{2}\alpha_{3}}^{\{\mu_{1}\mu_{2}\mu_{3}\}}\lambda_{\alpha_{1}}
^{\{\mu_{1}\}}\lambda_{\alpha_{2}}^{\{\mu_{2}\}}\lambda_{\alpha_{3}}^{\{\mu_{3}\}}\right.\\[6mm]
&&+\cdots
+\sum\limits_{\{\mu_{1}\mu_{2}\cdots\mu_{M}\}}\sum\limits_{\alpha_{1}\alpha_{2}\cdots\alpha_{M}}
{\cal{T}}_{\alpha_{1}\alpha_{2}\cdots\alpha_{M}}^{\{\mu_{1}\mu_{2}\cdots\mu_{M}\}}\lambda_{\alpha_{1}}
^{\{\mu_{1}\}}\lambda_{\alpha_{2}}^{\{\mu_{2}\}}\cdots\lambda_{\alpha_{M}}^{\{\mu_{M}\}}\\[4mm]
&&\left.+\cdots +\sum\limits_{\alpha_{1}\alpha_{2}\cdots\alpha_{N}}
{\cal{T}}_{\alpha_{1}\alpha_{2}\cdots\alpha_{M}}^{\{1,2,\cdots,N\}}\lambda_{\alpha_{1}}
^{\{1\}}\lambda_{\alpha_{2}}^{\{2\}}\cdots\lambda_{\alpha_{N}}^{\{N\}}\right).
\ea \ee \end{theorem}

{\sf[Proof]}\ Let $D_{1}, D_{2}, \cdots, D_{N}$ be the sets of
density matrices of the $N$ subsystems. The cartesian product $D_{1}
\times D_{2} \times \cdots \times D_{N}$ consisting of all product
density matrices $\rho_{1} \otimes \rho_{2} \otimes \cdots \otimes
\rho_{N}$ with normalization ${\rm Tr} [\rho_{i}]=1$, $i=1, 2,
\cdots, N$, is a compact set of matrices on the full Hilbert space
$\cal {H}$. For the given density matrix $\rho$ we define the
following function of $\rho_{i}$
\begin{eqnarray*}
f(\rho_{1}, \rho_{2}, \cdots, \rho_{N})=\frac{{\rm Tr}
[\rho(\rho_{1}\otimes \rho_{2} \otimes\cdots\otimes
\rho_{N})]}{\prod_{i=1}^{N}\det (\rho_{i})^{1/d_{i}}}.
\end{eqnarray*}
The function is well-defined on the interior of $D_{1} \times D_{2}
\times \cdots \times D_{N}$ where $\det \rho_{i}>0$. As $\rho$ is
assumed to be strictly positive, we have ${\rm Tr}
[\rho(\rho_{1}\otimes \rho_{2} \otimes\cdots\otimes \rho_{N})]>0$.
Since $D_{1} \times D_{2} \times \cdots \times D_{N}$ is compact, we
have ${\rm Tr} [\rho(\rho_{1}\otimes \rho_{2} \otimes\cdots\otimes
\rho_{N})]\geq C>0$ with a lower bound C depending on $\rho$.

It follows that $f\rightarrow \infty$ on the boundary of $D_{1}
\times D_{2} \times \cdots \times D_{N}$ where at least one of the
$\rho_{i}$s satisfies that $\det \rho_{i}=0$. It follows further
that $f$ has a positive minimum on the interior of $D_{1} \times
D_{2} \times \cdots \times D_{N}$ with the minimum value attained
for at least one product density matrix $\tau_{1} \otimes \tau_{2}
\otimes \cdots \otimes \tau_{N}$ with $\det \tau_{i}>0$, $i=1, 2,
\cdots, N$. Any positive density matrix $\tau_{i}$ with $\det
\tau_{i}>0$ can be factorized in terms of Hermitian matrices $F_{i}$
as
\begin{eqnarray}
\label{tau}\tau_{i}=F_{i}^{\dag}F_{i}
\end{eqnarray}
where $F_{i}\in GL(d_{i}, \Bbb{C})$. Denote $F=F_{1} \otimes F_{2}
\otimes \cdots \otimes F_{N}$, so that $\tau_{1} \otimes \tau_{2}
\otimes \cdots \otimes \tau_{N} =F^{\dag}F$. Set
$\widetilde{\rho}=F\rho F^{\dag}$ and define
\begin{eqnarray*}
\widetilde{f}(\rho_{1}, \rho_{2}, \cdots \rho_{N}) &=&\frac{{\rm Tr}
[\widetilde{\rho}(\rho_{1}\otimes \rho_{2} \otimes\cdots\otimes
\rho_{N})]}{\prod_{i=1}^{N}\det (\rho_{i})^{1/d_{i}}}\\
&=&\prod_{i=1}^{N}\det (\tau_{i})^{1/d_{i}}\cdot\frac{{\rm Tr}
[\rho(F_{1}^{\dag}\rho_{1}F_{1}\otimes F_{2}^{\dag}\rho_{2}F_{2}
\otimes\cdots\otimes
F_{N}^{\dag}\rho_{N}F_{N})]}{\prod_{i=1}^{N}\det (\tau_{i})^{1/d_{i}}\det (\rho_{i})^{1/d_{i}}}\\
&=&\prod_{i=1}^{N}\det (\tau_{i})^{1/d_{i}}\cdot
f(F_{1}^{\dag}\rho_{1}F_{1}, F_{2}^{\dag}\rho_{2}F_{2}, \cdots,
F_{N}^{\dag}\rho_{N}F_{N}).
\end{eqnarray*}

We see that when $F^{\dag}_{i}\rho_{i}F_{i}=\tau_{i}$,
$\widetilde{f}$ has a minimum and
\begin{eqnarray*}
\rho_{1} \otimes \rho_{2} \otimes \cdots \otimes
\rho_{N}=(F^{\dag})^{-1} \tau_{1} \otimes \tau_{2} \otimes \cdots
\otimes \tau_{N} F^{-1}=I.
\end{eqnarray*}

Since $\widetilde{f}$ is stationary under infinitesimal variations
about the minimum it follows that
\begin{eqnarray*}
{\rm Tr}[\widetilde{\rho}\delta(\rho_{1} \otimes \rho_{2} \otimes
\cdots \otimes \rho_{N})]=0
\end{eqnarray*}
for all infinitesimal variations,
\begin{eqnarray*}
\delta(\rho_{1} \otimes \rho_{2} \otimes \cdots \otimes
\rho_{N})=\delta \rho_{1} \otimes I_{d_{2}} \otimes \cdots \otimes
I_{d_{N}}+ I_{d_{1}} \otimes \delta_{\rho_{2}} \otimes I_{d_{3}}
\otimes \cdots \otimes I_{d_{N}}\\
 + \cdots \cdots + I_{d_{1}} \otimes
I_{d_{2}} \otimes \cdots \otimes I_{d_{N-1}} \otimes \delta
\rho_{N},
\end{eqnarray*}
subjected to the constraint $\det (I_{d_{i}}+\delta\rho_{i})=1$,
which is equivalent to ${\rm Tr} [\delta\rho_{i}]=0$, $i=1, 2,
\cdots, N$, using $\det (e^{A})=e^{{\rm Tr} [A]}$ for a given matrix
$A$. Thus, $\delta\rho_{i}$ can be represented by the $SU$
generators, $\delta\rho_{i}=\sum\limits_{k}\delta
c_{k}^{i}\lambda_{k}^{i}$. It follows that ${\rm
Tr}[\widetilde{\rho}\lambda_{\alpha_{k}}^{\{\mu_{k}\}}]=0$ for any
$\alpha_{k}$ and $\mu_{k}$. Hence the terms proportional to
$\lambda_{\alpha_{k}}^{\{\mu_{k}\}}$ in ($\ref{OS}$) disappear.
$\hfill\Box$

\begin{corollary}
The normal form of a product state in ${\cal {H}}$ must be
proportional to the identity.
\end{corollary}

{\sf[Proof]}\ Let $\rho$ be such a state. From (\ref{NF}), we get
that
\begin{eqnarray}
\label{RMONF}\widetilde{\rho}_{i}={\rm Tr}_{1, 2, \cdots, i-1, i+1,
\cdots, N}[\rho]=\frac{1}{d_{i}}I_{d_{i}}.
\end{eqnarray}
Therefore for a product state $\rho$ we have
\begin{eqnarray*}
\widetilde{\rho}=\rho_{1}\otimes\rho_{2}\otimes\cdots\otimes\rho_{N}=\frac{1}{\prod_{i=1}^{N}
d_{i}}\otimes_{i=1}^{N}I_{d_{i}}.
\end{eqnarray*}
$\hfill\Box$

As an example for separability of multipartite states in terms of their
normal forms ($\ref{NF}$), we consider the PPT entangled edge state
\cite{acin}
\begin{eqnarray}
\rho=\left(%
    \begin{array}{cccccccc}
      1 & 0 & 0 & 0 & 0 & 0 & 0 & 1\\
      0 & a & 0 & 0 & 0 & 0 & 0 & 0\\
      0 & 0 & b & 0 & 0 & 0 & 0 & 0\\
      0 & 0 & 0 & c & 0 & 0 & 0 & 0\\
      0 & 0 & 0 & 0 & \frac{1}{c} & 0 & 0 & 0\\
      0 & 0 & 0 & 0 & 0 & \frac{1}{b} & 0 & 0\\
      0 & 0 & 0 & 0 & 0 & 0 & \frac{1}{a} & 0\\
      1 & 0 & 0 & 0 & 0 & 0 & 0 & 1\\
    \end{array}%
    \right)
\end{eqnarray}
mixed with noises:
\begin{eqnarray*}
\rho_{p}=p\rho+\frac{(1-p)}{8}I_{8}.
\end{eqnarray*}
Select $a=2, b=3$, and $c=0.6$. Using the criterion in \cite{hassan}
we get that $\rho_{p}$ is entangled for $0.92744< p \leq 1$. But
after transforming $\rho_{p}$ to its normal form (\ref{NF}), the
criterion can detect entanglement for $0.90285< p \leq 1$.

Here we indicate that the filtering transformation does not change
the PPT property. Let $\rho\in {\cal {H}}_{A}\otimes{\cal {H}}_{B}$
be PPT, i.e. $\rho^{T_{A}}\geq 0,$ and $ \rho^{T_{B}}\geq 0$. Let
$\widetilde{\rho}$ be the normal form of $\rho$. From ($\ref{FT}$)
we have
\begin{eqnarray*}
\widetilde{\rho}^{T_{A}}=\frac{(F_{A}^{*}\otimes F_{B}) \rho^{T_{A}}
(F_{A}^{T}\otimes F_{B}^{\dag})}{{\rm Tr}[(F_{A}\otimes
F_{B})\rho(F_{A}\otimes F_{B})^{\dag}]}.
\end{eqnarray*}
For any vector $|\psi \rangle$, we have
\begin{eqnarray*}
\langle \psi | \widetilde{\rho}^{T_{A}}| \psi \rangle
&=&\frac{\langle \psi |(F_{A}^{*}\otimes F_{B}) \rho^{T_{A}}
(F_{A}^{T}\otimes F_{B}^{\dag})| \psi \rangle}{{\rm
Tr}[(F_{A}\otimes F_{B})\rho(F_{A}\otimes F_{B})^{\dag}]}
\equiv\langle \psi^{'} |\rho^{T_{A}} |\psi^{'} \rangle\geq 0,
\end{eqnarray*}
where $|\psi^{'} \rangle=\frac{(F_{A}^{T}\otimes F_{B}^{\dag})| \psi
\rangle}{\sqrt{{\rm Tr}[(F_{A}\otimes F_{B})\rho(F_{A}\otimes
F_{B})^{\dag}]}}.$ $\widetilde{{\rho}}^{T_{B}}\geq 0$ can be proved
similarly. This property is also valid for multipartite case. Hence
a bound entangled state will be bound entangled under filtering
transformations.

\subsection{Entanglement witness based on correlation matrix criterion}

Entanglement witness (EW) is another way to describe separability. Based
on CM criterion we can further construct entanglement witness
in terms of LOOs. EW \cite{sixia} is an observable of the
composite system such that (i) nonnegative expectation values in all
separable states, (ii) at least one negative eigenvalue
(can recognizes at least one entangled state). Consider bipartite systems in ${\cal {H}}_{A}^{M}\otimes
{\cal {H}}_{B}^{N}$ with $M\leq N$.

\begin{theorem} For any LOOs $G^{A}_{k}$ and $G^{B}_{k}$,
$$
W=I-\alpha\sum\limits^{N^{2}-1}_{k=0}G^{A}_{k}\otimes G^{B}_{k}
$$
is an EW, where $\alpha=\frac{\sqrt{MN}}{\sqrt{(M-1)(N-1)}+1}$ and
\begin{eqnarray}\label{th2c}
G^{A}_{0}=\frac{1}{\sqrt{M}}I_{M},~~~
G^{B}_{0}=\frac{1}{\sqrt{N}}I_{N}.
\end{eqnarray}
\end{theorem}

{\sf[Proof]}\ Let
$\rho=\sum\limits_{l,m=0}^{N^{2}-1}T_{lm}\lambda_{l}^{A}\otimes\lambda_{m}^{B}$
be a separable state, where $\lambda_{k}^{A/B}$ are normalized
generators of $SU(M/N)$ with
$\lambda^{A}_{0}=\frac{1}{\sqrt{M}}I_{M}$,
$\lambda^{B}_{0}=\frac{1}{\sqrt{N}}I_{N}$. Any other LOOs
$G^{A/B}_{k}$ fulfill ($\ref{th2c}$) can be obtained from these
$\lambda$s through orthogonal transformations ${\cal{O}}^{A/B}$,
$G^{A/B}_{k}=\sum\limits_{l=0}^{N^{2}-1}{\cal
{O}}^{A/B}_{kl}\lambda_{l}$, where
${\cal{O}}^{A/B}=\left(%
    \begin{array}{cc}
      1 & 0  \\
      0 & {\cal {R}}^{A/B} \\
    \end{array}%
    \right)$,
${\cal {R}}^{A/B}$ are $(N^{2}-1)\times (N^{2}-1)$ orthogonal matrices. We have
$$
\begin{array}{l}\displaystyle
{\rm Tr}  [\rho W]=1-\alpha \frac{1}{\sqrt{MN}}- \alpha
\sum_{k=1}^{N^{2}-1}\sum_{l,m=1}^{N^{2}-1}{\cal {R}}_{kl}^{A}{\cal
{R}}_{km}^{B}
{\rm Tr}  [\rho(\lambda^{A}_{l}\otimes \lambda^{B}_{m})]\\[3mm]\displaystyle
=\frac{\sqrt{(M-1)(N-1)}}{\sqrt{(M-1)(N-1)}+1}-\frac{1}{\sqrt{MN}(\sqrt{(M-1)(N-1)}+1)}
\sum_{k=1}^{N^{2}-1}\sum_{l,m=1}^{N^{2}-1}{\cal {R}}_{kl}^{A}T_{lm}{\cal {R}}_{km}^{B}\\[3mm]\displaystyle
\geq\frac{\sqrt{MN(M-1)(N-1)}-||T||_{KF}}{\sqrt{MN}(\sqrt{(M-1)(N-1)}+1)}\geq
0,
\end{array}
$$
where we have used ${\rm Tr}  [{\cal {R}}T] \leq ||T||_{KF}$ for any
unitary ${\cal {R}}$ in the first inequality and the CM criterion in
the second inequality.

Now let
$\rho=\frac{1}{MN}(I_{MN}+\sum\limits_{i=1}^{M^{2}-1}s_{i}\lambda_{i}^{A}\otimes
I_{N}+\sum\limits_{j=1}^{N^{2}-1}r_{j}I_{M}\otimes\lambda_{j}^{B}+\sum\limits_{i=1}^{M^{2}-1}
\sum\limits_{j=1}^{N^{2}-1}T_{ij}\lambda_{i}^{A}\otimes
\lambda_{j}^{B})$ be a state in ${\cal {H}}_{A}^{M}\otimes{\cal
{H}}_{B}^{N}$  which violates the CM criterion. Denote
$\sigma_{k}(T)$ the singular values of $T$. By singular value
decomposition, one has $T= U^{\dag} \Lambda V^{*}$, where $\Lambda$
is a diagonal matrix with $\Lambda_{kk}=\sigma_{k}(T)$. Now choose
LOOs to be $G_{k}^{A}=\sum_{l}U_{kl}\lambda^{A}_{l}$,
$G_{k}^{B}=\sum_{m}V_{km}\lambda^{B}_{m}$ for $k=1,2,\cdots,N^{2}-1$
and $G_{0}^{A}=\frac{1}{M}I_{M}, G_{0}^{B}=\frac{1}{N}I_{N}$. We
obtain
\begin{eqnarray*}
{\rm Tr}  [\rho W]&=&1-\alpha \frac{1}{\sqrt{MN}}- \alpha
\sum_{k=1}^{N^{2}-1}\sum_{l,m=1}^{N^{2}-1}U_{kl}V_{km}
{\rm Tr}  [\rho(\lambda^{A}_{l}\otimes \lambda^{B}_{m})]\\
&=&\frac{\sqrt{(M-1)(N-1)}}{\sqrt{(M-1)(N-1)}+1}-\frac{1}{\sqrt{MN}(\sqrt{(M-1)(N-1)}+1)}
{\rm Tr} [UTV^{T}]\\
&=&\frac{\sqrt{MN(M-1)(N-1)}-||T||_{KF}}{\sqrt{MN}(\sqrt{(M-1)(N-1)}+1)}<
0
\end{eqnarray*}
where the CM criterion has been used in the last step. \hfill$\Box$

As the CM criterion can be generalized to multipartite form
\cite{hassan}, we can also define entanglement witness for
multipartite system in ${\cal {H}}_{1}^{d_{1}}\otimes {\cal
{H}}_{2}^{d_{2}}\otimes\cdots\otimes {\cal {H}}_{N}^{d_{N}}$. Set
$d(M)=\max\{d_{\mu_{i}}, i=1, 2, \cdots, M\}$. Choose LOOs
$G^{\{\mu_{i}\}}_{k}$ for $0\leq k\leq d(M)^{2}-1$ with
$G^{\{\mu_{i}\}}_{0}=\frac{1}{d_{\mu_{i}}}I_{d_{\mu_{i}}}$ and
define
\begin{eqnarray}
\label{EWFM} W^{(M)} = I-\beta^{(M)}
\sum_{k=0}^{d(M)^{2}-1}G_{k}^{\{\mu_{1}\}}\otimes
G_{k}^{\{\mu_{2}\}}\otimes \cdots \otimes G_{k}^{\{\mu_{M}\}},
\end{eqnarray}
where
$\beta^{(M)}=\frac{\sqrt{\prod_{i=1}^{M}d_{\mu_{i}}}}{1+\sqrt{\prod_{i=1}^{M}(d_{\mu_{i}}-1)}},
2 \leq M\leq N$. One can prove that ($\ref{EWFM}$) is an EW
candidate for multipartite states. First we assume
$||{\cal{T}}^{(M)}||_{KF}=||{\cal{T}}_{(m_{0})}||_{KF}$. Note that
for any ${\cal{T}}_{(m_{0})}$, there must exist an elementary
transformation $P$ such that
$\sum\limits_{k=1}^{d(M)^{2}-1}{\cal{T}}_{kk\cdots
k}^{\{\mu_{1}\mu_{2}\cdots\mu_{M}\}} ={\rm Tr}
[{\cal{T}}_{(m_{0})}P]$. Then for an N-partite separable state we
have
\begin{eqnarray*}
{\rm Tr}  [\rho W^{(M)}]&=&1-\beta^{(M)}
\frac{1}{\sqrt{\prod_{i=1}^{M}d_{\mu_{i}}}}-\beta^{(M)}
\frac{1}{\prod_{i=1}^{M}d_{\mu_{i}}}{\rm Tr} [{\cal{T}}_{(m_{0})}P]\\
&\geq&1-\beta^{(M)}
\frac{1}{\sqrt{\prod_{i=1}^{M}d_{\mu_{i}}}}-\beta^{(M)}
\frac{1}{\prod_{i=1}^{M}d_{\mu_{i}}}||{\cal{T}}_{(m_{0})}||_{KF}\\
&\geq&1-\beta^{(M)}
\frac{1}{\sqrt{\prod_{i=1}^{M}d_{\mu_{i}}}}-\beta^{(M)}
\frac{1}{\prod_{i=1}^{M}d_{\mu_{i}}}\sqrt{\prod_{k=1}^{M}d_{\mu_{k}}(d_{\mu_{k}}-1)}\\
&=&0
\end{eqnarray*}
for any $2\leq M\leq N$, where we have taken into account that $P$ is
orthognal and ${\rm Tr} [MU]\leq ||M||_{KF}$ for any unitary
$U$ at the first inequality. The second inequality is due to the
generalized CM criterion.

By choosing proper LOOs it is also easy to show that $W^{(M)}$ has
negative eigenvalues. For example for three qubits case , taking the
normalized pauli matrices as LOOs, one find a negative eigenvalue of
$W^{(M)}$, $(1-\sqrt{3})/2$.

\section{Concurrence and Tangle}\label{consec}

In this section, we focus on two important measures:
concurrence and tangle (see, \cite{contang}).
An elegant formula for concurrence of two-qubit states is derived analytically by Wootters
\cite{wotters,HillWootters}. This quantity has recently been shown
to play an essential role in describing quantum phase transition in
various interacting quantum many-body systems \cite{osterloh} and
may affect macroscopic properties of solids significantly
\cite{ghosh}. Furthermore, concurrence also provides an
estimation \cite{mintert} for the entanglement of formation (EOF)
\cite{bennett}, which quantifies the required minimally physical
resources to prepare a quantum state.

Let ${\cal H}_A$ (resp. ${\cal H}_B$) be an $M$ (resp.
$N$)-dimensional complex vector space with $|i\ra$, $i=1, \cdots, M$
(resp. $|j\ra$, $j=1, \cdots, N$), as an orthonormal basis. A
general pure state on ${\cal H}_A\otimes {\cal H}_B$ is of the form
\be\label{Psi22} |\Psi\ra=\sum_{i=1}^M\sum_{j=1}^N
a_{ij}|i\ra\otimes | j\ra, \ee where $a_{ij}\in\Bbb C$ satisfy the
normalization $\sum_{i=1}^M\sum_{j=1}^N a_{ij}a_{ij}^\ast=1$.

The concurrence of $(\ref{Psi22})$ is defined
by \cite{rungta,anote}
\be C(|\psi\ra)=\sqrt{2(1-{\rm
Tr}[\rho_{A}^{2}])} , \ee
where $\rho_{A}={\rm
Tr}_{B}[|\psi\ra\la\psi|]$. The definition is
extended to general mixed states $\rho=\sum_{i}p_{i}|\psi_{i}\ra\la\psi_{i}|$ by the convex roof,
\begin{eqnarray}\label{cmix}
C(\rho)=\min\limits_{\{p_{i},|\psi_{i}\ra\}}\sum_{i}p_{i}C(|\psi_{i}\ra).
\end{eqnarray}

For two qubits systems, the concurrence of $|\Psi\ra$ is given by:
\be \label{concurrence22}
C(|\Psi\ra)=|\la\Psi|\tilde{\Psi}\ra|=2|a_{11}a_{22}-a_{12}a_{21}|,\ee
where $|\tilde{\Psi}\ra=\sigma_{y}\otimes \sigma_{y}|\Psi^{*}\ra$,
$|\Psi^{*}\ra$ is the complex conjugate of $|\Psi\ra$, $\sigma_{y}$
is the Pauli matrix,
$\sigma_{y}=\left(\begin{array}{cc}0&-i\\i&0\end{array}\right).$

For a mixed two-qubit quantum state $\rho$, the entanglement of formation
$E(\rho)$ has a simple relation with the concurrence \cite{wotters,HillWootters}
$$E(\rho)=h(\frac{1+\sqrt{1-C(\rho)^2}}{2}),$$
where $h(x) = -x\log_2 x - (1-x)\log_2 (1-x)$,
\be
C(\rho)=\max{\{\lambda_1-\lambda_2-\lambda_3-\lambda_4, 0\}},\ee
where the $\lambda_i$s are the eigenvalues, in decreasing order, of
the Hermitian matrix $\sqrt{\sqrt{\rho}\widetilde{\rho}\sqrt{\rho}}$
and
$\widetilde{\rho}=(\sigma_y\otimes\sigma_y)\rho^\ast(\sigma_y\otimes\sigma_y)$.

Another entanglement measure called tangle is defined by
\be\label{tangle}\tau(|\psi\ra)=C^{2}(|\psi\ra)=2(1-{\rm
Tr}[\rho_{A}^{2}])
\ee
for a pure state $|\psi\ra$. For mixed state $\rho=\sum_{i}p_{i}|\psi_{i}\ra\la\psi_{i}|$,
the definition is given by
\begin{eqnarray}\label{t33}
\tau(\rho)=\min\limits_{\{p_{i},|\psi_{i}\ra\}}\sum_{i}p_{i}\tau(|\psi_{i}\ra).
\end{eqnarray}

For multipartite state $|\psi\ra\in {\cal {H}}_{1}\otimes{\cal
{H}}_{2}\otimes\cdots\otimes{\cal {H}}_{N}$, $dim {\cal
{H}}_{i}=d_i$, $i=1,...,N$, the
concurrence of $|\psi\ra$ is defined by \cite{multicon}
\begin{eqnarray}\label{xxxx}
C_{N}(|\psi\ra\la\psi|)=2^{1-\frac{N}{2}}\sqrt{(2^{N}-2)-\sum_{\alpha}{\rm
Tr}[\rho_{\alpha}^{2}]},
\end{eqnarray}
where $\alpha$ labels all different reduced density matrices.

Up to constant factor (\ref{xxxx}) can be also expressed in another way. Let $H$
denotes a $d$-dimensional vector space with basis $|i\ra$,
$i=1,2,...,d$. An $N$-partite pure state in
${H}\otimes\cdots\otimes{H}$ is generally of the form,
\begin{eqnarray}\label{purestate}
|\Psi\ra=\sum\limits_{i_{1},i_{2},\cdots
i_{N}=1}^{d}a_{i_{1},i_{2},\cdots i_{N}}|i_{1},i_{2},\cdots
i_{N}\ra,\quad a_{i_{1},i_{2},\cdots i_{N}}\in \Cb.
\end{eqnarray}

Let $\alpha$ and $\alpha^{'}$ (resp.$\beta$ and $\beta^{'}$) be
subsets of the subindices of $a$, associated to the same sub Hilbert
spaces but with different summing indices. $\alpha$ (or
$\alpha^{'}$) and $\beta$ (or $\beta^{'}$) span the whole space of
the given sub-indix of $a$. The generalized concurrence of
$|\Psi\ra$ is then given by \cite{anote}
\begin{eqnarray}\label{defmulticon}
C_{d}^{N}(|\Psi\ra)=\sqrt{\frac{d}{2m(d-1)}\sum\limits_{p}
\sum\limits_{\{\alpha,\alpha^{'},\beta,\beta^{'}\}}^{d}
|a_{\alpha\beta}a_{\alpha^{'}\beta^{'}}-a_{\alpha\beta^{'}}a_{\alpha^{'}\beta}|^{2}},
\end{eqnarray}
where $m=2^{N-1}-1$, $\sum\limits_{p}$ stands for the summation over
all possible combinations of the indices of $\alpha$ and $\beta$.

For a mixed multipartite quantum state,
$\rho=\sum_{i}p_{i}|\psi_{i}\ra\la\psi_{i}|$ in ${\cal
{H}}_{1}\otimes{\cal {H}}_{2}\otimes\cdots\otimes{\cal {H}}_{N}$,
the corresponding concurrence is given by the convex roof:
\begin{eqnarray}\label{defe}
C_{N}(\rho)=\min_{\{p_{i},|\psi_{i}\}\ra}\sum_{i}p_{i}C_{N}(|\psi_{i}\ra).
\end{eqnarray}

\subsection{Lower and upper bounds of concurrence}

Calculation of the concurrence for general mixed states are extremely difficult.
However, one can try to
find the lower and the upper bounds to estimate the exact values of
the concurrence \cite{chen,gao,vicente,optloo}.

\subsubsection{Lower bound of concurrence from covariance matrix
criterion}

In \cite{chen} a lower bound of $C(\rho)$ has been obtained,
\begin{eqnarray}\label{con0}
C(\rho)\geq
\sqrt{\frac{2}{M(M-1)}}\left[Max(||T_{A}(\rho)||,||R(\rho)||)-1\right],
\end{eqnarray}
where $T_A$ and $R$ stand for partial transpose with respect to
subsystem $A$ and the realignment respectively. This bound is further
improved based on local uncertainty relations \cite{vicente},
\begin{eqnarray}\label{con81}
C(\rho)\geq
\frac{M+N-2-\sum_{i}\triangle_{\rho}^{2}(G_{i}^{A}\otimes I+
I\otimes G_{i}^{B})}{\sqrt{2M(M-1)}},
\end{eqnarray}
where ${G_{i}^{A}}$ and $G_{i}^{B}$ are any set of local orthonormal
observables, $\triangle_{\rho}^2(X)={\rm Tr} [X^2\rho]-({\rm
Tr}[X\rho])^2$.

Bound (\ref{con81}) again depends on the choice of the local
orthonormal observables. This bound
can be optimized, in the sense that a local orthonormal
observable-independent up bound of the right hand side of
(\ref{con81}) can be obtained.

\begin{theorem} Let $\rho$ be a bipartite state in ${\cal
{H}}_{M}^{A}\otimes{\cal {H}}_{N}^{B}$. Then $C(\rho)$ satisfies
\begin{eqnarray}\label{con2}
C(\rho)\geq \frac{2||C||_{KF}-(1-{\rm Tr}[\rho_{A}^{2}])-(1-{\rm
Tr}[\rho_{B}^{2}])}{\sqrt{2M(M-1)}}.
\end{eqnarray}
\end{theorem}

{\sf[Proof]}\ The other orthonormal normalized basis of the local
orthonormal observable space can be obtained from $A_{i}$ and
$B_{i}$ by unitary transformations $U$ and $V$:
${\widetilde{A}}_{i}= \sum\limits_{l}U_{il}A_{l}$ and
${\widetilde{B}}_{j}= \sum\limits_{m}V_{jm}^{*}B_{m}$. Select $U$
and $V$ so that $C=U^{\dag}\Lambda V$ is the singular value
decomposition of $C$. Then the new observables can be written as
${\widetilde{A}}_{i}= \sum\limits_{l}U_{il}A_{l}$,
${\widetilde{B}}_{j}=-\sum\limits_{m}V_{jm}^{*}B_{m}$. We have
\begin{eqnarray*}
\sum_{i}\triangle_{\rho}^{2}({\widetilde{A}}_{i}\otimes I+ I\otimes
{\widetilde{B}}_{i})
&=&\sum_{i}[\triangle_{\rho_{A}}^{2}({\widetilde{A}}_{i})
+\triangle_{\rho_{A}}^{2}({\widetilde{B}}_{i})
+2(\la{\widetilde{A}}_{i}\otimes{\widetilde{B}}_{i}\ra
-\la{\widetilde{A}}_{i}\ra\la{\widetilde{B}}_{i}\ra)]\\
&=&M-{\rm Tr}[\rho_{A}^{2}]+N-{\rm Tr}[\rho_{B}^{2}]-2\sum_{i}(UCV^{\dag})_{ii}\\
&=&M-{\rm Tr}[\rho_{A}^{2}]+N-{\rm Tr}[\rho_{B}^{2}]-2||C||_{KF}.
\end{eqnarray*}
Substituting above relation to (\ref{con81}) one gets (\ref{con2}).
$\hfill\Box$

Bound $(\ref{con2})$ does not depend on the choice of local
orthonormal observables. It can be easily applied and realized by
direct measurements in experiments. It is in accord with the result
in \cite{optloo} where optimization of entanglement witness based on
local uncertainty relation has been taken into account. As an
example let us consider the $3\times 3$ bound entangled state
\cite{bennett}, \be\label{3x3}
\rho=\frac{1}{4}(I_{9}-\sum\limits_{i=0}^{4}|\xi_{i}\ra\la\xi_{i}|),
\ee where $I_9$ is the $9\times 9$ identity matrix, $|\xi_{0}\ra
=\frac{1}{\sqrt{2}}|0\ra(|0\ra-|1\ra)$, $|\xi_{1}\ra =
\frac{1}{\sqrt{2}}(|0\ra-|1\ra)|2\ra$, $|\xi_{2}\ra =
\frac{1}{\sqrt{2}}|2\ra(|1\ra-|2\ra)$, $|\xi_{3}\ra =
\frac{1}{\sqrt{2}}(|1\ra-|2\ra)|0\ra$, $|\xi_{4}\ra =
\frac{1}{3}(|0\ra+|1\ra+|2\ra)(|0\ra+|1\ra+|2\ra)$. We simply choose
the local orthonormal observables to be the normalized generators of
$SU(3)$. Formula $(\ref{con0})$ gives $C(\rho)\geq 0.050$. Formula
$(\ref{con81})$ gives $C(\rho)\geq 0.052$ \cite{vicente}, while
formula $(\ref{con2})$ yields a better lower bound $C(\rho)\geq
0.0555$.

\begin{figure}[h]
\begin{center}
\resizebox{10cm}{!}{\includegraphics{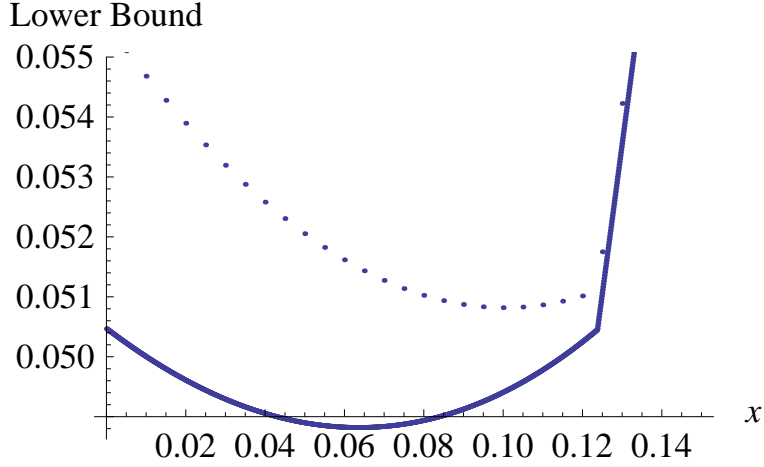}}
\end{center}
\caption{Lower bounds from $(\ref{con2})$ (dashed line) and
$(\ref{con0})$ (solid line)} \label{fig111}
\end{figure}

If we mix the bound entangled state (\ref{3x3}) with
$|\psi\ra=\frac{1}{\sqrt{3}}\sum\limits_{i=0}^{2}|ii\ra$,
$\rho^{'}=(1-x)\rho+x|\psi\ra\la\psi|$, it is easily seen that
$(\ref{con2})$ gives a better lower bound of concurrence than
formula $(\ref{con0})$ (Fig. \ref{fig111}).

\subsubsection{Lower bound of concurrence from ``two-qubit" decomposition}

In \cite{ou} the authors derived an analytical lower bound of
concurrence for arbitrary bipartite quantum states by decomposing
the joint Hilbert space into many $2\otimes 2$ dimensional
subspaces, which does not involve any optimization procedure and
gives an effective evaluation of entanglement together with an
operational sufficient condition for the distill ability of any
bipartite quantum states.

\begin{enumerate}
{\sf \item[{(1)}] Lower bound of concurrence for bipartite states}
\end{enumerate}

The lower bound $\tau_{2}$ of concurrence for
bipartite states has been obtained in \cite{ou}. For a bipartite
quantum state $\rho$ in $H\otimes H$, the concurrence $C(\rho)$
satisfies
\begin{eqnarray}\label{ou}
\tau_{2}(\rho)\equiv\frac{d}{2(d-1)}\sum_{m,n=1}^{\frac{d(d-1)}{2}}C_{mn}^{2}(\rho)\leq
C^{2}(\rho),
\end{eqnarray}
where $C_{mn}(\rho)=\max\{0,\lambda_{mn}^{(1)}-\lambda_{mn}^{(2)}
-\lambda_{mn}^{(3)}-\lambda_{mn}^{(4)}\}$ with $\lambda_{mn}^{(1)}, ...,
\lambda_{mn}^{(4)}$ the square roots of the four nonzero
eigenvalues, in decreasing order, of the non-Hermitian matrix
$\rho\widetilde{\rho}_{mn}$ with
$\widetilde{\rho}_{mn}=(L_{m}\otimes L_{n})\rho^{*}(L_{m}\otimes
L_{n})$, $L_{m}$ and $L_{n}$  are the generators of $SO(d)$.

The lower bound $\tau_2$ in Eq.(\ref{ou}) in fact
characterizes all two-qubit's entanglement in a high dimensional bipartite state.
One can directly verify that there are at most $4\times
4=16$ nonzero elements in each matrix $\rho\widetilde{\rho}_{mn}$.
These elements constitute a $4\times
4$ matrix $\varrho(\sigma_y\otimes \sigma_y)\varrho^*(\sigma_y%
\otimes \sigma_y)$,  where $\sigma_y$ is the Pauli matrix, the matrix $\varrho$ is a
submatrix of the original $\rho$:
\begin{equation}\label{c11}
\varrho=\left(\begin{array}{cccc}
               \rho_{ik,ik} &  \rho_{ik,il} &  \rho_{ik,jk} &  \rho_{ik,jl}
               \\
               \rho_{il,ik} &  \rho_{il,il} &  \rho_{il,jk} &  \rho_{il,jl}
               \\
               \rho_{jk,ik} &  \rho_{jk,il} &  \rho_{jk,jk} &  \rho_{jk,jl}
               \\
               \rho_{jl,ik} &  \rho_{jl,il} &  \rho_{jl,jk} &  \rho_{jl,jl} \\
             \end{array}\right),
\end{equation}
$i\neq j$ and $k \neq l$, with subindices $i$ and $j$ associated with
the first space, $k$ and $l$ with the second space.
The two-qubit submatrix $\varrho$
is not normalized but positive semidefinite.
$\mathcal {C}_{mn}$ are just the concurrences of these states (\ref{c11}).

The bound $\tau_2$ provides a much clearer structure of entanglement,
which not only yields an effective separability criterion and an
easy evaluation of entanglement, but also helps one to classify
mixed-state entanglement.

\begin{enumerate}
{\sf \item[{(2)}] Lower bound of concurrence for multipartite states}
\end{enumerate}

We first consider tripartite case. A general pure state on $H\otimes
H\otimes H$ is of the form
\begin{eqnarray}\label{pure}
|\Psi\ra=\sum\limits_{i,j,k=1}^{d}a_{ijk}|ijk\ra,\quad a_{ijk}\in
\Cb,\quad \sum\limits_{i,j,k=1}^{d}a_{ijk}a_{ijk}^{*}=1
\end{eqnarray}
with
\begin{eqnarray*}
&&C_{d}^{3}(|\Psi\ra)=\sqrt{\frac{d}{6(d-1)}}\times \\
&&\sqrt{\sum(|a_{ijk}a_{pqm}-a_{ijm}a_{pqk}|^{2}
+|a_{ijk}a_{pqm}-a_{iqk}a_{pjm}|^{2}+|a_{ijk}a_{pqm}-a_{pjk}a_{iqm}|^{2})}
\end{eqnarray*}
or equivalently
\begin{eqnarray}\label{def12}
C_{d}^{3}(|\Psi\ra)=\sqrt{\frac{d}{6(d-1)}(3-({\rm
Tr}[\rho_{1}^{2}]+{\rm Tr}[\rho_{2}^{2}]+{\rm Tr}[\rho_{3}^{2}]))},
\end{eqnarray}
where $\rho_{1}={\rm Tr}_{23}[\rho], \rho_{2}={\rm Tr}_{13}[\rho],
\rho_{3}={\rm Tr}_{12}[\rho]$ are the reduced density matrices of
$\rho=|\Psi\ra\la\Psi|$.

Define $C_{\alpha\beta}^{12|3}(|\Psi\ra)=|a_{ijk}a_{pqm}-a_{ijm}a_{pqk}|$,
$C_{\alpha\beta}^{13|2}(|\Psi\ra)=|a_{ijk}a_{pqm}-a_{iqk}a_{pjm}|$,
$C_{\alpha\beta}^{23|1}(|\Psi\ra)=|a_{ijk}a_{pqm}-a_{pjk}a_{iqm}|$,
where $\alpha$ and $\beta$ of $C_{\alpha\beta}^{12|3}$ (resp.
$C_{\alpha\beta}^{13|2}$ resp. $C_{\alpha\beta}^{23|1}$) stand for
the sub-indices of $a$ associated with the subspaces $1,2$ and $3$
(resp. $1,3$ and $2$ resp. $2,3$ and $1$). Let $L^{i_{1}i_{2}\cdots
i_{N}}$ denote the generators of group $SO(d_{i_{1}}d_{i_{2}}\cdots
d_{i_{N}})$ associated to the subsystems $i_{1},i_{2},\cdots,
i_{N}$. Then for a tripartite pure state ($\ref{pure}$), one has
\begin{eqnarray*}
C_{d}^{3}(|\Psi\ra)&=&\sqrt{\frac{d}{6(d-1)}
\sum_{\alpha}^{\frac{d^{2}(d^{2}-1)}{2}}\sum_{\beta}^{\frac{d(d-1)}{2}}
[({C_{\alpha\beta}^{12|3}(|\Psi\ra)})^{2}
+({C_{\alpha\beta}^{13|2}}(|\Psi\ra))^{2}+({C_{\alpha\beta}^{23|1}(|\Psi\ra)})^{2}]}\nonumber\\
&=&\sqrt{\frac{d}{6(d-1)} \sum_{\alpha\beta}
[(|\la\Psi|S_{\alpha\beta}^{12|3}|\Psi^{*}\ra|)^{2}
+(|\la\Psi|S_{\alpha\beta}^{13|2}|\Psi^{*}\ra|)^{2}+(|\la\Psi|S_{\alpha\beta}^{23|1}|\Psi^{*}\ra|)^{2}]},
\end{eqnarray*}
where $S_{\alpha\beta}^{12|3}=(L_{\alpha}^{12}\otimes
L^{3}_{\beta})$, $S_{\alpha\beta}^{13|2}=(L_{\alpha}^{13}\otimes
L^{2}_{\beta})$ and $S_{\alpha\beta}^{23|1}=(L_{\beta}^{1}\otimes
L^{23}_{\alpha})$.

\begin{theorem}\label{oubound} For an arbitrary mixed state $\rho$ in
$H\otimes H\otimes H$, the concurrence $C(\rho)$ satisfies
\begin{eqnarray}\label{11}
\tau_{3}(\rho)\equiv\frac{d}{6(d-1)}\sum_{\alpha}^{\frac{d^{2}(d^{2}-1)}{2}}\sum_{\beta}^{\frac{d(d-1)}{2}}
[({C_{\alpha\beta}^{12|3}(\rho)})^{2}
+({C_{\alpha\beta}^{13|2}(\rho)})^{2}+({C_{\alpha\beta}^{23|1}(\rho)})^{2}]\leq
C^{2}(\rho),
\end{eqnarray}
where $\tau_{3}(\rho)$ is a lower bound of $C(\rho)$,
\begin{eqnarray}
C_{\alpha\beta}^{12|3}(\rho)=\max\{0,\lambda(1)_{\alpha\beta}^{12|3}-\lambda(2)_{\alpha\beta}^{12|3}
-\lambda(3)_{\alpha\beta}^{12|3}-\lambda(4)_{\alpha\beta}^{12|3}\},
\end{eqnarray}
$\lambda(1)_{\alpha\beta}^{12|3}, \lambda(2)_{\alpha\beta}^{12|3},
\lambda(3)_{\alpha\beta}^{12|3}, \lambda(4)_{\alpha\beta}^{12|3}$
are the square roots of the four nonzero eigenvalues, in decreasing
order, of the non-Hermitian matrix
$\rho\widetilde{\rho}_{\alpha\beta}^{12|3}$ with
$\widetilde{\rho}_{\alpha\beta}^{12|3}=S_{\alpha\beta}^{12|3}\rho^{*}S_{\alpha\beta}^{12|3}$.
$C_{\alpha\beta}^{13|2}(\rho)$ and $C_{\alpha\beta}^{23|1}(\rho)$
are defined in a similar way to $C_{\alpha\beta}^{12|3}(\rho)$.
\end{theorem}

{\sf[Proof]}\ Set $|\xi_{i}\ra=\sqrt{p_{i}}|\psi_{i}\ra$,
$x_{\alpha\beta}^{i}=|\la\xi_{i}|S_{\alpha\beta}^{12|3}|\xi_{i}^{*}\ra|$,
$y_{\alpha\beta}^{i}=|\la\xi_{i}|S_{\alpha\beta}^{13|2}|\xi_{i}^{*}\ra|$
and
$z_{\alpha\beta}^{i}=|\la\xi_{i}|S_{\alpha\beta}^{1|23}|\xi_{i}^{*}\ra|$.
We have, from Minkowski inequality
\begin{eqnarray}
C(\rho)&=&\min\sum_{i}\sqrt{\frac{d}{6(d-1)} \sum_{\alpha\beta}
\left[(x_{\alpha\beta}^{i})^{2} +(y_{\alpha\beta}^{i})^{2}
+(z_{\alpha\beta}^{i})^{2}\right]}\nonumber\\
&\geq&\min\sqrt{\frac{d}{6(d-1)}
\sum_{\alpha\beta}\left(\sum_{i}[(x_{\alpha\beta}^{i})^{2}
+(y_{\alpha\beta}^{i})^{2}+(z_{\alpha\beta}^{i})^{2}]^{\frac{1}{2}}\right)^{2}}.\nonumber
\end{eqnarray}

Noting that for nonnegative real variables $x_{\alpha}$,
$y_{\alpha}$, $z_{\alpha}$ and given
$X=\sum\limits_{\alpha=1}^{N}x_{\alpha}$,
$Y=\sum\limits_{\alpha=1}^{N}Y_{\alpha}$ and
$Z=\sum\limits_{\alpha=1}^{N}z_{\alpha}$, by using Lagrange
multipliers one obtains that the following inequality holds,
\begin{eqnarray}
\sum\limits_{\alpha=1}^{N}(x_{\alpha}^{2}+y_{\alpha}^{2}+z_{\alpha}^{2})^{\frac{1}{2}}\geq
(X^{2}+Y^{2}+Z^{2})^{\frac{1}{2}}.
\end{eqnarray}
Therefore we have
\begin{eqnarray}\label{14p}
C(\rho)&\geq&\min\sqrt{\frac{d}{6(d-1)}
\sum_{\alpha\beta}[(\sum_{i}x_{\alpha\beta}^{i})^{2}
+(\sum_{i}y_{\alpha\beta}^{i})^{2}+(\sum_{i}z_{\alpha\beta}^{i})^{2}]}\nonumber\\
&\geq&\sqrt{\frac{d}{6(d-1)}
\sum_{\alpha\beta}[(\min\sum_{i}x_{\alpha\beta}^{i})^{2}
+(\min\sum_{i}y_{\alpha\beta}^{i})^{2}+(\min\sum_{i}z_{\alpha\beta}^{i})^{2}]}.\nonumber\\
\end{eqnarray}

The values of
$C_{\alpha\beta}^{12|3}(\rho)\equiv\min\sum\limits_{i}x_{\alpha\beta}^{i}$,
$C_{\alpha\beta}^{13|2}(\rho)\equiv\min\sum\limits_{i}y_{\alpha\beta}^{i}$
and
$C_{\alpha\beta}^{23|1}(\rho)\equiv\min\sum\limits_{i}z_{\alpha\beta}^{i}$
can be calculated by using the similar procedure in \cite{wotters}.
Here we compute the value of $C_{\alpha\beta}^{12|3}(\rho)$ in
detail. The values of $C_{\alpha\beta}^{13|2}(\rho)$ and
$C_{\alpha\beta}^{23|1}(\rho)$ can be obtained analogously.

Let $\lambda_{i}$ and $|\chi_{i}\ra$ be eigenvalues and eigenvectors
of $\rho$ respectively. Any decomposition of $\rho$ can be obtained
from a unitary $d^{3}\times d^{3}$ matrix $V_{ij}$,
$|\xi_{j}\ra=\sum\limits_{i=1}^{d^{3}}V^{*}_{ij}(\sqrt{\lambda_{i}}|\chi_{i}\ra)$.
Therefore one has
$\la\xi_{i}|S^{12|3}_{\alpha\beta}|\xi_{j}^{*}\ra=(VY_{\alpha\beta}V^{T})_{ij}$,
where the matrix $Y_{\alpha\beta}$ is defined by
$(Y_{\alpha\beta})_{ij}=\la\chi_{i}|S^{12|3}_{\alpha\beta}|\chi_{j}^{*}\ra$.
Namely
$C_{\alpha\beta}^{12|3}(\rho)=\min\sum_{i}|[VY_{\alpha\beta}V^{T}]_{ii}|$,
which has an analytical expression \cite{wotters},
$C_{\alpha\beta}^{12|3}(\rho)=\max\{0,\lambda(1)_{\alpha\beta}^{12|3}
-\sum_{j>1}\lambda(j)_{\alpha\beta}^{12|3}\}$, where
$\lambda^{12|3}_{\alpha\beta}(k)$ are the square roots of the
eigenvalues of the positive Hermitian matrix
$Y_{\alpha\beta}Y_{\alpha\beta}^{\dag}$, or equivalently the
non-Hermitian matrix $\rho\widetilde{\rho}_{\alpha\beta}$, in
decreasing order. Here as the matrix $S_{\alpha\beta}^{12|3}$ has
$d^{2}-4$ rows and $d^{2}-4$ columns that are identically zero, the
matrix $\rho\widetilde{\rho}_{\alpha\beta}$ has a rank no greater
than 4, i.e., $\lambda_{\alpha\beta}^{12|3}(j)=0$ for $j\geq 5$.
From Eq.($\ref{14p}$) we have Eq.($\ref{11}$).$\hfill\Box$

Theorem $\ref{oubound}$ can be directly generalized to arbitrary
multipartite case.

\begin{theorem} For an arbitrary $N$-partite state $\rho\in
{{H}}\otimes {H}\otimes\cdots\otimes {H}$, the concurrence defined
in ($\ref{def}$) satisfies:
\begin{eqnarray}
\tau_{N}(\rho)\equiv\frac{d}{2m(d-1)}\sum_{p}\sum_{\alpha\beta}(C_{\alpha\beta}^{p}(\rho))^{2}\leq
C^{2}(\rho),
\end{eqnarray}
where $\tau_{N}(\rho)$ is the lower bound of $C(\rho)$,
$\sum\limits_{p}$ stands for the summation over all possible
combinations of the indices of $\alpha,\beta$,
$C_{\alpha\beta}^{p}(\rho)=\max\{0,
\lambda(1)_{\alpha\beta}^{p}-\lambda(2)_{\alpha\beta}^{p}
-\lambda(3)_{\alpha\beta}^{p}-\lambda(4)_{\alpha\beta}^{p}\}$,
$\lambda(i)_{\alpha\beta}^{p}$, $i=1, 2, 3, 4$, are the square roots
of the four nonzero eigenvalues, in decreasing order, of the
non-Hermitian matrix $\rho\widetilde{\rho}_{\alpha\beta}^{p}$ where
$\widetilde{\rho}_{\alpha\beta}^{p}=S_{\alpha\beta}^{p}\rho^{*}S_{\alpha\beta}^{p}$.
\end{theorem}

\begin{enumerate}
{\sf \item[{(3)}] Lower bound and separability}
\end{enumerate}

An N-partite quantum state $\rho$ is fully separable if and only if
there exist $p_{i}$ with $p_{i}\geq0, \sum\limits_{i}p_{i}=1$ and
pure states $\rho_{i}^{j}=|\psi_{i}^{j}\ra\la\psi_{i}^{j}|$ such
that
\begin{eqnarray}
\rho=\sum_{i}p_{i}\rho_{i}^{1}\otimes\rho_{i}^{2}\otimes\cdots\otimes\rho_{i}^{N}.
\end{eqnarray}

It is easily verified that for a fully separable multipartite state
$\rho$, $\tau_{N}(\rho)=0$. Thus $\tau_{N}(\rho)>0$ indicates that
there must be some kinds of entanglement inside the quantum state,
which shows that the lower bound $\tau_{N}(\rho)$ can be used to
recognize entanglement.

As an example we consider a tripartite quantum state \cite{acin},
$\rho=\frac{1-p}{8}I_{8}+p|W\ra\la W|$, where $I_{8}$ is the
$8\times8$ identity matrix, and
$|W\ra=\frac{1}{\sqrt{3}}(|100\ra+|010\ra+|001\ra)$ is the
tripartite W-state. Select an entanglement witness operator to be
${\cal {W}}=\frac{1}{2}I_{8}-|GHZ\ra\la GHZ|$, where
$|GHZ\ra=\frac{1}{\sqrt{2}}(|000\ra+|111\ra)$ to be the tripartite
GHZ-state. From the condition ${\rm Tr}[{\cal {W}}\rho]<0$, the
entanglement of $\rho$ is detected for $\frac{3}{5} < p \leq 1$ in
\cite{acin}. In \cite{hassan} the authors have obtained the
generalized correlation matrix criterion which says if an N-qubit
quantum state is fully separable then the inequality $||{\cal
{T}}^{N}||_{KF}\leq 1$ must hold, where $||{\cal
{T}}^{N}||_{KF}=\max\{||{\cal {T}}_{n}^{N}||_{KF}\}$, ${\cal
{T}}_{n}^{N}$ is a kind of matrix unfold of
$t_{\alpha_{1}\alpha_{2}\cdots\alpha_{N}}$ defined by
$t_{\alpha_{1}\alpha_{2}\cdots\alpha_{N}}={\rm
Tr}[\rho\sigma_{\alpha_{1}}^{(1)} \sigma_{\alpha_{2}}^{(2)}\cdots
\sigma_{\alpha_{N}}^{(N)}]$ and $\sigma_{\alpha_{i}}^{(i)}$ stands
for the pauli matrix. Now using the generalized correlation matrix
criterion the entanglement of $\rho$ is detected for $0.3068 < p
\leq 1$. From theorem \ref{oubound}, we have that the lower bound
$\tau_{3}(\rho)>0$ for $0.2727 < p \leq 1$. Therefore the bound (\ref{ll})
detects entanglement better than these two criteria in this case. If
we replace W with GHZ state in $\rho$, the criterion in
\cite{hassan} detects the entanglement of $\rho$ for $0.35355 < p
\leq 1$, while $\tau_{3}(\rho)$ detects, again better, the
entanglement for $0.2 < p \leq 1$.

Nevertheless for PPT states $\rho$, we have $\tau_{3}(\rho)=0$,
which can be seen in the following way. A density matrix $\rho$ is
called PPT if the partial transposition of $\rho$ over any
subsystem(s) is still positive. Let $\rho^{T_{i}}$ denote the
partial transposition with respect to the $i$-th subsystem. Assume
that there is a PPT state $\rho$ with $\tau(\rho)>0$. Then at least
one term in ($\ref{11}$), say
$C_{\alpha_{0}\beta_{0}}^{12|3}(\rho)$, is not zero. Define
$\rho_{\alpha_{0}\beta_{0}}=L_{\alpha_{0}}^{12}\otimes
L_{\beta_{0}}^{3}\rho (L_{\alpha_{0}}^{12}\otimes
L_{\beta_{0}}^{3})^{\dag}$. By using the PPT property of $\rho$, we
have:
\begin{eqnarray}\label{19}
\rho_{\alpha_{0}\beta_{0}}^{T_{3}}=L_{\alpha_{0}}^{12}\otimes
(L_{\beta_{0}}^{3})^{*}\rho^{T_{3}}
(L_{\alpha_{0}}^{12})^{\dag}\otimes (L_{\beta_{0}}^{3})^{T}\geq 0.
\end{eqnarray}
Noting that both $L_{\alpha_{0}}^{12}$ and $ L_{\beta_{0}}^{3}$ are
projectors to two-dimensional subsystems,
$\rho_{\alpha_{0}\beta_{0}}$ can be considered as a $4\times 4$
density matrix. While a PPT $4\times 4$ density matrix
$\rho_{\alpha_{0}\beta_{0}}$ must be a separable state, which
contradicts with $C_{\alpha_{0}\beta_{0}}^{12|3}(\rho)\neq 0$.

\begin{enumerate}
{\sf \item[{(4)}] Relation between lower bounds of bi- and tripartite concurrence}
\end{enumerate}

$\tau_{3}$ is basically different from $\tau_{2}$ as $\tau_{3}$
characterizes also genuine tripartite entanglement that can not be
described by bipartite decompositions. Nevertheless, there are
interesting relations between them.

\begin{theorem} For any pure tripartite state ($\ref{pure}$), the
following inequality holds:
\begin{eqnarray}
\tau_{2}(\rho_{12})+\tau_{2}(\rho_{13})+\tau_{2}(\rho_{23}) \leq
3\tau_{3}(\rho),
\end{eqnarray}
where $\tau_{2}$ is the lower bound of bipartite concurrence
($\ref{ou}$), $\tau_{3}$ is the lower bound of tripartite
concurrence ($\ref{11}$) and $\rho_{12}={\rm Tr}_{3}[\rho]$,
$\rho_{13}={\rm Tr}_{2}[\rho]$, $\rho_{23}={\rm Tr}_{1}[\rho]$,
$\rho=|\Psi\ra_{123}\la\Psi|$. \end{theorem}

{\sf[Proof]}\ Since
$C_{\alpha\beta}^{2}\leq(\lambda_{\alpha\beta}(1))^{2}
\leq\sum_{i=1}^{4}(\lambda_{\alpha\beta}(i))^{2}={\rm Tr}[\rho
\widetilde{\rho}_{\alpha\beta}]$ for $\rho=\rho_{12}$,
$\rho=\rho_{13}$ and $\rho=\rho_{23}$, we have
\begin{eqnarray}
&&\tau_{2}(\rho_{12})+\tau_{2}(\rho_{13})+\tau_{2}(\rho_{23})
\nonumber\\
&\leq& \frac{d}{2(d-1)}(\sum_{\alpha,\beta=1}^{\frac{d(d-1)}{2}}{\rm
Tr}[\rho_{12}
(\widetilde{\rho}_{12})_{\alpha\beta}]+\sum_{\alpha,\beta=1}^{\frac{d(d-1)}{2}}{\rm
Tr}[\rho_{13}
(\widetilde{\rho}_{13})_{\alpha\beta}]+\sum_{\alpha,\beta=1}^{\frac{d(d-1)}{2}}{\rm
Tr}[\rho_{23}
(\widetilde{\rho}_{23})_{\alpha\beta}]) \nonumber\\
&=&\frac{d}{2(d-1)}(3-{\rm Tr}[\rho_{1}^{2}]-{\rm
Tr}[\rho_{2}^{2}]-{\rm
Tr}[\rho_{3}^{2}])=3C^{2}(\rho)=3\tau_{3}(\rho),
\end{eqnarray}
where we have used the similar analysis in \cite{ou,ckw} to obtain
the equality $\sum\limits_{\alpha,\beta}{\rm Tr}[\rho_{12}
(\widetilde{\rho}_{12})_{\alpha\beta}]=1-{\rm Tr}[\rho_{1}^{2}]-{\rm
Tr}[\rho_{2}^{2}]+{\rm Tr}[\rho_{3}^{2}]$,
$\sum\limits_{\alpha,\beta}{\rm Tr}[\rho_{13}
(\widetilde{\rho}_{13})_{\alpha\beta}]=1-{\rm Tr}[\rho_{1}^{2}]+{\rm
Tr}[\rho_{2}^{2}]-{\rm Tr}[\rho_{3}^{2}]$,
$\sum\limits_{\alpha,\beta}{\rm Tr}[\rho_{23}
(\widetilde{\rho}_{23})_{\alpha\beta}]=1+{\rm Tr}[\rho_{1}^{2}]-{\rm
Tr}[\rho_{2}^{2}]-{\rm Tr}[\rho_{3}^{2}]$. The last equality is due
to that $\rho$ is a pure state.\hfill$\Box$

In fact, the bipartite entanglement inside a tripartite state is
useful for distilling maximally entangled states. Assume that there
are two of the qualities $\{\tau(\rho_{12}), \tau(\rho_{13}),
\tau(\rho_{23})\}$ larger than zero, say $\tau(\rho_{12})>0$ and
$\tau(\rho_{13})>0$. According to \cite{ou}, one can distill two
maximal entangled states $|\psi_{12}\ra$ and $|\psi_{13}\ra$ which
belong to ${\cal {H}}_{1}\otimes{\cal {H}}_{2}$ and ${\cal
{H}}_{1}\otimes{\cal {H}}_{3}$ respectively. In terms of the result
in \cite{zukowski}, one can use them to produce a GHZ state.

\subsubsection{Estimation of multipartite entanglement}

For a pure N-partite quantum state $|\psi\ra\in {\cal
{H}}_{1}\otimes{\cal {H}}_{2}\otimes\cdots\otimes{\cal {H}}_{N}$,
$dim {\cal {H}}_{i}=d_i$, $i=1,...,N$, the concurrence of bipartite
decomposition between subsystems $12\cdots M$ and $M+1\cdots N$ is
defined by
\begin{eqnarray}\label{xx}
C_{2}(|\psi\ra)=\sqrt{2(1-{\rm Tr}[\rho_{{1}{2}\cdots
{M}}^{2}])}
\end{eqnarray}
where $\rho_{{1}{2}\cdots {M}}^{2}={\rm Tr}_{{M+1}\cdots
{N}}[|\psi\ra\la\psi|]$ is the reduced density matrix of
$\rho=|\psi\ra\la\psi|$ by tracing over subsystems $M+1\cdots{N}$.
On the other hand, the concurrence of $|\psi\ra$ is defined by
(\ref{xxxx}).

For a mixed multipartite quantum state,
$\rho=\sum_{i}p_{i}|\psi_{i}\ra\la\psi_{i}| \in {\cal
{H}}_{1}\otimes{\cal {H}}_{2}\otimes\cdots\otimes{\cal {H}}_{N}$,
the corresponding concurrence of (\ref{xx}) and (\ref{xxxx}) are then
given by the convex roof:
\begin{eqnarray}\label{def1}
C_{2}(\rho)=\min_{\{p_{i},|\psi_{i}\}\ra}\sum_{i}p_{i}C_{2}(|\psi_{i}\ra\la\psi_{i}|),
\end{eqnarray}
and (\ref{defe}). We now investigate the relation between these two
kinds of concurrences.

\begin{lemma} For a bipartite density matrix $\rho\in {\cal
{H}}_{A}\otimes{\cal {H}}_{B}$, one has
\begin{eqnarray}\label{yy}
1-{\rm Tr}[\rho^{2}]\leq 1-{\rm Tr}[\rho_{A}^{2}]+ 1-{\rm
Tr}[\rho_{B}^{2}],
\end{eqnarray}
where $\rho_{A/B}={\rm Tr}_{B/A}[\rho]$ be the reduced density
matrices of $\rho$.
\end{lemma}

{\sf[Proof]}\ Let $\rho=\sum\limits_{ij}\lambda_{ij}|ij\ra\la ij|$
be the spectral decomposition, where $\lambda_{ij}\geq 0,
\sum_{ij}\lambda_{ij}=1$. Then
$\rho_{1}=\sum_{ij}\lambda_{ij}|i\ra\la i|$,
$\rho_{2}=\sum_{ij}\lambda_{ij}|j\ra\la j|$. Therefore
\begin{eqnarray*}
&&1-{\rm Tr}[\rho_{A}^{2}]+1-{\rm Tr}[\rho_{B}^{2}]-1+{\rm
Tr}[\rho^{2}]
=1-{\rm Tr}[\rho_{A}^{2}]-{\rm Tr}[\rho_{B}^{2}]+{\rm Tr}[\rho^{2}]\\
&=&(\sum_{ij}\lambda_{ij})^{2}-\sum_{i,j,j^{'}}\lambda_{ij}\lambda_{ij^{'}}
-\sum_{i,i^{'},j}\lambda_{ij}\lambda_{i^{'}j}+\sum_{ij}\lambda_{ij}^{2}\\
&=&(\sum_{i=i^{'},j=j^{'}}\lambda_{ij}^{2}+\sum_{i=i^{'},j\neq
j^{'}}\lambda_{ij}\lambda_{ij^{'}}+\sum_{i\neq i^{'},j=
j^{'}}\lambda_{ij}\lambda_{i^{'}j}+\sum_{i\neq i^{'},j\neq
j^{'}}\lambda_{ij}\lambda_{i^{'}j^{'}})\\
&=&\sum_{i\neq i^{'},j\neq j^{'}}\lambda_{ij}\lambda_{i^{'}j^{'}}\geq 0.
\end{eqnarray*}
$\hfill\Box$

This lemma can be also derived in another way
\cite{optloo,cai}.

\begin{theorem} For a multipartite quantum state $\rho\in
{\cal {H}}_{1}\otimes{\cal {H}}_{2}\otimes\cdots\otimes{\cal
{H}}_{N}$ with $N\geq 3$, the following inequality holds,
\begin{eqnarray}
C_{N}(\rho)\geq\max 2^{\frac{3-N}{2}}C_{2}(\rho),
\end{eqnarray}
where the maximum is taken over all kinds of bipartite concurrence.
\end{theorem}

{\sf[Proof]}\ Without lose of generality, we suppose that the
maximal bipartite concurrence is attained between subsystems
$12\cdots M$ and $(M+1)\cdots N$.

For a pure multipartite state $|\psi\ra\in {\cal
{H}}_{1}\otimes{\cal {H}}_{2}\otimes\cdots\otimes{\cal {H}}_{N}$,
${\rm Tr}[\rho_{12\cdots M}^{2}]={\rm Tr}[\rho_{(M+1)\cdots
N}^{2}]$. From (\ref{yy}) we have
\begin{eqnarray*}
C_{N}^{2}(|\psi\ra\la\psi|)&=&2^{2-N}((2^{N}-2)-\sum_{\alpha}{\rm
Tr}[\rho_{\alpha}^{2}])\geq
2^{3-N}(N-\sum_{k=1}^{N}{\rm Tr}[\rho_{k}^{2}])\\
&\geq& 2^{3-N}(1-{\rm Tr}[\rho_{12\cdots M}^{2}]+1-{\rm
Tr}[\rho_{(M+1)\cdots
N}^{2}])\\
&=&2^{3-N}*2(1-{\rm Tr}[\rho_{12\cdots
M}^{2}])=2^{3-N}C_{2}^{2}(|\psi\ra\la\psi|),
\end{eqnarray*}
i.e. $C_{N}(|\psi\ra\la\psi|)\geq
2^{\frac{3-N}{2}}C_{2}(|\psi\ra\la\psi|)$.

Let $\rho=\sum\limits_{i}p_{i}|\psi_{i}\ra\la\psi_{i}|$ attain the
minimal decomposition of the multipartite concurrence. One has
\begin{eqnarray*}
C_{N}(\rho)=\sum_{i}p_{i}C_{N}(|\psi_{i}\ra\la\psi_{i}|)\geq
2^{\frac{3-N}{2}}\sum_{i}p_{i}C_{2}(|\psi_{i}\ra\la\psi_{i}|)\\
\geq2^{\frac{3-N}{2}}\min_{\{p_{i},|\psi_{i}\}}
\sum_{i}p_{i}C_{2}(|\psi_{i}\ra\la\psi_{i}|)=2^{\frac{3-N}{2}}C_{2}(\rho).
\end{eqnarray*}
$\hfill\Box$

\begin{corollary} For a tripartite quantum state $\rho\in{\cal
{H}}_{1}\otimes{\cal {H}}_{2}\otimes{\cal {H}}_{3}$, the following
inequality holds:
\begin{eqnarray}
C_{3}(\rho)\geq\max C_{2}(\rho)
\end{eqnarray}
where the maximum is taken over all kinds of bipartite concurrence.
\end{corollary}

In \cite{vicente,optloo}, from the separability criteria related to
local uncertainty relation, covariance matrix and correlation
matrix, the following lower bounds for bipartite concurrence are
obtained:
\begin{eqnarray}\label{lowerbound2}
C_{2}(\rho)\geq\frac{2||C(\rho)||-(1-{\rm Tr}[\rho_{A}^{2}])-(1-{\rm
Tr}[\rho_{B}^{2}])}{\sqrt{2d_A(d_A-1)}}
\end{eqnarray}
and
\begin{eqnarray}\label{lowerbound3}
C_{2}(\rho)\geq\sqrt{\frac{8}{d_A^{3}d_B^{2}(d_A-1)}}(||T(\rho)||-\frac{\sqrt{d_Ad_B(d_A-1)(d_B-1)}}{2}),
\end{eqnarray}
where the entries of the matrix $C$,
$C_{ij}=\la\lambda^{A}_{i}\otimes\lambda^{B}_{j}\ra-\la\lambda^{A}_{i}\otimes
I_{d_B}\ra\la I_{d_A}\otimes\lambda^{B}_{j}\ra$,
$T_{ij}=\frac{d_Ad_B}{2}\la\lambda^{A}_{i}\otimes\lambda^{B}_{j}\ra$,
$\lambda^{A/B}_{k}$ stands for the normalized generator of
$SU(d_A/d_B)$, i.e. ${\rm
Tr}[\lambda^{A/B}_{k}\lambda^{A/B}_{l}]=\delta_{kl}$ and $\la
X\ra={\rm Tr}[\rho X]$. It is shown that the lower bounds
$(\ref{lowerbound2})$ and $(\ref{lowerbound3})$ are independent of
$(\ref{con0})$.

Now we consider a multipartite quantum state $\rho\in{\cal
{H}}_{1}\otimes{\cal {H}}_{2}\otimes\cdots\otimes{\cal {H}}_{N}$ as
a bipartite state belonging to ${\cal {H}}^{A}\otimes{\cal {H}}^{B}$
with the dimensions of the subsystems A and B being
$d_A=d_{s_{1}}d_{s_{2}}\cdots d_{s_{m}}$ and
$d_B=d_{s_{m+1}}d_{s_{m+2}}\cdots d_{s_{N}}$ respectively. By using
the corollary, $(\ref{con0})$, $(\ref{lowerbound2})$ and
$(\ref{lowerbound3})$ one has the following lower bound:

\begin{theorem}\label{46} For any N-partite quantum state $\rho$,
\begin{eqnarray}\label{newlowerbound}
C_{N}(\rho)\geq2^{\frac{3-N}{2}}\max\{B_1,B_2,B_3\},
\end{eqnarray}
where
\begin{eqnarray*}
B_1&=&\max_{\{i\}}\sqrt{\frac{2}{M_{i}(M_{i}-1)}}\left[\max(||{\cal
{T}}_{A}(\rho^{i})||,||R(\rho^{i})||)-1\right],\\
B_2&=&\max_{\{i\}}\frac{2||C(\rho^{i})||- (1-{\rm
Tr}[(\rho^{i}_{A})^{2}])-(1-{\rm Tr}[(\rho^{i}_{B})^{2}])}
{\sqrt{2M_{i}(M_{i}-1)}},\\
B_3&=&\max_{\{i\}}\sqrt{\frac{8}{M_{i}^{3}N_{i}^{2}(M_{i}-1)}}
(||T(\rho^{i})||-\frac{\sqrt{M_{i}N_{i}(M_{i}-1)(N_{i}-1)}}{2}),
\end{eqnarray*}
$\rho^i$s are all possible bipartite decompositions of $\rho$, and
\begin{eqnarray*}
M_{i}=\min{\{d_{s_{1}}d_{s_{2}}\cdots d_{s_{m}},
d_{s_{m+1}}d_{s_{m+2}}\cdots d_{s_{N}}\}},\\
N_{i}=\max{\{d_{s_{1}}d_{s_{2}}\cdots d_{s_{m}},
d_{s_{m+1}}d_{s_{m+2}}\cdots d_{s_{N}}\}}.
\end{eqnarray*}
\end{theorem}

In \cite{optloo,mintert,aolita}, it is shown that the upper and lower
bound of multipartite concurrence satisfy
\begin{eqnarray}\label{upperlowerbound}
\sqrt{(4-2^{3-N}){\rm Tr}[\rho^{2}]-2^{2-N}\sum_{\alpha}{\rm
Tr}[\rho_{\alpha}^{2}]}\leq
C_{N}(\rho)\leq\sqrt{2^{2-N}[(2^{N}-2)-\sum_{\alpha}{\rm
Tr}[\rho_{\alpha}^{2}]]}\nonumber\\.
\end{eqnarray}

In fact one can obtain a more effective upper bound for multi-partite
concurrence. Let $\rho=\sum\limits_{i}\lambda_{i}|\psi_{i}\ra\la
\psi_{i}|\in {\cal
{H}}_{1}\otimes{\cal{H}}_{2}\otimes\cdots\otimes{\cal {H}}_{N}$,
where $|\psi_{i}\ra$s are the  orthogonal pure states and
$\sum\limits_{i}\lambda_{i}=1$. We have
\begin{eqnarray}\label{newupperbound}
C_{N}(\rho)=\min_{\{p_{i},|\varphi_{i}\}\ra}\sum_{i}p_{i}C_{N}(|\varphi_{i}\ra\la\varphi_{i}|)
\leq\sum_{i}\lambda_{i}C_{N}(|\psi_{i}\ra\la\psi_{i}|).
\end{eqnarray}
The right side of $(\ref{newupperbound})$ gives a new upper bound of
$C_{N}(\rho)$. Since
\begin{eqnarray*}
\sum_{i}\lambda_{i}C_{N}(|\psi_{i}\ra\la\psi_{i}|)
&=&2^{1-\frac{N}{2}}\sum_{i}\lambda_{i}\sqrt{(2^{N}-2)-\sum_{\alpha}{\rm Tr}[(\rho^{i}_{\alpha})^{2}]}\\
&\leq&
2^{1-\frac{N}{2}}\sqrt{(2^{N}-2)-\sum_{\alpha}{\rm Tr}[\sum_{i}\lambda_{i}(\rho^{i}_{\alpha})^{2}]}\\
&\leq& 2^{1-\frac{N}{2}}\sqrt{(2^{N}-2)-\sum_{\alpha}{\rm
Tr}[(\rho_{\alpha})^{2}]},
\end{eqnarray*}
the upper bound obtained in $(\ref{newupperbound})$ is better than
that in $(\ref{upperlowerbound})$.

\subsubsection{Bounds of concurrence and tangle}

In \cite{vicentejpa}, a lower bound for tangle defined in
$(\ref{t33})$ has been derived:
\be\label{lbt}
\tau(\rho)\geq\frac{8}{MN(M+N)}(||T(\rho)||_{HS}^2-\frac{MN(M-1)(N-1)}{4}),
\ee
where $||X||_{HS}=\sqrt{{\rm Tr}[XX^{\dag}]}$ denotes the Frobenius
or Hilbert-Schmidt norm. Experimentally measurable lower and upper
bounds for concurrence have been also given by Mintert and Zhang
et.al. in \cite{mintert,optloo}:
\be\label{mb}\sqrt{2({\rm
{Tr}}[\rho^2]-{\rm{Tr}}[\rho_A^2])}\leq C(\rho)\leq
\sqrt{2(1-{\rm{Tr}}[\rho_A^2])}.
\ee

Since the convexity of $C^{2}(\rho)$, we have that $\tau(\rho)\geq
C^{2}(\rho)$ always holds. For two qubits quantum systems, tangle
$\tau$ is always equal to the square of concurrence $C^2$
\cite{Rungta03,ckw}, as a decomposition $\{p_i,|\psi_i\ra\}$
achieving the minimum in Eq. $(\ref{cmix})$ has the property that
$C(|\psi_i\ra)=C(|\psi_j\ra)$ $\forall i,j$. For higher dimensional
systems we do not have similar relations. Thus it is meaningful to
derive valid upper bound for tangle and lower bound for concurrence.

\begin{theorem} For any quantum state $\rho\in{\mathcal
{H}}_A\otimes{\mathcal {H}}_B$, we have
\be\label{u}
\tau(\rho)\leq\min\{1-{\rm Tr}[\rho^2_A],1-{\rm
Tr}[\rho^2_B]\},
\ee
\be\label{lcb}
C(\rho)\geq\sqrt{\frac{8}{MN(M+N)}}(||T(\rho)||_{HS}-\frac{\sqrt{MN(M-1)(N-1)}}{2}),
\ee
where $\rho_A$ is the reduced matrix of $\rho$, and $T(\rho)$ is
the correlation matrix of $\rho$ defined in $(\ref{lowerbound3})$.
\end{theorem}

{\sf[Proof]}\ We assume $1-{\rm Tr}[\rho^2_A]\leq 1-{\rm
Tr}[\rho^2_B]$ for convenience. By the definition of $\tau$, we have
that for a pure state $|\psi\ra, \tau(|\psi\ra)=2(1-{\rm
Tr}[({\rho_A^{|\psi\ra}})^2])$. Let $\rho=\sum_i p_i\rho_i$ be the
optimal decomposition such that $\tau(\rho)=\sum_i p_i \tau(\rho_i).$
We get \be\tau(\rho)=\sum_i p_i\tau(\rho_i)=\sum_i
p_i2[1-{\rm{Tr}}[({\rho_A^{|\psi_i\ra}})^2]]=2[1-{\rm{Tr}}[\sum_i
p_i({\rho_A^{|\psi_i\ra}})^2]]\leq 2[1-{\rm{Tr}}[\rho_A^2]].\ee

Note that for pure state
$|\psi\ra\in{\mathcal {H}}_A\otimes{\mathcal {H}}_B$
\cite{vicentejpa}, \be
C(|\psi\ra)=\sqrt{\frac{8}{MN(M+N)}(||T(|\psi\ra)||^2-\frac{MN(M-1)(N-1)}{4})}.\ee
Using the inequality $\sqrt{a-b}\geq\sqrt{a}-\sqrt{b}$ for any
$a\geq b$, we get \be
C(|\psi\ra)\geq\sqrt{\frac{8}{MN(M+N)}}(||T(|\psi\ra)||_{HS}-\frac{\sqrt{MN(M-1)(N-1)}}{2}).\ee
Now let $\rho=\sum_i p_i\rho_i$ be the optimal decomposition such that
$C(\rho)=\sum_i p_i C(\rho_i).$ We get
\begin{eqnarray*}
C(\rho)&=&\sum_ip_iC(\rho_i)\geq\sum_i p_i
\sqrt{\frac{8}{MN(M+N)}}(||T(\rho_i)||_{HS}-\frac{\sqrt{MN(M-1)(N-1)}}{2})\\
&=&\sqrt{\frac{8}{MN(M+N)}}(\sum_i
p_i||T(\rho_i)||_{HS}-\frac{\sqrt{MN(M-1)(N-1)}}{2})\\
&\geq&\sqrt{\frac{8}{MN(M+N)}}(||T(\rho)||_{HS}-\frac{\sqrt{MN(M-1)(N-1)}}{2})
\end{eqnarray*} which ends the proof. $\hfill\Box$

The upper bound $(\ref{u})$, together with the lower
bound (\ref{lcb}), (\ref{lowerbound2}), (\ref{lowerbound3}), (\ref{lbt}) and (\ref{mb}),
can allow for estimations of entanglement for arbitrary quantum states.
Moreover, since the upper bound is exactly the value of tangle for
pure states, the upper bound can be a good estimation when the state
is very weakly mixed.

\subsection{Concurrence and tangle of two entangled states are strictly larger than that of one}

In this subsection we show that although bound entangled states can
not be distilled, the concurrence and tangle of two entangled states
will be always strictly larger than that of one, even the two
entangled states are both bound entangled.

Let $\rho=\sum_{ijkl}\rho_{ij,kl}|ij\ra\la kl|\in{\cal
H}_{A}\otimes{\cal H}_{B}$ and
$\sigma=\sum_{i^{'}j^{'}k^{'}l^{'}}\sigma_{i^{'}j^{'},k^{'}l^{'}}|i^{'}j^{'}\ra\la
k^{'}l^{'}|\in{\cal H}_{A^{'}}\otimes{\cal H}_{B^{'}}$ be two
quantum states shared by subsystems $AA^{'}$ and $BB^{'}$. We use
$\rho\otimes\sigma=\sum_{ijkl,i^{'}j^{'}k^{'}l^{'}}\rho_{ij,kl}\sigma_{i^{'}j^{'},k^{'}l^{'}}|ii^{'}\ra_{AA^{'}}\la
kk^{'}|\otimes|jj^{'}\ra_{BB^{'}}\la ll^{'}|$ to denote the state of
the whole system.

\begin{lemma}\label{lll1} For pure states $|\psi\ra\in{\cal
H}_A\otimes{\cal H}_B$ and $|\varphi\ra\in{\cal
H}_{A^{'}}\otimes{\cal H}_{B^{'}}$, the inequalities
\be\label{pureco}
C(|\psi\ra\otimes|\varphi\ra)\geq\max\{C(|\psi\ra),C(|\varphi\ra)\}
\ee and \be\label{puretangle}
\tau(|\psi\ra\otimes|\varphi\ra)\geq\max\{\tau(|\psi\ra),\tau(|\varphi\ra)\}
\ee always hold, and $"="$ in the two inequalities hold if and only
if at least one of $\{|\psi\ra, |\varphi\ra\}$ is separable.
\end{lemma}

{\sf[Proof]}\ Without loss of generality we assume $C(|\psi\ra)\geq
C(|\varphi\ra)$. Fist note that
\be\label{1}\rho_{AA^{'}}^{|\psi\ra\otimes|\varphi\ra}=\rho_{A}^{|\psi\ra}\otimes\rho_{A^{'}}^{|\varphi\ra}.\ee
Let $\rho_{A}^{|\psi\ra}=\sum_{i}\lambda_{i}|i\ra\la i|$ and
$\rho_{A^{'}}^{|\varphi\ra}=\sum_{j}\pi_{j}|j\ra\la j|$ be the
spectral decomposition of $\rho_{A}^{|\psi\ra}$ and
$\rho_{A^{'}}^{|\varphi\ra}$, with $\sum_{i}\lambda_{i}=1$ and
$\sum_{j}\pi_{j}=1$ respectively. By using $(\ref{1})$ one obtains
that \be {\rm
Tr}[(\rho_{AA^{'}}^{|\psi\ra\otimes|\varphi\ra})^{2}]=\sum\lambda_{i}\pi_{j}\lambda_{i^{'}}\pi_{j^{'}}
|ij\ra\la ij|i^{'}j^{'}\ra\la
i^{'}j^{'}|=\sum\lambda_{i}^{2}\pi_{j}^{2}\ee while \be {\rm
Tr}[(\rho_{A}^{|\psi\ra})^{2}]=\sum_{i}\lambda_{i}^{2}.\ee

Now using the definition of concurrence and the normalization
conditions of $\lambda_{i}$ and $\pi_{j}$ one immediately gets
\be\label{inep} C(|\psi\ra\otimes|\varphi\ra)=\sqrt{2(1-{\rm
Tr}[(\rho_{AA^{'}}^{|\psi\ra\otimes|\varphi\ra})^{2}])} \geq
\sqrt{2(1-{\rm Tr}[(\rho_{A}^{|\psi\ra})^{2}])} =C(|\psi\ra).\ee

If one of $\{|\psi\ra, |\varphi\ra\}$ is separable, say
$|\varphi\ra$, then the rank of $\rho_{A^{'}}^{|\varphi\ra}$ must be
one, which means that there is only one item in the spectral
decomposition in $\rho_{A^{'}}^{|\varphi\ra}$. Using the
normalization condition of $\pi_{j}$ we obtain ${\rm
Tr}[(\rho_{AA^{'}}^{|\psi\ra\otimes|\varphi\ra})^{2}]={\rm
Tr}[(\rho_{A}^{|\psi\ra})^{2}]$. Then the inequality $(\ref{inep})$
becomes an equality.

On the other hand, if both $|\psi\ra$ and $|\varphi\ra$ are
entangled (not separable), there must be at least two items in the
decomposition of their reduced density matrices
$\rho_{A}^{|\psi\ra}$ and $\rho_{A^{'}}^{|\varphi\ra}$, which means
that ${\rm Tr}[(\rho_{AA^{'}}^{|\psi\ra\otimes|\varphi\ra})^{2}]$ is
strictly larger than ${\rm Tr}[(\rho_{A}^{|\psi\ra})^{2}]$.

The inequality $(\ref{puretangle})$ also holds because that for pure
quantum state $\rho$, $\tau(\rho)=C^{2}(\rho)$. $\hfill\Box$

From the lemma, we have, for mixed states,

\begin{theorem}\label{larger} For any quantum states $\rho\in{\cal
H}_{A}\otimes{\cal H}_{B}$ and $\sigma\in{\cal
H}_{A^{'}}\otimes{\cal H}_{B^{'}}$, the inequalities \be\label{t1}
C(\rho\otimes\sigma)\geq\max\{C(\rho),C(\sigma)\}\ee and
\be\label{t2}
\tau(\rho\otimes\sigma)\geq\max\{\tau(\rho),\tau(\sigma)\}\ee always
hold, and the $"="$ in the two inequalities hold if and only if at
least one of $\{\rho, \sigma\}$ is separable, i.e. if both $\rho$
and $\sigma$ are entangled (even bound entangled),
$C(\rho\otimes\sigma)>\max\{C(\rho),C(\sigma)\}$ and
$\tau(\rho\otimes\sigma)>\max\{\tau(\rho),\tau(\sigma)\}$ always
hold.
\end{theorem}

{\sf[Proof]}\ We still assume $C(\rho)\geq C(\sigma)$ for
convenience. Let $\rho=\sum_i p_i\rho_i$ and $\sigma=\sum_j
q_j\sigma_j$ be the optimal decomposition such that
$C(\rho\otimes\sigma)=\sum_i p_i q_j C(\rho_i\otimes\sigma_j).$ By
using the inequality obtained in lemma \ref{lll1} we have
\be\label{14} C(\rho\otimes\sigma)=\sum_i p_i q_j
C(\rho_i\otimes\sigma_j)\geq\sum_i p_i q_j
C(\rho_i)=\sum_ip_iC(\rho_i)\geq C(\rho).\ee

Case 1: Now let one of $\{\rho, \sigma\}$ be separable, say
$\sigma$, with ensemble representation $\sigma=\sum_j q_j\sigma_j$,
where $\sum_jq_j=1$ and $\sigma_j$ is the density matrix of
separable pure state. Suppose $\rho=\sum_i p_i \rho_i$ be the
optimal decomposition of $\rho$ such that $C(\rho)=\sum_i p_i
C(\rho_i)$. Using lemma \ref{lll1} we have \be\label{16}
C(\rho\otimes\sigma)\leq\sum_i p_i q_j
C(\rho_i\otimes\sigma_j)=\sum_i p_i q_j
C(\rho_i)=\sum_ip_iC(\rho_i)= C(\rho).\ee The inequalities
$(\ref{14})$ and $(\ref{16})$ show that if $\sigma$ is separable,
then $C(\rho\otimes\sigma)=C(\rho)$.

Case 2: If both $\rho$ and $\sigma$ are inseparable, i.e. there is
at least one pure state in the ensemble decomposition of $\rho$ (and
$\sigma$ respectively), using lemma \ref{lll1} we have \be
C(\rho\otimes\sigma)=\sum_i p_i q_j C(\rho_i\otimes\sigma_j)>\sum_i
p_i q_j C(\rho_i)=\sum_ip_iC(\rho_i)\geq C(\rho).\ee

The inequality for tangle $\tau$ can be proved in a similar way.
$\hfill\Box$

${\bf{Remark:}}$ In \cite{l150501} it is shown that any
entangled state $\rho$ can enhance the teleportation power of
another state $\sigma$. This holds even if the state $\rho$ is bound
entangled. But if $\rho$ is bound entangled, the corresponding
$\sigma$ must be free entangled (distillable). By theorem
$\ref{larger}$, we can see that even two entangled quantum
states $\rho$ and $\sigma$ are bound entangled, their concurrence and tangle are
strictly larger than that of one state.

\subsection{Subadditivity of concurrence and tangle}

We now give a proof of the subadditivity of
concurrence and tangle, which illustrates that concurrence and
tangle may be proper entanglement measurements.

\begin{theorem} Let $\rho$ and $\sigma$ be quantum states in
${\mathcal {H}}_A\otimes{\mathcal {H}}_B$, we have \be\label{lll}
C(\rho\otimes\sigma)\leq
C(\rho)+C(\sigma)\quad{\rm{and}}\quad\tau(\rho\otimes\sigma)\leq
\tau(\rho)+\tau(\sigma).\ee \end{theorem}

{\bf{Proof:}}\ We first prove that the theorem holds for pure
states, i.e. for $|\psi\ra$ and $|\phi\ra$ in ${\mathcal
{H}}_A\otimes{\mathcal {H}}_B$,
\be
C(|\psi\ra\otimes|\phi\ra)\leq
C(|\psi\ra)+C(|\phi\ra)\quad{\rm{and}}\quad\tau(|\psi\ra\otimes|\phi\ra)\leq
\tau(|\psi\ra)+\tau(|\phi\ra).\ee Assume that
$\rho_A^{|\psi\ra}=\sum_{i}\lambda_i|i\ra\la i|$ and
$\rho_A^{|\phi\ra}=\sum_{j}\pi_j|j\ra\la j|$ be the spectral
decomposition of the reduced matrices $\rho_A^{|\psi\ra}$ and
$\rho_A^{|\phi\ra}$. One has
\begin{eqnarray}\label{ll}
&&\frac{1}{2}[C(|\psi\ra)+C(|\phi\ra)]^2\geq1-Tr[(\rho_A^{|\psi\ra})^2]+1-Tr[(\rho_A^{|\phi\ra})^2]\nonumber\\
&=&1-\sum_i\lambda_i^2+1-\sum_j\pi_j^2\geq1-\sum_{ij}\lambda_i^2\pi_j^2=\frac{1}{2}C^2(|\psi\ra\otimes|\phi\ra).
\end{eqnarray}

Now we prove that $(\ref{lll})$ holds for any mixed quantum states
$\rho$ and $\sigma$. Let $\rho=\sum_i p_i\rho_i$ and $\sigma=\sum_j
q_j \sigma_j$ be the optimal decomposition such that
$C(\rho)=\sum_ip_iC(\rho_i)$ and $C(\sigma)=\sum_jq_jC(\sigma_j)$.
We have \be
C(\rho)+C(\sigma)=\sum_{ij}p_iq_j[C(\rho_i)+C(\sigma_j)]\geq
\sum_{ij}p_iq_jC(\rho_i\otimes\sigma_j)\geq C(\rho\otimes\sigma).\ee
The inequality for $\tau$ can be derived in a similar way.
$\hfill\Box$

\section{Fidelity of teleportation and distillation of entanglement}\label{fefsec}

Quantum teleportation is an important subject
in quantum information processing. In terms of a classical
communication channel and a quantum resource (a nonlocal entangled
state like an EPR-pair of particles), the teleportation protocol
gives ways to transmit an unknown quantum state from a sender
traditionally named ``Alice" to a receiver ``Bob" who are spatially
separated. These teleportation processes can be viewed as quantum
channels. The nature of a quantum channel is determined by the
particular protocol and the state used as a teleportation resource.
The standard teleportation protocol $T_0$ proposed by Bennett et.al
in 1993 uses {\it Bell} measurements and {\it Pauli} rotations. When
the maximally entangled pure state
$|\phi>=\frac{1}{\sqrt{n}}\sum_{i=0}^{n-1}|ii>$ is used as the
quantum resource, it provides an ideal noiseless quantum channel
$\Lambda^{(|\phi><\phi|)}_{T_0}(\cal \rho)=\cal \rho$. However in
realistic situation, instead of the pure maximally entangled states,
Alice and Bob usually share a mixed entangled state due to the
decoherence. Teleportation using mixed state as an entangled
resource is, in general, equivalent to having a noisy quantum
channel. An explicit expression for the output state of the quantum
channel associated with the standard teleportation protocol $T_0$
with an arbitrary  mixed state resource has been obtained
\cite{Bow01,Alb02}.

It turns out that by local quantum operations (including collective
actions over all members of pairs in each lab) and classical
communication (LOCC) between Alice and Bob, it is possible to obtain
a number of pairs in nearly maximally entangled state $|\psi_+\ra$
from many pairs of non-maximally entangled states. Such a procedure
proposed in \cite{distill1,distill2,distill3,bennett,gdp} is called
distillation. In \cite{distill1} the authors give operational
protocol to distill an entangled two-qubit state whose single
fraction $F$, defined by $F(\rho)=\la\psi_+|\rho|\psi_+\ra$, is
larger than $\frac{1}{2}$. The protocol is then generalized in
\cite{gdp} to distill any d-dimensional bipartite entangled quantum
states with $F(\rho)>\frac{1}{d}$. It is shown that a quantum state
$\rho$ violating the reduction criterion can always be distilled.
For such states if their single fraction of entanglement
$F(\rho)=\la\psi_{+}|\rho|\psi_{+}\ra$ is greater than
$\frac{1}{d}$, one can distill these states directly by using the
generalized distillation protocol, otherwise a proper filtering
operation has to be used at first to transform $\rho$ to another
state $\rho^{'}$ so that $F(\rho^{'})>\frac{1}{d}$.

\subsection{Fidelity of quantum teleportation}

Let ${\cal {H}}$ be a $d$-dimensional complex vector space with
computational basis $|i\ra$, $i=1,...,d$. The fully entangled
fraction (FEF) of a density matrix $\rho\in{\cal {H}}\otimes{\cal
{H}}$ is defined by
\begin{eqnarray}\label{def}
{\cal {F}}(\rho)=\max_{U}\la\psi_{+}|(I\otimes U^{\dag})\rho
(I\otimes U)|\psi_{+}\ra
\end{eqnarray}
under all unitary transformations $U$, where
$|\psi_{+}\ra=\frac{1}{\sqrt{d}}\sum\limits_{i=1}^{d}|ii\ra$ is the
maximally entangled state and $I$ is the corresponding identity
matrix.

In \cite{fefandtel}, the authors give a optimal teleportation
protocol by using a mixed entangled quantum state. The optimal
teleportation fidelity is given by
\begin{eqnarray}
f_{\max}(\rho)=\frac{d{\cal {F}}(\rho)}{d+1}+\frac{1}{d+1},
\end{eqnarray}
which solely depends the FEF of the entangled resource state
$\rho$.

In fact the fully entangled fraction is tightly related to many
quantum information processing such as dense coding \cite{dense},
teleportation \cite{teleportation}, entanglement swapping
\cite{swapping}, and quantum cryptography (Bell inequalities)
\cite{schemes}. As the optimal fidelity of teleportation is given by
$FEF$ \cite{fefandtel}, experimentally measurement of FEF can be
also used to determine the entanglement of the non-local source used
in teleportation. Thus an analytic formula for FEF is of great
importance. In \cite{grondalski} an elegant formula of FEF for two-qubit
system is derived analytically by using the method of Lagrange
multipliers. For high dimensional quantum states the analytical computation of
FEF remains formidable and less results have been known.
In the following we give an estimation on the values of FEF
by giving some upper bounds of FEF.

Let $\lambda_{i}$, $i=1,...,d^2-1$, be the generators of the $SU(d)$
algebra. A bipartite state $\rho\in{\cal {H}}\otimes{\cal {H}}$ can
be expressed as
\begin{eqnarray}\label{rho}
\rho=\frac{1}{d^{2}}I\otimes
I+\frac{1}{d}\sum\limits_{i=1}^{d^{2}-1}r_{i}(\rho)\lambda_{i}\otimes
I+\frac{1}{d}\sum\limits_{j=1}^{d^{2}-1}s_{j}(\rho)I\otimes
\lambda_{j}+\sum\limits_{i,j=1}^{d^{2}-1}m_{ij}(\rho)\lambda_{i}\otimes
\lambda_{j},
\end{eqnarray}
where $r_{i}(\rho)=\frac{1}{2}{\rm Tr}[\rho\lambda_{i}(1)\otimes
I]$, $s_{j}(\rho)=\frac{1}{2}{\rm Tr}[\rho I\otimes \lambda_{j}(2)]$
and $m_{ij}(\rho)=\frac{1}{4}{\rm Tr}[\rho \lambda_{i}(1)\otimes
\lambda_{j}(2)]$. Let $M(\rho)$ denote the correlation matrix with
entries $m_{ij}(\rho)$.

\begin{theorem}\label{ufef} For any $\rho\in{\cal {H}}\otimes{\cal
{H}}$, the fully entangled fraction ${\cal {F}}(\rho)$ satisfies
\begin{eqnarray}\label{inequ}
{\cal {F}}(\rho)\leq \frac{1}{d^{2}}+4||M^{T}(\rho)M(P_{+})||_{KF},
\end{eqnarray}
where $M^{T}$ stands for the transpose of $M$ and $||M||_{KF}={\rm
Tr}[\sqrt{MM^{\dag}}]$ is the Ky Fan norm of $M$.
\end{theorem}

{\sf[Proof]}\ First, we note that
\begin{eqnarray*}
P_{+}=\frac{1}{d^{2}}I\otimes
I+\sum\limits_{i,j=1}^{d^{2}-1}m_{ij}(P_{+}) \lambda_{i}\otimes
\lambda_{j},
\end{eqnarray*}
where $m_{ij}(P_{+})=\frac{1}{4}{\rm Tr}[P_{+} \lambda_{i}\otimes
\lambda_{j}]$.
By definition ($\ref{def}$), one obtains
\begin{eqnarray*}\label{pro}
{\cal {F}}(\rho)&=&\max_{U}\la\psi_{+}|(I\otimes U^{\dag})\rho
(I\otimes U)|\psi_{+}\ra
=\max_{U}{\rm Tr}[\rho (I\otimes U) P_{+} (I\otimes U^{\dag})]\\
&=&\max_{U}\{\frac{1}{d^{2}}{\rm
Tr}[\rho]+\sum\limits_{i,j=1}^{d^{2}-1}m_{ij}(P_{+}) {\rm
Tr}[\rho\lambda_{i}\otimes U\lambda_{j}U^{\dag}]\}.
\end{eqnarray*}

Since $U\lambda_{i}U^{\dag}$ is a traceless Hermitian operator, it
can be expanded according to the $SU(d)$ generators,
\begin{eqnarray}\label{rr}
U\lambda_{i}U^{\dag}=\sum\limits_{j=1}^{d^{2}-1}\frac{1}{2}{\rm
Tr}[U\lambda_{i}U^{\dag}\lambda_{j}]\lambda_{j}\equiv
\sum\limits_{j=1}^{d^{2}-1}O_{ij}\lambda_{j}.
\end{eqnarray}
Entries $O_{ij}$ defines a real $(d^{2}-1)\times(d^{2}-1)$ matrix
$O$. From the completeness relation of $SU(d)$ generators
\begin{eqnarray}
\sum\limits_{j=1}^{d^{2}-1}(\lambda_{j})_{ki}(\lambda_{j})_{mn}=2\delta_{im}\delta_{kn}
-\frac{2}{d}\delta_{ki}\delta_{mn},
\end{eqnarray}
one can show that $O$ is an orthonormal matrix. Using (\ref{rr}) we
have
\begin{eqnarray*}\label{propp}
{\cal {F}}(\rho)&\leq&
\frac{1}{d^{2}}+\max_{O}\sum\limits_{i,j,k}m_{ij}(P_{+})O_{jk}{\rm
Tr}[\rho\lambda_{i}\otimes
\lambda_{k}]\\
&=&\frac{1}{d^{2}}+4\max_{O}\sum\limits_{i,j,k}m_{ij}(P_{+})O_{jk}m_{ik}(\rho)
=\frac{1}{d^{2}}+4\max_{O}{\rm Tr}[M(\rho)^{T}M(P_{+})O]\\
&=&\frac{1}{d^{2}}+4||M(\rho)^{T}M(P_{+})||_{KF}.
\end{eqnarray*}
\hfill$\Box$

For the case $d=2$, we can get an exact result from (\ref{inequ}):

\begin{corollary} For two qubits system, we have
\be\label{corollary}
{\cal{F}}(\rho)=\frac{1}{4}+4||M(\rho)^{T}M(P_{+})||_{KF}, \ee i.e.
the upper bound derived in Theorem $\ref{ufef}$ is exactly the
$FEF$.
\end{corollary}

{\sf[Proof]}\ We have shown in ($\ref{rr}$) that given an arbitrary
unitary $U$, one can always obtain an orthonormal matrix $O$. Now we
show that in two-qubit case, for any $3\times3$ orthonormal matrix
$O$ there always exits $2\times2$ unitary matrix $U$ such that
(\ref{rr}) holds.

For any vector ${\bf{t}}=\{t_{1},t_{2},t_{3}\}$ with unit norm,
define an operator $X\equiv\sum\limits_{i=1}^{3}t_{i}\sigma_{i},$
where $\sigma_{i}$s are Pauli matrices. Given an orthonormal matrix
$O$ one obtains a new operator
$X^{'}\equiv\sum\limits_{i=1}^{3}t_{i}^{'}\sigma_{i}=\sum\limits_{i,j=1}^{3}O_{ij}t_{j}\sigma_{i}$.

$X$ and $X^{'}$ are both hermitian traceless matrices. Their
eigenvalues are given by the norms of the vectors ${\bf{t}}$ and
${\bf{t^\prime}}=\{t_{1}^\prime,t_{2}^\prime,t_{3}^\prime\}$
respectively. As the norms are invariant under orthonormal
transformations $O$, they have the same eigenvalues:
$\pm\sqrt{t_{1}^{2}+t_{2}^{2}+t_{3}^{2}}$. Thus there must be a
unitary matrix $U$ such that $X^{'}=UXU^{\dag}$. Hence the
inequality in the proof of Theorem $\ref{ufef}$ becomes an equality.
The upper bound ($\ref{inequ}$) then becomes exact at this
situation, which is in accord with the result in \cite{grondalski}.
$\hfill\Box$

${\bf{Remark:}}$ The upper bound of FEF (\ref{inequ}) and the
FEF (\ref{corollary}) depend on the correlation
matrices $M(\rho)$ and $M(P_+)$. They can be calculated directly
according to a given set of $SU(d)$ generators $\lambda_{i}$,
$i=1,...,d^2-1$. As an example, for $d=3$, if we choose $\lambda_{1}=\left(%
    \begin{array}{ccc}
      1 & 0 & 0\\
      0 & -1 & 0\\
      0 & 0 & 0\\
    \end{array}%
    \right), \lambda_{2}=\left(%
    \begin{array}{ccc}
      \frac{1}{\sqrt{3}} & 0 & 0\\
      0 & \frac{1}{\sqrt{3}} & 0\\
      0 & 0 & -\frac{2}{\sqrt{3}}\\
    \end{array}%
    \right), \lambda_{3}=\left(%
    \begin{array}{ccc}
      0 & 1 & 0\\
      1 & 0 & 0\\
      0 & 0 & 0\\
    \end{array}%
    \right), \lambda_{4}=\left(%
    \begin{array}{ccc}
      0 & 0 & 1\\
      0 & 0 & 0\\
      1 & 0 & 0\\
    \end{array}%
    \right), \lambda_{5}=\left(%
    \begin{array}{ccc}
      0 & 0 & 0\\
      0 & 0 & 1\\
      0 & 1 & 0\\
    \end{array}%
    \right), \lambda_{6}=\left(%
    \begin{array}{ccc}
      0 & i & 0\\
      -i & 0 & 0\\
      0 & 0 & 0\\
    \end{array}%
    \right), \lambda_{7}=\left(%
    \begin{array}{ccc}
      0 & 0 & i\\
      0 & 0 & 0\\
      -i & 0 & 0\\
    \end{array}%
    \right),$ and $\lambda_{8}=\left(%
    \begin{array}{ccc}
      0 & 0 & 0\\
      0 & 0 & i\\
      0 & -i & 0\\
    \end{array}%
    \right)$, then
we have \be
M(P_{+})=Diag\{\frac{1}{6},\frac{1}{6},\frac{1}{6},\frac{1}{6},\frac{1}{6},
-\frac{1}{6},-\frac{1}{6},-\frac{1}{6}\}.\ee Nevertheless the $FEF$
and its upper bound do not depend on the choice of the $SU(d)$
generators.

The usefulness of the bound depends on detailed states. In the
following we give two new upper bounds which is different from
theorem $\ref{ufef}$. These bounds work for different states.

Let $h$ and $g$ be $n\times n$ matrices such that $h|j>=|(j+1){\rm
mod}\, n>$, $g|j>=\omega^j|j>$, with
$\omega=exp\{\frac{-2i\pi}{n}\}$. We can introduce  $n^2$
linear-independent $n\times n$-matrices $U_{st}=h^{t}g^s$, which
satisfy \be U_{st}U_{s't'}=\omega^{st'-ts'}U_{s't'}U_{st},{\rm
Tr}[U_{st}]=n\delta_{s0}\delta_{t0}. \label{Mat} \ee One can also
check that $\left\{U_{st}\right\}$ satisfy the condition of {\it
bases of the unitary operators} in the sense of \cite{Wer00}, i.e.
\be\label{Wer}
{\rm Tr} [U_{st}U^+_{s't'}]=n \delta_{tt'}\delta_{ss'},~~~
U_{st}U_{st}^+=I_{n\times n},
\ee
where $I_{n\times n}$ is the $n\times n$ identity matrix.
$\left\{U_{st}\right\}$ form a complete basis of $n\times
n$-matrices, namely, for any $n\times n$ matrix $W$, $W$ can be
expressed as \be\label{3e} W=\frac{1}{n}\sum_{s,t}{\rm
Tr}[U_{st}^+W]U_{st}. \ee

From $\{U_{st}\}$, we can introduce the generalized Bell-states,
\be\label{bellbas}
|\Phi_{st}>=(I\otimes
U^*_{st})|\psi_+>=\frac{1}{\sqrt{d}}\sum_{i,j}(U_{st})^*_{ij}|ij>
,~{\rm and }~~|\Phi_{00}>=|\psi_{+}>,
\ee
$|\Phi_{st}>$ are all
maximally entangled states and form a complete orthogonal normalized
basis  of ${\mathcal {H}}_d\otimes{\mathcal {H}}_d$.

\begin{theorem} For any quantum state $\rho\in{\mathcal {H}}_d\otimes{\mathcal
{H}}_d$, the fully entangled fraction defined in $(\ref{def})$
fulfills the following inequality: \be\label{newfef} {\mathcal
{F}}(\rho)\leq\max_j\{\lambda_j\},\ee where $\lambda_j$s are the
eigenvalues of the real part of matrix
$M=\left(%
    \begin{array}{cc}
      T & iT \\
      -iT & T \\
    \end{array}%
    \right)$, $T$ is a $d^{2}\times d^{2}$ matrix with entries
$T_{n,m}=\la\Phi_n|\rho|\Phi_{m}\ra$ and $\Phi_j$ are the maximally
entangled basis states defined in $(\ref{bellbas})$.
\end{theorem}

{\sf[Proof]}\ From $(\ref{3e})$, any $d\times d$ unitary matrix $U$
can be represented by $U=\sum_{k=1}^{d^2}z_{k}U_{k},$ where
$z_{k}=\frac{1}{d}{\rm Tr}[U_k^{\dag}U]$. Define
\be
x_l=\left\{\begin{array}{l}
{\rm Re}[z_l], 1\leq l\leq d^2;\\
{\rm Im}[z_l], d^{2}< l\leq 2d^2
\end{array}\right. {\rm{and \quad}}U^{'}_l=\left\{\begin{array}{l}
U_l, 1\leq l\leq d^2;\\
i*U_l, d^{2}< l\leq 2d^2.
\end{array}\right.
\ee
Then the unitary matrix $U$ can be rewritten as
$U=\sum_{k=1}^{2d^2}z_{k}U^{'}_{k}$.  The necessary
condition for the unitary property of $U$ implies that
$\sum_k x_k^2=1$. Thus we have
\be\label{beinsert} F(\rho)\equiv\la\psi_{+}|(I\otimes U^{\dag})\rho
(I\otimes U)|\psi_{+}\ra=\sum_{m,n=1}^{2d^2}x_m x_n M_{mn},\ee where
$M_{mn}$ is defined in the theorem. One can deduce that
\be\label{hhh} M_{mn}^{*}=M_{nm}\ee from the hermiticity of $\rho$.

Taking into account the constraint with an undetermined
Lagrange multiplier $\lambda$, we have
\be\frac{\partial}{\partial
x_k}\{F(\rho)+\lambda(\sum_l x_l^2-1)\}=0.
\ee
Accounting to
$(\ref{hhh})$ we have the eigenvalue equation \be\label{insert}
\sum_{n=1}^{2d^2}{\rm{Re}}[M_{k,n}]x_n=-\lambda x_k.\ee

Inserting $(\ref{insert})$ into $(\ref{beinsert})$ results in \be
{\mathcal {F}}(\rho)=\max_U F \leq\max_j\{\eta_j\},\ee where
$\eta_j=-\lambda_j$ is the corresponding eigenvalues of the real
part of the matrix $M$. $\hfill\Box$

\begin{figure}[tbp]
\begin{center}
\resizebox{10cm}{!}{\includegraphics{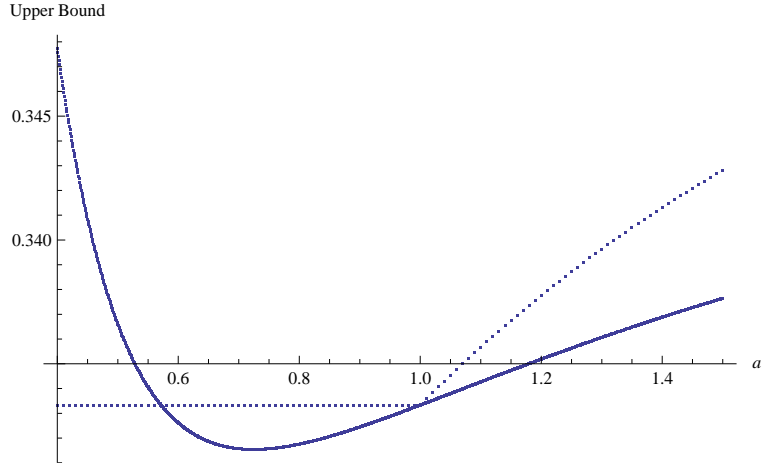}}
\end{center}
\caption{Upper bound of ${\mathcal {F}}(\rho(a))$ from
($\ref{newfef}$) (solid line) and upper bound from
$(\ref{inequ})$(dashed line). \label{fig11}}
\end{figure}

Example: Horodecki gives a very interesting bound entangled state
in  \cite{ho232},
\be\label{exam}\rho(a)=\frac{1}{8a+1}\left(%
    \begin{array}{ccccccccc}
      a & 0 & 0 & 0 & a & 0 & 0 & 0 & a\\
      0 & a & 0 & 0 & 0 & 0 & 0 & 0 & 0\\
      0 & 0 & a & 0 & 0 & 0 & 0 & 0 & 0\\
      0 & 0 & 0 & a & 0 & 0 & 0 & 0 & 0\\
      a & 0 & 0 & 0 & a & 0 & 0 & 0 & a\\
      0 & 0 & 0 & 0 & 0 & a & 0 & 0 & 0\\
      0 & 0 & 0 & 0 & 0 & 0 & \frac{1+a}{2} & 0 & \frac{\sqrt{1-a^{2}}}{2}\\
      0 & 0 & 0 & 0 & 0 & 0 & 0 & a & 0\\
      a & 0 & 0 & 0 & a & 0 & \frac{\sqrt{1-a^{2}}}{2} & 0 & \frac{1+a}{2}\\
    \end{array}%
    \right).\ee One can easily compare the upper bound obtained in
$(\ref{newfef})$ and that in $(\ref{inequ})$. From Fig.
$\ref{fig11}$ we see that for $0\leq a<0.572$, the upper bound in
$(\ref{newfef})$ is larger than that in $(\ref{inequ})$. But for
$0.572<a<1$ the upper bound in $(\ref{newfef})$ is always lower than
that in $(\ref{inequ})$, which means the upper bound
$(\ref{newfef})$ is tighter than $(\ref{inequ})$.

In fact, we can drive another upper bound for FEF which will be very
tight for weakly mixed quantum states.

\begin{theorem}\label{dalidada} For any bipartite quantum state $\rho\in{\mathcal
{H}}_{d}\otimes{\mathcal {H}}_{d}$, the following inequality holds:
\be\label{fefm} {\mathcal {F}}(\rho)\leq
\frac{1}{d}({\rm{Tr}}[\sqrt{\rho_A}])^2,\ee where $\rho_A$ is the
reduced matrix of $\rho$.
\end{theorem}

{\sf[Proof]}\ Note that in \cite{gdp} the authors have obtained the
FEF for pure state $|\psi\ra$,
\be\label{fefp}{\mathcal
{F}}(|\psi\ra)=\frac{1}{d}({\rm{Tr}}[\sqrt{\rho_A^{|\psi\ra}}])^2,\ee
where $\rho_A^{|\psi\ra}$ is the reduced matrix of $|\psi\ra\la\psi|$.

For mixed state $\rho=\sum_ip_i \rho^i$, we have
\begin{eqnarray}\label{33}
{\mathcal {F}}(\rho)&=&\max_{U}\la\psi_{+}|(I\otimes U^{\dag})\rho
(I\otimes U)|\psi_{+}\ra
\leq\sum_i p_i \max_U\la\psi_{+}|(I\otimes U^{\dag})\rho^i
(I\otimes U)|\psi_{+}\ra\nonumber\\
&=&\frac{1}{d}\sum_ip_i({\rm{Tr}}[\sqrt{\rho_A^i}])^2=\frac{1}{d}\sum_i({\rm{Tr}}[\sqrt{p_i\rho_A^i}])^2.
\end{eqnarray}
Let $\lambda_{ij}$ be the real and nonnegative eigenvalues of
the matrix $p_i\rho_A^i$. Recall that for any function
$F=\sum_i(\sum_jx_{ij}^2)^{\frac{1}{2}}$ subjected to the
constraints $z_j=\sum_ix_{ij}$ with $x_{ij}$ being real and
nonnegative, the inequality $\sum_jz_j^2\leq F^2$ holds, from which
it follows that
\begin{eqnarray}
{\mathcal
{F}}(\rho)\leq\frac{1}{d}\sum_i(\sum_j\sqrt{\lambda_{ij}})^2\leq\frac{1}{d}(\sum_j\sqrt{\sum_i\lambda_{ij}})^2
=\frac{1}{d}({\rm{Tr}}[\sqrt{\rho_A}])^2,
\end{eqnarray}
which ends the proof. $\hfill\Box$

\subsection{Fully entangled fraction and concurrence}

The upper bound of $FEF$ has also interesting relations to the
entanglement measure concurrence. As shown in \cite{grondalski}, the
concurrence of a two-qubit quantum state has some kinds of relation
with the optimal teleportation fidelity. For quantum state with high
dimension, we have the similar relation between them too.

\begin{theorem} For any bipartite quantum state $\rho\in{\mathcal
{H}}_{d}\otimes{\mathcal {H}}_{d}$, we have
\be\label{limit} C(\rho)\geq\sqrt{\frac{2d}{d-1}}[{\mathcal
{F}}(\rho)-\frac{1}{d}].\ee
\end{theorem}

{\sf[Proof]}\  In \cite{062330}, the authors show that for any pure
state $|\psi\ra\in{\mathcal {H}}_{A}\otimes{\mathcal {H}}_{B}$, the
following inequality holds: \be C(|\psi\ra)\geq
\sqrt{\frac{2d}{d-1}}(max_{|\phi\ra\in
\varepsilon}|\la\psi|\phi\ra|^{2}-\frac{1}{d}),\ee where
$\varepsilon$ denotes the set of $d \times d$-dimensional
maximally entangled states.

Let $\rho=\sum_{i}p_i|\phi_i\ra\la\phi_i|$ be the optimal
decomposition such that $C(\rho)=\sum_{i}p_i C(|\psi_i\ra)$. We have
\begin{eqnarray*}
C(\rho)&=&\sum_{i}p_i C(|\psi_i\ra)\geq \sum_i p_i
\sqrt{\frac{2d}{d-1}}(max_{|\phi\ra\in
\varepsilon}|\la\psi_{i}|\phi\ra|^{2}-\frac{1}{d}) \\
&&\geq \sqrt{\frac{2d}{d-1}}(max_{|\phi\ra\in \varepsilon}\sum_i
p_i|\la\psi_{i}|\phi\ra|^{2}-\frac{1}{d})\\
&&=\sqrt{\frac{2d}{d-1}}(max_{|\phi\ra\in
\varepsilon}\la\phi|\rho|\phi\ra-\frac{1}{d})=\sqrt{\frac{2d}{d-1}}
({\mathcal{F}}(\rho)-\frac{1}{d}),
\end{eqnarray*} which ends the proof.
$\hfill\Box$

The inequality $(\ref{limit})$ has demonstrated the relation between
the lower bound of concurrence and the fully entangled fraction
(thus the optimal teleportation fidelity), i.e. the fully entangled
fraction of a quantum state $\rho$ is limited by it's concurrence.

We now consider tripartite case. Let $\rho_{ABC}$ be a state of
three-qubit systems denoted by $A$, $B$ and $C$. We study the upper
bound of the $FEF$, ${\cal F}(\rho_{AB})$, between qubits $A$ and
$B$, and its relations to the concurrence under bipartite partition
$AB$ and $C$. For convenience we normalize ${\cal F}(\rho_{AB})$ to
be
\begin{eqnarray}\label{frho}
{\cal {F}}_{N}(\rho_{AB})=\max\{2{\cal {F}}(\rho_{AB})-1,0\}.
\end{eqnarray}
Let $C({\rho_{AB|C}})$ denote the concurrence between subsystems
$AB$ and $C$.

\begin{theorem}\label{fefcon} For any triqubit state $\rho_{ABC}$, ${\cal
{F}}_{N}(\rho_{AB})$ satisfies \be\label{inequformix} {\cal
{F}}_{N}(\rho_{AB})\leq\sqrt{1-C^{2}({\rho_{AB|C}})}. \ee
\end{theorem}

{\sf[Proof]}\ We first consider the case that $\rho_{ABC}$ is pure,
$\rho_{ABC}=|\psi\ra_{ABC}\la\psi|$. By using the Schmidt
decomposition between qubits $A, B$ and $C$, $|\psi\ra_{ABC}$ can be
written as:
\begin{eqnarray}\label{schmidt}
|\psi\ra_{AB|C}=\sum\limits_{i=1}^{2}\eta_{i}|i_{AB}\ra|i_{C}\ra,~~
\eta_{1}^{2}+\eta_{2}^{2}=1,~~~\eta_{1}\geq \eta_{2}
\end{eqnarray}
for some othonormalized bases $|i_{AB}\ra$, $|i_{C}\ra$ of
subsystems $AB$, $C$ respectively. The reduced density matrix
$\rho_{AB}$ has the form
\begin{eqnarray*}
\rho_{AB}&=&{\rm
Tr}_{C}[\rho_{ABC}]=\sum\limits_{i=1}^{2}\eta_{i}^{2} |i_{AB}\ra\la
i_{AB}|=U^{T}\Lambda U^{*},
\end{eqnarray*}
where $\Lambda$ is a $4\times 4$ diagonal matrix with diagonal
elements $\{\eta_{1}^{2}, \eta_{2}^{2}, 0, 0\}$, $U$ is a unitary
matrix and $U^{*}$ denotes the conjugation of $U$.

The $FEF$ of the two-qubit state $\rho_{AB}$ can be calculated by
using formula (\ref{corollary}) or the one in \cite{grondalski}. Let
$$M=\frac{1}{\sqrt{2}}\left(%
    \begin{array}{cccc}
      1 & 0 & 0 & i \\
      0 & i & -1 & 0 \\
      0 & i & 1 & 0 \\
      1 & 0 & 0 & -i \\
    \end{array}%
    \right)
    $$
be the $4\times 4$ matrix constituted by the four Bell bases. The
$FEF$ of $\rho_{AB}$ can be written as
\be\ba{rcl}\label{pppp} \displaystyle{\cal
{F}}(\rho_{AB})&=&\eta_{max}(Re\{M^{\dag}\rho_{AB} M\})
=\frac{1}{2}\eta_{max}(M^{\dag}\rho_{AB} M+M^{T}\rho_{AB}^{*} M^{*}) \\[3mm]\displaystyle
&\leq&\frac{1}{2}[\eta_{max}(M^{\dag}U^{T}\Lambda
U^{*}M)+\eta_{max}(M^{T}U^{\dag}\Lambda UM^{*})] =\eta_{1}^{2} \ea
\ee
where $\eta_{max}(X)$ stands for the maximal eigenvalues of the
matrix $X$.

For pure state ($\ref{schmidt}$) in bipartite partition $AB$ and
$C$, we have
\begin{eqnarray}\label{con}
C(|\psi\ra_{AB|C})=\sqrt{2(1-{\rm
Tr}[\rho_{AB}^{2}])}=2\eta_{1}\eta_{2}.
\end{eqnarray}
From (\ref{frho}), (\ref{pppp}) and (\ref{con}) we get
\begin{eqnarray}\label{inequality}
{\cal {F}}_{N}(\rho_{AB})\leq\sqrt{1-C^{2}(|\psi\ra_{AB|C})}.
\end{eqnarray}

We now prove that the above inequality ($\ref{inequality}$) also
holds for mixed state $\rho_{ABC}$. Let
$\rho_{ABC}=\sum\limits_{i}p_{i}|\psi_{i}\ra_{ABC}\la\psi_{i}|$ be
the optimal decomposition of $\rho_{ABC}$ such that
$C(\rho_{AB|C})=\sum\limits_{i}p_{i}C(|\psi_{i}\ra)_{AB|C}$. We have
\begin{eqnarray*}
{\cal {F}}_{N}(\rho_{AB}) &\leq&\sum\limits_{i}p_{i}{\cal
{F}}_{N}({\rho^{i}_{AB}})
\leq\sum\limits_{i}p_{i}\sqrt{1-C^{2}({\rho^{i}_{AB|C}})}\\
&\leq&\sqrt{1-\sum\limits_{i}p_{i}C^{2}({\rho^{i}_{AB|C}})}
\leq\sqrt{1-C^{2}({\rho_{AB|C}})},
\end{eqnarray*}
where $\rho^{i}_{AB|C}=|\psi_{i}\ra_{ABC}\la\psi_{i}|$ and
$\rho^{i}_{AB}={\rm Tr}_{C}[\rho^{i}_{AB|C}]$. $\hfill\Box$

\begin{figure}[tbp]
\begin{center}
\resizebox{10cm}{!}{\includegraphics{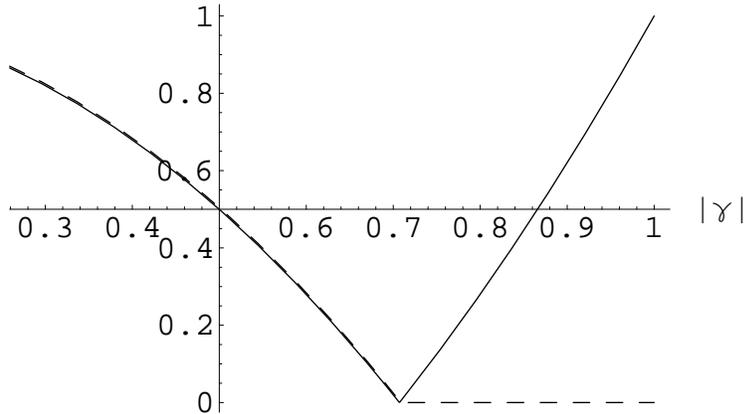}}
\end{center}
\caption{${\cal {F}}_{N}(\rho_{AB}^{W^{'}})$ (dashed line) and Upper
bound $\sqrt{1-C^{2}(|W^{'}\ra_{AB|C})}$ (solid line) of state
$|W^{'}\ra_{AB|C}$ at $|\alpha|=|\beta|$. \label{fig22}}
\end{figure}

From Theorem $\ref{fefcon}$ we see that the $FEF$ of quibts $A$ and
$B$ are bounded by the concurrence between qubits $A$, $B$ and qubit
$C$. The upper bound of $FEF$ for $\rho_{AB}$ decreases when the
entanglement between qubits $A, B$ and $C$  increases. As an
example, we consider the generalized W state defined by
$|W^{'}\ra=\alpha|100\ra+\beta|010\ra+\gamma|001\ra$,
$|\alpha|^{2}+|\beta|^{2}+|\gamma|^{2}=1$. The reduced density
matrix is given by
$$\rho_{AB}^{W^{'}}=\left(%
    \begin{array}{cccc}
      |\gamma|^{2} & 0 & 0 & 0 \\
      0 & |\beta|^{2} & \alpha^{*}\beta & 0 \\
      0 & \alpha\beta^{*} & |\alpha|^{2} & 0 \\
      0 & 0 & 0 & 0 \\
    \end{array}%
    \right).
    $$
The $FEF$ of $\rho_{AB}^{W^{'}}$ is given by
$$
{\cal
{F}}_{N}(\rho_{AB}^{W^{'}})=-\frac{1}{2}+2|\alpha||\beta|+\frac{1}{2}||\alpha|^{2}+|\beta|^{2}-|\gamma|^{2}|.
$$
While the concurrence of $|W^{'}\ra$ has the from
$C_{AB|C}(|W^{'}\ra)=2|\gamma|\sqrt{|\alpha|^{2}+|\beta|^{2}}$. We
see that (\ref{inequformix}) always holds. In particular for
$|\alpha|=|\beta|$ and $|\gamma|\leq \frac{\sqrt{2}}{2}$, the
inequality (\ref{inequformix}) is saturated (see Fig.
$\ref{fig22}$).

\subsection{Improvement of entanglement distillation protocol}

The upper bound can give rise to not only an estimation of the
fidelity in quantum information processing such as teleportation,
but also an interesting application in entanglement distillation of
quantum states.  In \cite{gdp} a generalized distillation protocol
has been presented. It is shown that a quantum state $\rho$
violating the reduction criterion can always be distilled. For such
states if their single fraction of entanglement
$F(\rho)=\la\psi_{+}|\rho|\psi_{+}\ra$ is greater than
$\frac{1}{d}$, then one can distill these states directly by using
the generalized distillation protocol. If the $FEF$ (the largest
value of single fraction of entanglement under local unitary
transformations) is less than or equal to $\frac{1}{d}$, then a
proper filtering operation has to be used at first to transform
$\rho$ to another state $\rho^{'}$ so that
$F(\rho^{'})>\frac{1}{d}$. For $d=2$, one can compute $FEF$
analytically according to the corollary. For $d \geq 3$ our upper
bound ($\ref{inequ}$) can supply a necessary condition in the
distillation:

\begin{figure}[h]
\begin{center}
\resizebox{10cm}{!}{\includegraphics{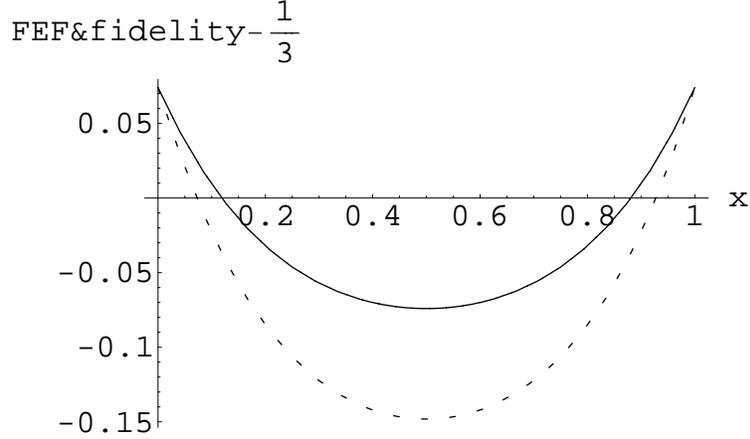}}
\end{center}
\caption{Upper bound of ${\cal {F}}(\rho)-\frac{1}{3}$ from
($\ref{inequ}$) (solid line) and fidelity $F(\rho)-\frac{1}{3}$
(dashed line). \label{figlast}}
\end{figure}

\begin{theorem} For an entangled state $\rho\in
{\cal{H}}\otimes{\cal {H}}$ violating the reduction criterion, if
the upper bound ($\ref{inequ}$) is less than or equal to
$\frac{1}{d}$, then the filtering operation has to be applied before
using the generalized distillation protocol. \end{theorem}

As an example we consider a $3\times 3$ state
\begin{eqnarray}
\rho=\frac{8}{9}\sigma+\frac{1}{9}|\psi_{+}\ra\la\psi_{+}|,
\end{eqnarray}
where $\sigma=(x|0\ra\la 0|+(1-x)|1\ra\la 1|)\otimes(x|0\ra\la
0|+(1-x)|1\ra\la 1|)$. It is direct to verify that $\rho$ violates
the reduction criterion for $0\leq x\leq 1$, as $(\rho_{1}\otimes
I)-\rho$ has a negative eigenvalue $-\frac{2}{27}$. Therefore the
state is distillable. From Fig. $\ref{figlast}$ we see that for
$0\leq x<0.0722$ and $0.9278<x\leq 1$, the fidelity is already
greater than $\frac{1}{3}$, thus the generalized distillation
protocol can be applied without the filtering operation. However for
$0.1188\leq x\leq 0.8811$, even the upper bound of the fully
entangled fraction is less than or equal to $\frac{1}{3}$, hence the
filtering operation has to be applied first, before using the
generalized distillation protocol.

Moreover, the lower bounds of concurrence can be also used to
study the distillability of quantum states.
Based on the positive partial transpose
(PPT) criterion, a necessary and sufficient condition for the
distillability was proposed in \cite{ho1}, which is not
operational in general. An alternative distillability
criterion based on the bound $\tau_2$ in (\ref{ou}) can be obtained to improve the operationality.

\begin{theorem}\label{ou1}
A bipartite quantum state $\rho$ is distillable
if and only if $\tau_2(\rho^{\otimes N})>0$ for some number $N$.
\end{theorem}

{\sf[Proof]}\ It was shown in \cite{ho1} that a density matrix $\rho$
id distillable if and only if there are some
projectors $P$, $Q$ that map high dimensional spaces to two-dimensional ones
and some number $N$ such that the state $P\otimes Q\rho^{\otimes N}P\otimes Q$ is
entangled \cite{ho1}. Thus if $\tau_2(\rho^{\otimes N})>0$, there exists
one submatrix of matrix $\rho^{\otimes N}$, similar to Eq. (\ref{c11}), which has nonzero
$\tau_2$ and is entangled in a $2\otimes 2$ space, hence $\rho$ is distillable.
$\hfill$$\Box$

\begin{corollary}\label{ouc1}
The lower bound $\tau_2(\rho)>0$ is a sufficient
condition for the distillability of any bipartite state $\rho$.
\end{corollary}

\begin{corollary}\label{ouc2}
The lower bound $\tau_2(\rho)=0$ is a necessary
condition for separability of any bipartite state $\rho$.
\end{corollary}

\emph{Remark:} Corollary \ref{ouc1} directly follows from Theorem \ref{ou1}
and this case is referred to as one-distillable \cite{ou21}.
The problem of whether non-PPT (NPPT) nondistillable states exist is studied
numerically in \cite{ou21,ou20}. By using Theorem \ref{ou1}, although it
seems impossible to solve the problem completely,
it is easy to judge the distillability of a state under condition
that it is one-distillable.

The lower bound $\tau_2$, PPT criterion,
separability and distillability for any bipartite quantum state
$\rho$ have the following relations: if $\tau_2(\rho)>0$, $\rho$ is entangled.
If $\rho$ is separable, it is PPT. If
$\tau_2(\rho)>0$, $\rho$ is distillable. If $\rho$ is distillable, it
is NPPT. From the last two propositions it follows that if $\rho$ is
PPT, $\tau_2(\rho)=0$, i.e., if $\tau_2(\rho)>0$, $\rho$ is NPPT.

\begin{theorem}\label{ou2} For any pure tripartite state
$|\phi\rangle_{ABC}$ in arbitrary $d\otimes d \otimes d$
dimensional spaces, bound $\tau_2$ satisfies
\begin{equation}\label{c18}
\tau_2(\rho_{AB})+\tau_2(\rho_{AC})\leq\tau_2(\rho_{A:BC}),
\end{equation}
where $\rho_{AB}={Tr}_{C}(|\phi\rangle_{ABC}\langle\phi|)$,
$\rho_{AC}={Tr}_{B}(|\phi\rangle_{ABC}\langle\phi|)$, and
$\rho_{A:BC}={Tr}_{BC}(|\phi\rangle_{ABC}\langle\phi|)$.
\end{theorem}

{\sf[Proof]}\  Since ${C}_{mn}^2\leq
\left(\lambda^{(1)}_{mn}\right)^2\leq\sum^{4}_{i=1}\left(\lambda^{(i)}_{mn}\right)^2=
\texttt{Tr}(\rho\widetilde{\rho}_{mn})$, one can derive the inequality:
\be\label{rhs}
  \tau(\rho_{AB})+\tau(\rho_{AC})\leq  \sum_{l,k}^{D}
  {Tr}\left[\rho_{AB}(\widetilde{\rho}_{AB})_{lk}\right]
 +\sum_{p,q}^{D}{Tr}\left[\rho_{AC}(\widetilde{\rho}_{AC})_{pq}\right],
 \ee
where $D=d(d-1)/2$. Note that
$\sum_{lk}\texttt{Tr}\left[\rho_{AB}(\widetilde{\rho}_{AB})_{lk}\right]\leq
1-{Tr}\rho_A^2- {Tr}\rho_B^2+{Tr}\rho_C^2$ and
$\sum_{pq}{Tr}\left[\rho_{AC}(\widetilde{\rho}_{AC})_{pq}\right]\leq
1-{Tr}\rho_A^2+{Tr}\rho_B^2-{Tr}\rho_C^2$, where $l,pk,q,=1,...,D$.
By using the similar analysis in \cite{ckw} one has that the
right-hand side of Eq. (\ref{rhs}) is equal to
$2(1-{Tr}\rho^2_A)={C}^2(\rho_{A:BC})$. Taking into account that
$\tau_2(\rho_{A:BC})=\mathcal {C}^2(\rho_{A:BC})$ for a pure state,
one obtains the inequality (\ref{c18}). $\hfill$$\Box$

Generally for any pure multipartite quantum state $\rho_{AB_1B_2...b_n}$,
one has the following monogamy inequality:
$$
\tau_2(\rho_{AB_1})+\tau_2(\rho_{AB_2})+...+\tau_2(\rho_{AB_n})\leq\tau_2(\rho_{A:B_1B_2...B_n}).
$$

\section{Summary and Conclusion}

We have introduced some recent results on three aspects in quantum
information theory. The first one is the separability
of quantum states. New criteria to detect more entanglements
have been discussed. The normal form of quantum states have been also studied, which helps in
investigating the separability of quantum states.
Moreover, since many kinds of quantum states can be transformed into
the same normal forms, quantum states can be classified in terms of the normal forms.
For the well known entanglement measure concurrence, we have discussed the tight lower and upper
bounds. It turns out that although one can not distill
a singlet from many pairs of bound entangled states, the concurrence
and tangle of two entangled quantum states are always larger than
that of one, even both two entangled quantum states are bound
entangled. Related to the optimal
teleportation fidelity, upper bounds for
the fully entangled fraction have been studied, which can be used to
improve the distillation protocol. Interesting
relations between fully entangled fraction and concurrence
have been also introduced. All these related problems in the theory of quantum
entanglement have not been completely solved yet.
Many problems remain open concerning the physical properties and mathematical structures
of quantum entanglement, and the applications of entangled states in information processing.

\end{document}